\documentclass[preprint2,appendix]{emulateapj}
\usepackage{color}
\usepackage{graphicx}
\usepackage{longtable}
\usepackage{lineno}
\usepackage{draftcopy}
\usepackage{threeparttable}
\usepackage{rotating}
\usepackage{multirow}
\slugcomment{version \today: fm}
\shorttitle{}
\shortauthors{}

\begin{document}
\title{Spatial Structures in the Globular Cluster Distribution of the Ten Brightest Virgo Galaxies}
\author{R.~D'Abrusco\altaffilmark{1}, G.~Fabbiano\altaffilmark{1}, A.~Zezas\altaffilmark{1,2,3}}

\altaffiltext{1}{Harvard-Smithsonian Astrophysical Observatory, 60 Garden Street, Cambridge, MA 02138, USA}

%\author{R.~D'Abrusco\altaffilmark{1}, G.~Fabbiano\altaffilmark{1}, J.~Strader\altaffilmark{2}, 
%A.~Zezas\altaffilmark{1,3,4}, S.~Mineo\altaffilmark{1}, T.~Fragos\altaffilmark{1}, P.~Bonfini\altaffilmark{3}, 
%B.~Luo\altaffilmark{5}, D.-W.~Kim\altaffilmark{1}, A.~King\altaffilmark{6}}
%
%\altaffiltext{1}{Harvard-Smithsonian Astrophysical Observatory, 60 Garden Street, Cambridge, MA 02138, USA}
%\altaffiltext{2}{Department of Astronomy, Michigan State University, 567 Wilson Road, East Lansing, 
%MI  48824-2320, USA}
%\altaffiltext{3}{Physics Department and Institute of Theoretical and Computational Physics, 
%University of Crete, 71003 Heraklion, Crete, Greece}
%\altaffiltext{4}{Foundation of Research and Technology, 71003, Heraklion, Crete, Greece}
\altaffiltext{2}{Physics Department \& Institute of Theoretical \& Computational Physics, 
University of Crete, 71003 Heraklion, Crete, Greece} 
\altaffiltext{3}{Foundation for Research and Technology-Hellas, 71110 Heraklion, Crete, Greece}

\begin{abstract}

We report the discovery of significant localized structures in the projected two-dimensional 
(2D) spatial distributions of the Globular Cluster (GC) systems of the ten brightest galaxies in the 
Virgo Cluster. We use catalogs of GCs extracted from 
the HST ACS Virgo Cluster Survey (ACSVCS) imaging data, complemented, when available, 
by additional archival ACS data. These structures have projected sizes ranging from $\sim\!5\arcsec$ 
to few arc-minutes ($\sim\!1$ to $\sim\!25$ kpc). Their morphologies range from localized, circular,
to coherent, complex shapes resembling arcs and 
streams. The largest structures are preferentially aligned with the major axis of the host galaxy. 
A few relatively smaller structures follow the minor axis. Differences in the shape and significance
of the GC structures can be noticed by investigating the spatial distribution of GCs grouped by
color and luminosity. The largest coherent
GC structures are located in low-density regions within the Virgo cluster. This trend is more evident in 
the red GC population, believed to form in mergers involving late-type galaxies. We suggest that GC 
over-densities may be driven by either 
accretion of satellite galaxies, major dissipationless mergers or wet dissipation mergers. 
We discuss caveats to these scenarios, and estimate the masses of the potential progenitors
galaxies. These masses range in the interval $10^{8.5}\!-\!10^{9.5}\!M_{\sun}$, larger than those of 
the Local Group dwarf galaxies.

\end{abstract}


\section{Introduction}
\label{sec:intro}

The large scale distribution of GCs has been studied in depth in several elliptical galaxies. Radially, 
GCs are known to
extend to larger radii than the diffuse stellar light of the galaxies~\citep{harris1986}, with the most
luminous galaxies having relatively more centrally concentrated GCs distributions~\citep{ashman1998}. 
The radial profiles of red (metal-rich) and blue (metal-poor) GCs differ, with red GCs matching more
closely the light profile of the stellar spheroid - except for central flattened cores in the GC 
distribution - and blue GCs presenting a flatter extended distribution~\citep{rhode2001,dirsch2003,
bassino2006,mineo2014}. Studies of the 2D distributions of GC systems show that the ellipticity and
position angles are consistent, on average, with those of the stellar spheroid of the host galaxy
(e.g. NGC4471 in~\citealt{rhode2001}, NGC1399 in~\citealt{dirsch2003,bassino2006}, 
NGC3379, NGC4406 and NGC4594 in \citealt{rhode2004}, NGC4636 in~\citealt{dirsch2005}, multiple 
galaxies in~\citealt{hargis2012}, NGC3585 and NGC5812 in~\citealt{lane2013}, NGC4621 
in~\citealt{bonfini2012} and~\citealt{dabrusco2013}). However, more recent work using the 
large sample of ACSVCS 
galaxies~\citep{wang2013} has uncovered highly anisotropic azimuthal distributions of both red and 
blue GCs; these GCs are preferentially aligned with the major axis in host galaxies with significant 
elongation and intermediate to high luminosity. 

These characteristics of the GC systems provide insight on the evolution of the host galaxy, 
as well as on 
the formation of GCs. The relative importance of dissipative merging or of general galaxy
collapse has been debated in the literature, mostly based on analytical models and observational 
inferences (\citealt{searle1978}; see~\citealt{ashman1992} and references 
therein;~\citealt{forbes1997};~\citealt{bekki1997}). Recently,  N-body smoothed particle 
hydrodynamics simulations~\citep{bekki1998,bekki2002} have reproduced the general radial 
spatial distributions and metallicity properties of 
red and blue GC systems (but not the observed ratio of red and blue GCs)
under the assumption that the elliptical galaxy was the result of major 
dissipative 
merging, as long as the merging event happened early enough in the life of the galaxy. 
Tidal stripping and GC accretion in interacting elliptical galaxies could help 
reconcile the simulations with the data~\citep{bekki2003}. Dissipationless merging of spirals could also give
rise to these GC systems, but should retain kinematic signatures~\citep{bekki2005}.

While the results discussed above still do not provide a unique scenario of GC and 
galaxy formation, it is clear that both major merging (dissipative and more likely dissipationless) and 
tidal interactions play important roles. More recently, localized 2D 
inhomogeneities and streams have been 
found in GC systems of elliptical galaxies, emphasizing the role of minor merging and companion 
galaxy tidal disruption and 
accretion. These results include a spatially extended spiral-arm shaped over-density in the spatial 
distribution of GCs in NGC4261~\citep{bonfini2012,dabrusco2013}, and more localized anisotropies in NGC4649 
and NGC4278~\citep{dabrusco2014a,dabrusco2014b}. These features were observed within the 
$D_{25}$ isophotal contour 
of the host galaxy (1.3-4.5 $r_{e}$), using uniform and complete GC samples 
from Hubble Space Telescope (HST) observations. At larger galactocentric radii, streams 
in the GC distributions have been observed with ground-based coverage in M87~\citep{romanowsky2012}, 
NGC4365~\citep{blom2012b} and NGC4278~\citep{dabrusco2014b}.  

In this fourth paper in our series~\citep{dabrusco2013,dabrusco2014a,dabrusco2014b}, 
we investigate how common 
these 2D anisotropies and streams of GC may be, by studying the 2D GC distributions of 
the ten brightest galaxies in the 
Virgo cluster observed in the ACSVCS survey (Section~\ref{sec:data}). This sample 
provides a unique catalog of GCs 
in the central regions of massive Virgo early-type galaxies. 
As in our previous works, we analyze the 2D distribution of GC systems in elliptical galaxies with 
a K-Nearest-Neighbor (KNN) technique supplemented by Monte Carlo simulations.

The Virgo cluster is close enough to permit the detection of GCs 
with HST imaging, down to relatively low flux levels with limited 
integration times. While the dense central environment of Virgo may discourage accretion of 
companions, Virgo provides a range of local densities
for its galaxies. Even in its densest environment there is observational evidence of accretion of a 
companion galaxy, suggested by the kinematically
detected GC shell in M87~\citep{romanowsky2012}. With our sample, we can therefore begin to investigate
{\it observationally} the past accretion history of galaxies in a range of environments, 
using the GC population as a potential marker of these events. Moreover, our method~\citep{dabrusco2013} 
is more accurate for larger samples of GCs, such as those of the more
luminous elliptical galaxies of the Virgo cluster~\citep{harris1981,brodie2006,peng2008,georgiev2010}.

The paper is structured as follows: in Section~\ref{sec:data} and Section~\ref{sec:method} we
describe the data and the method. In Section~\ref{sec:results} 
we present the results of the analysis of the spatial distribution of GCs of each individual 
Virgo galaxy. In Section~\ref{sec:discussion} we discuss the 
results obtained, focusing on the spatial distribution of the GCs structures,
and we infer the properties of accreted companion galaxies that could
have originated the observed GCs over-densities. We summarize our results 
in Section~\ref{sec:conclusions}. We have used galaxy distances from~\cite{blakeslee2009}.

\section{The Data}
\label{sec:data}

We used data from the observations of the ten brightest galaxies in the HST Advanced 
Camera for Surveys (ACS) Virgo Cluster Survey~\citep[ACSVCS][]{cote2004}, which observed 
the central regions of the 100 brightest early-type galaxies in the Virgo cluster, although with 
incomplete spatial coverage.
For two galaxies (NGC4649 and NGC4365) we were able to use additional 
HST data to cover the full area of the galaxy.

The ACSVCS galaxies were observed in two bands, with the F475W and F850LP 
HST ACS filters, corresponding to the $g$ and $z$ Sloan Digital Sky Survey filters. Seven of the
galaxies in our sample were covered with a single central ACS pointing; 
the NGC4472 observations 
were arranged in a triangle-shaped three-pointing mosaic reaching the $D_{25}$ elliptical
isophote along the major axis. NGC4649 and NGC4365 were covered with a single pointing in the 
ACSVCS, but subsequent observations produced full ACS coverage~\citep{strader2012,blom2012b}.
The GC catalogs for our work were extracted from~\cite{jordan2009}, which includes
12763 {\it bona fide} GC candidates, identified on the basis of their photometric 
properties and spatial extent~\citep{jordan2004}. 

The number of GCs in the ACSVCS catalog associated to the ten brightest Virgo 
galaxies is 7053, representing $\sim\!55\%$ of the total~\cite{jordan2009} catalog. This number
represents the fraction of the GC population of these galaxies that could be detected in ACSVCS observations.
The basic information about the ten brightest galaxies in the ACSVCS list of targets
are summarized in Table~1. Visual magnitudes $B_{T}$, numbers of GC candidates 
and optical morphology are all based on ACSVCS 
data~\citep[see][for $B_{T}$ magnitudes, GC numbers and morphology, 
respectively]{cote2004,jordan2009,ferrarese2006}. The details of the additional data used for NGC4649 
and NGC4365 are given below.

\underline{NGC4649}: a joint {\it Chandra}-HST program (P.I. Fabbiano) completed the 
single central ACSVCS pointing with a mosaic of five additional ACS pointings 
in the same F475W and F850LP filters, with comparable integration times. The final mosaic
covers the whole area of the galaxy comprised within the $D_{25}$ ($\sim\!4.5\ r_{e}$), 
with the exclusion of the region occupied by the companion spiral galaxy NGC4647 (i.e. excluding $\sim15\%$ 
of the NGC4649 area). From these data,~\cite{strader2012} extracted a catalog of 1516 GC candidates 
that was used to characterize 
the spatial distribution of the GCs in NGC4649~\citep{dabrusco2014a}. 

\underline{NGC4365}:~\cite{blom2012b} used six separate ACS pointings in the same F475W and 
F850LP (P.I. Sivakoff) 
publicly available on the Hubble Legacy Archive to complement the single central ACSVCS
pointing, and produced a catalog of 2163 GCs covering the entire galaxy out to $\sim\!1.3$ $D_{25}$
$(\sim\!3$ $r_{e}$). The magnitudes and sizes of the
GCs in the central pointing measured by~\cite{blom2012b} are compatible with the same properties 
from~\cite{jordan2009}. Here we use the catalog of GCs extracted by~\cite{blom2012b}. 

\begin{table*}
	\centering	
	\resizebox{\textwidth}{!}{	
	\begin{threeparttable}	
	\caption{Properties of the ten brightest Virgo cluster galaxies.}
	\begin{tabular}{lcccccccccccccc}
	\tableline
	Galaxy			&	$B_{T}$ 	& 	$D_{L}$ [Mpc]	&	\% $D_{25}$ area 			& 	$N_{\mathrm{Tot}}$ 	&	$[N_{\mathrm{red}},N_{\mathrm{blue}}]$ 	& 	
						$[N_{\mathrm{HighL}},N_{\mathrm{LowL}}]$	 &	Optical	&	r$_{\mathrm{e}}$ $[\arcsec]$	&  $D_{25}$	[r$_{\mathrm{e}}$]	&	$\Delta$(R.A.)	& $\Delta$(Dec.)	&	
						$(g\!-\!z)^{\mathrm{thresh}}$	&	$M_{g}^{\mathrm{thresh}}$	& $M_{\mathrm{lim}}(GCs)$ \\
		 			&			&				& covered			&					&				& 			&   	morphology		& 			&		&	 $[^{\circ}]$	& $[^{\circ}]$	&	 [mag]	& 
					 [Mag]		&	 [Mag]	\\
					&	(a)		&		(b)		&			(c)		&			(d)		&		(e)		&	(f)		&	(g)			&	(h)		&	(i)	&	(l)	&	(m) 	&	(n)	& (o)		& (p)	\\
	\tableline
	NGC4472	 (M49)	&	9.31		& 		16.2		&	$\sim\!35\%$		&	1206				&	$[708,498]$	&	$[751,455]$	& 	E2/S0\_1\_(2)	&	209.15	&	2.9	&	$\sim0.004^{\circ}$(14.7$\arcsec$)& $\sim0.004^{\circ}$(13.4$\arcsec$)		&		1.14		&	 -8.10			&	[-6.4, -7.3]			\\
	NGC4486	 (M87)	&	9.58		& 		16.8		&	$\sim\!40\%$		&	1745				&	$[1006,739]$	&	$[698,1047]$	&	E0			&	171.71	&	2.9	&	$\sim0.002^{\circ}$(6.1$\arcsec$)& $\sim0.002^{\circ}$(7.2$\arcsec$)		&		1.21		&	-8.11				&	 [-6.4, -7.3]				\\
	NGC4649	 (M60)$^{1}$&	9.81		& 		16.6		&	$\sim\!40\%$($\sim\!85\%)$	&	807(1603)	&	$[841,762]$	&	$[1136,467]$	&      S0\_1\_(2)		&	99.34	&	4.5	&	$\sim0.005^{\circ}$(18.6$\arcsec$)& $\sim0.004^{\circ}$(15.7$\arcsec$)		&		1.18		&	-8.09			&	{ [-6.3, -7.2]}				\\
	NGC4406	 (M86)	&	10.06	& 		16.4		&	$\sim\!40\%$		&	367				&	$[140,227]$	&	$[218,149]$	&	{ S0\_1\_(3)/E3}&	411.84	&	1.3	&	$\sim0.005^{\circ}$(18.6$\arcsec$)& $\sim0.004^{\circ}$(15.7$\arcsec$)		&		1.15		&	{-8.26}			&	{ [-6.5, -7.4]}				\\
	NGC4382	 (M85)	&	10.09	& 		17.1		&	$\sim\!45\%$		&	507				&	$[283,224]$	&	$[241,266]$	&	{ S0\_1\_(3) pec} &	170.82	&	2.5	&	$\sim0.002^{\circ}$(6.7$\arcsec$)	& $\sim0.002^{\circ}$(5.8$\arcsec$)		&		1.14		&	{-8.26}			&	{ [-6.5, -7.4]}				\\
	NGC4374	 (M84)	&	10.26	& 		17.0		&	$\sim\!45\%$		&	506				&	$[251,255]$	&	$[296,210]$	&	E1			&	149.73	&	2.6	&	$\sim0.002^{\circ}$(7.1$\arcsec$)	& $\sim0.002^{\circ}$(6.4$\arcsec$)		&		1.12		&	{-8.36}			&	{ [-6.6, -7.5]}				\\
	NGC4526 		& 	10.61	& 		15.3		&	$\sim\!75\%$		&	244				&      	$[134,110]$	&	$[124,120]$	&	S0\_3\_(6)		&		-	&	-	&	$\sim0.002^{\circ}$(6.9$\arcsec$)& $\sim0.002^{\circ}$(6.1$\arcsec$)		&		1.10		&	{-8.14}			&	{ [-6.4, -7.3]}				\\
	NGC4365$^{2}$	&	10.51	& 		22.0		&	$\sim\!45\%$($100\%$)	&	907(2163)		&	$[1149,1014]$	&	$[883,1280]$	&	E3			&	116.14	&	3.6	&	$\sim0.005^{\circ}$(17.6$\arcsec$)& $\sim0.005^{\circ}$(18.9$\arcsec$)		&		1.16		&	{-8.82}			&	{ [-7.1, -8.0]}				\\
	NGC4621	(M59) 	&	10.76	& 		15.7		&	$\sim\!75\%$		&	308				&	$[158,150]$	&	$[184,124]$	&	{ E4}		&	106.75	&	3	&	$\sim0.002^{\circ}$(7$\arcsec$)& 
$\sim0.002^{\circ}$(6.3$\arcsec$)		&		1.23		&	{-7.87}			&	{ [-6.1, -7.0]}				\\
	NGC4552	(M89)	&	10.78	& 		15.7		&	$\sim\!60\%$		&	456				&	$[204,252]$	&	$[213,243]$	&	{ S0\_1\_(0)} 	&	85.93	&	3.6	&	$\sim0.002^{\circ}$(7.4$\arcsec$)& $\sim0.002^{\circ}$(6.4$\arcsec$)		&		1.25		&	{-8.02}			&	{ [-6.3, -7.2]}				\\
	\tableline
	\end{tabular}
	\begin{tablenotes}[para]
 Ê Ê Ê Ê Ê	\item {Column description:}\\
 Ê Ê Ê Ê Ê	\item[a] $B_{T}$ magnitude from~\cite{cote2004}.\\
 Ê Ê Ê Ê Ê	\item[b] Luminosity distance of the galaxy (from NED).\\ 
		\item[c] fraction of the $D_{25}$ area of each galaxy covered by the ACSVCS observations.\\
		\item[d] total number of GC candidates for each galaxy according to the catalog of~\cite{jordan2009}\\
		\item[e] number of GCs in the red/blue classes\\
		\item[f] number of GCs in the high-L/low-L classes\\
		\item[g] morphological classification of the host galaxy from~\cite{ferrarese2006}\\
		\item[h] effective radius $r_{e}$ of the host galaxy from~\cite{ferrarese2006}\\	
		\item[i] $D_{25}$ expressed in units of the effective radius $r_{e}$\\ 
		\item[l] size of pixels in the residual maps along the R.A. axis\\
		\item[m] size of the pixels in the residual maps along the Dec. axis\\	
		\item[n] threshold $g-z$ color used to define the red and blue GC classes\\			
		\item[o] threshold $M_{g}$ absolute magnitude used to separate high-L and low-L GC classes\\		
		\item[p] Limiting absolute magnitude of the GC catalog in the $g$ filter, based on the completeness 
		estimated by~\cite{jordan2007} using simulations. The limiting magnitudes are shown for two values of 
		the half-light radii $r_{h}\!=\!0.0128\arcsec$ and $r_{h}\!=\!0.1283\arcsec$, assuming an 
		intermediate background level ($b_{g}\!=\!0.4023$ mag/arcsec$^{2}$).		
		
		\item[$^{1}$] The fraction of area covered by observations and the number of GCs in 
		brackets have been obtained from the imaging data and GCs catalog presented 
		by~\cite{strader2012}, consisting in the central ACSVCS pointing plus five complementary 
		HST pointings of the outskirts of the galaxy obtained through a joint {\it Chandra}-HST proposal 
		(P.I. Fabbiano).\\
		\item[$^{2}$] The fraction of area covered by the observations and the number of GCs in 
		brackets have been obtained from the imaging data and GCs catalog used by~\cite{blom2012b}, 
		consisting in the central ACSVCS pointing plus six complementary HST pointings of the outskirts of 
		the galaxy (P.I. Sivakoff).\\
	\end{tablenotes}	
	\end{threeparttable}	
	}	
	\label{tab:galaxies}
\end{table*}

For each galaxy in Table~1, we investigate the spatial distribution of the entire sample 
of GCs for each galaxy and of color and luminosity classes, following the procedure outlined 
by~\cite{dabrusco2013}. The red and blue GC classes were determined by 
fitting the histogram of the $g\!-\!z$ color distribution with a double Gaussian model, 
and choosing the threshold corresponding to the 50\% probability for a 
GC to belong to the red or blue class. Although for some of the galaxies the color histograms
are better fit by a single Gaussian model, we adopted the color threshold based on the 
double-Gaussian fit to define the color classes in all cases. 

For our analysis, we have also considered a split of the GC samples in luminosity using, 
for all galaxies, the fixed magnitude threshold $g\!\leq\!23$ mag, corresponding to an 
average absolute magnitude $<\!M_{g}^{\mathrm{thresh}}$ ranging between $\sim\!-7.9$ and
$\sim\!-8.3$ Mag (see Table~\ref{tab:galaxies} for the absolute magnitude threshold for each galaxy). 
These luminosity thresholds produce 
high-L and low-L GC subsamples of similar size for all the galaxies in our sample. While setting the
brightness thresholds as the turnoff luminosity on the GC luminosity function 
may be more physically meaningful, as it would correspond to a mass threshold, 
that choice would produce highly unbalanced high-L and low-L classes, making our analysis 
statistically unreliable. The definitions used to define color and luminosity classes 
produce somewhat redundant samples. For instance, the overlap between red and high-L 
classes varies between 56\% to 87\% according to the host galaxy. For this reason, similarities
in the residual maps of these classes are to be expected. We present the results for both 
color and luminosity classes for the sake of completeness.

\section{Method}
\label{sec:method}

Following~\cite{dabrusco2013}~\citep[see also][]{dabrusco2014a,dabrusco2014b}, 
for each sample of GCs, we generated density maps using the K-Nearest 
Neighbor (KNN) method of~\cite{dressler1980}. These maps are based on the {\it local} 
distribution of GCs, i.e. on the distance of the $K$-th closest GCs from the points where 
the density is estimated. The density is evaluated in each knot of a regular grid covering 
the region of the sky where GCs are found, and is given by: 

\begin{equation}
	D_{K}\!=\!\frac{K}{A_{D}(d_{K})} 
	\label{eq:knn}
\end{equation}

where $K$ is the index of the nearest neighbor used to calculate the density, 
and $A_{D}(d_{K})\!=\!\pi\cdot d_{K}^2$ is the area of the circle with radius equal 
to the distance $d_{K}$ of the $K$-th nearest neighbor. The uncertainty on the KNN density scales with 
the square root of K, so that the relative fractional error is inversely proportional
to the square root of $K$. For this reason, the fractional accuracy of the method 
increases with increasing $K$ at the expense of spatial resolution. In this paper, 
we reconstruct the spatial distribution of GCs using values of $K$ ranging from 
2 to 10. The size of the spatial structures depends on a combination of the value of $K$ 
and the density of sources. Therefore, the optimal value for $K$ depends on how 
extended and densely populated are these structures. In our experiments 
we consider a range of $K$ values that are roughly proportional 
to the average density of sources 
of the samples for which the density maps are reconstructed. Moreover, very large values 
of $K$ would smooth over spatially smaller structures thus losing spatial information. 

For all the GC distributions, we produce residual maps by subtracting smooth azimuthally homogeneous 
density maps from the observed 
density maps. These smooth maps were generated to have the radial number density 
profiles of the observed GCs. Using MonteCarlo simulations, we then estimate the 
statistical significance of each pixel in the residual map 
in terms of deviation from the mean value of the simulated density distribution in the same 
pixels (after making sure that such distribution of simulated densities can be modeled
with a Gaussian). Finally, we estimate the statistical significance of spatial structures containing multiple
pixels{, by} taking into account the spatial features of the structure, 
i.e. the number of member pixels and the peculiar shape of the pixel aggregation.

The density maps for all galaxies in our sample of galaxies were evaluated
on a regular spatial grid with spacing varying between $\Delta\!(\mathrm{R.A})$ $\sim\!0.002^{\circ}$ 
($\sim6\arcsec$) and $\Delta\!(\mathrm{R.A})$ $\sim\!0.005^{\circ}$ ($\sim18\arcsec$)
along the R.A. axis, and $\Delta\!(\mathrm{Dec})$ $\sim\!0.002^{\circ}$ ($\sim\!6\arcsec$) and 
$\Delta\!(\mathrm{R.A})$ $\sim\!0.005^{\circ}$ ($\sim19\arcsec$) along the Dec. axis.
The sizes of pixels for each target (columns (i) and (l) of Table~1) were 
chosen to produce a similar number of pixels in the grid covering the regions occupied by the GCs
in all the galaxies investigated. This strategy permits to meaningfully compare the statistical 
reliability and the area measured in pixels of each spatial feature 
for different galaxies without taking into account the different physical scales
of the host galaxies. For each galaxy, we have also compared the 
residual maps obtained by randomly varying the pixel sizes along both axes within a $\pm\!30\%$
interval centered on the values shown in Table~1, to rule out systematic effects on 
the determination of the spatial features. The density in the boundary pixels was 
weighted according to the fraction of the pixel within the observed region.

In Appendix A we compare the KNN method with other density reconstruction 
techniques, and show how it is optimal for our analysis.

\section{Structures in the spatial distribution of GCs}
\label{sec:results}

For each galaxy, we have performed experiments for the entire GC sample and for the
red/blue and low-L/high-L classes. Each experiment involved 50,000 simulations. 
Table~2 shows the percentage of simulated distributions of GCs with ``extreme'' number 
of pixels, i.e. with values exceeding those of the 90\% of pixels in the 
observed residual map for $K\!=\!\{7,8,9\}$, for each galaxy. The $K$ values for which 
these fractions are small ($\leq\!1\%$) correspond to the spatial scale at which the 
observed structures become statistically significant. The parameter $K$ is a measure 
of the expected spatial scale of the investigated structures and of the density
contrast of these structures over the average local density.  
Higher values of $K$ are more suitable to detect structures located within the $D_{25}$
of the galaxies, where the overall density of GCs is larger, since only high-contrast structures
can be reliably detected over this high-density background. 
Smaller $K$ values, on the other hand, are more apt at detecting structures in regions where 
the total number of GCs is smaller, as in the outskirts of the host galaxies. Since the data used
in this paper all cover the central regions of the galaxies, we have focused on large values of $K$.
For the six galaxies NGC4472, NGC4486, NGC4649, 
NGC4406, NGC4526 and NGC4621, we find significant inhomogeneities for $K\!=\![9,9,8,9,8,9]$, 
respectively. For NGC4382 and NGC4552 we find marginally significant inhomogeneity for 
$K\!=\!9$. The GC systems of NGC4374 and NGC4365 are not consistent with an 
overall inhomogeneous distribution for the range of $K$ values explored (see the non-negligible 
fraction of simulated distributions with similar overall number of ``extreme'' pixels in column (1) 
in Table~2), however, even in these 
cases localized statistically significant structures can still be observed.

\begin{table*}
	\centering	
	\resizebox{\textwidth}{!}{	
	\begin{threeparttable}	
	\caption{Fractions of simulated residual maps with number of
	``extreme'' pixels exceeding the thresholds.}
	\begin{tabular}{lcccccc}
	\tableline
	Galaxy			&			& 	All GCs		&	Red GCs	 	& 	Blue GCs		&	High-L GCs		&	Low-L GCs	\\
					&			&		(a)		&		(b)		&		(c)		&		(d)			&		(e)		\\
	\tableline
	NGC4472 (M49)	&	$K\!=\!7$	& 	12.1\%(3.1\%)	&	13.7\%(4.0\%) 	& 	9.2\%(3.8\%)	&	9.9\%(4.0\%)		&	12.4\%(2.1\%)	\\
					&	$K\!=\!8$	& 	4.2\%(0.7\%)	&	8.6\%(0.5\%)	& 	6.6\%(0.5\%)	&	4.3\%(0.2\%)		&	3.9\%(1\%)	\\	
					&	$K\!=\!9$	& 	1.1\%(0\%)	&	2.3\%(0\%)	& 	0.9\%(0\%)	&	0.1\%(0\%)		&	0.3\%(0.1\%)	\\						
	NGC4486 (M87)	&	$K\!=\!7$	& 	10\%(2.3\%)	&	12\%(4.1\%)	& 	8.8\%(3.5\%)	&	11\%(4.4\%)		&	7.5\%(4.4\%)	\\
					&	$K\!=\!8$	& 	3.4\%(0.7\%)	&	8.7\%(1.9\%)	& 	6.6\%(1.5\%)	&	9.1\%(2.1\%)		&	1.5\%(0.9\%)	\\	
					&	$K\!=\!9$	& 	0.6\%(0\%)	&	1.4\%(0.2\%)	& 	2.9\%(0.2\%)	&	4.3\%(0.1\%)		&	0.2\%(0\%)	\\						
	NGC4649 (M60)	&	$K\!=\!7$	& 	3.3\%(0.2\%)	&	0.2\%(0.3\%) 	& 	1.2\%(0\%)	&	3.6\%(0.5\%)		&	1.9\%(0.7\%)	\\
					&	$K\!=\!8$	& 	0\%(0\%)		&	0\%(0\%)		& 	0\%(0\%)		&	0\%(0\%)			&	0.1\%(0\%)	\\	
					&	$K\!=\!9$	& 	0\%(0\%)		&	0\%(0\%)		& 	0\%(0\%)		&	0\%(0\%)			&	0\%(0\%)		\\						
	NGC4406 (M86)	&	$K\!=\!7$	& 	4.1\%(0.5\%)	&	5.2\%(0.9\%) 	& 	9.3\%(3.1\%)	&	3.1\%(0.4\%)		&	1.9\%(0.4\%)	\\
					&	$K\!=\!8$	& 	1.4\%(0\%)	&	1.7\%(0\%)	& 	5.5\%(1.2\%)	&	0.2\%(0\%)		&	0\%(0\%)		\\	
					&	$K\!=\!9$	& 	0\%(0\%)		&	0.1\%(0\%)	& 	1.4\%(0.1\%)	&	0\%(0\%)			&	0\%(0\%)		\\						
	NGC4382 (M85)	&	$K\!=\!7$	& 	13.4\%(5.4\%)	&	12\%(3.3\%)	& 	18.9\%(9.1\%)	&	7.9\%(2.1\%)		&	9.3\%(3.1\%)	\\
					&	$K\!=\!8$	& 	8.9\%(2\%)	&	8.6\%(0.8\%)	& 	11.1\%(3.7\%)	&	4\%(0.1\%)		&	4.5\%(1.2\%)	\\	
					&	$K\!=\!9$	& 	3.3\%(0.2\%)	&	2.2\%(0\%)	& 	9.2\%(0.9\%)	&	0.2\%(0\%)		&	0.6\%(0.1\%)	\\						
	NGC4374 (M84)	&	$K\!=\!7$	& 	24.4\%(11.2\%)	&	11.6\%(7.8\%)	& 	14.5\%(5.5\%)	&	19.9\%(8.8\%)		&	13.5\%(8.1\%)	\\
					&	$K\!=\!8$	& 	21\%(9.7\%)	&	9\%(5.1\%)	& 	3.9\%(1.1\%)	&	12.3\%(4.6\%)		&	9.7\%(4.4\%)	\\	
					&	$K\!=\!9$	& 	12.5\%(2.9\%)	&	4.9\%(0.8\%) 	& 	2.1\%(0.3\%)	&	7.8\%(1.9\%)		&	5.4\%(3.2\%)	\\						
	NGC4526			&	$K\!=\!7$	& 	0.1\%(0\%)	&	0\%(0\%)		& 	0\%(0\%)		&	0\%(0\%)			&	0.3\%(0\%)	\\
					&	$K\!=\!8$	& 	0\%(0\%)		&	0\%(0\%)		& 	0\%(0\%)		&	0\%(0\%)			&	0\%(0\%)		\\	
					&	$K\!=\!9$	& 	0\%(0\%)		&	0\%(0\%)		& 	0\%(0\%)		&	0\%(0\%)			&	0\%(0\%)		\\						
	NGC4365			&	$K\!=\!7$	& 	36.6\%(25.7\%)	&	30.8\%(19.5\%)	& 	21.3\%(14.4\%)	&	18.3\%(10.4\%)		&	40.6\%(31.1\%)	\\
					&	$K\!=\!8$	& 	31.4\%(23.1\%)	&	28.2\%(14.5\%) & 	15.6\%(9.1\%)	&	12.7\%(7.6\%)		&	26.9\%(19.1\%)	\\	
					&	$K\!=\!9$	& 	24.3\%(18.9\%)	&	17.6\%(8.3\%)	& 	5.9\%(0.9\%)	&	5.7\%(0.5\%)		&	22.2\%(12\%)	\\						
	NGC4621 (M59)	&	$K\!=\!7$	& 	8.4\%(0.4\%)	&	3.9\%(0.1\%)	& 	4.8\%(0.5\%)	&	6.5\%(1.7\%)		&	1.9\%(0\%)	\\
					&	$K\!=\!8$	& 	1.1\%(0\%)	&	0.5\%(0\%)	& 	2\%(0\%)		&	4.3\%(0.2\%)		&	0.3\%(0\%)	\\	
					&	$K\!=\!9$	& 	0\%(0\%)		&	0\%(0\%)		& 	0.1\%(0\%)	&	0.2\%(0\%)		&	0\%(0\%)		\\						
	NGC4552 (M89)	&	$K\!=\!7$	& 	12.9\%(7.6\%)	&	13.6\%(6.5\%)	& 	7.7\%(1.9\%)	&	6.4\%(2\%)		&	13.7\%(8.8\%)	\\
					&	$K\!=\!8$	& 	6.5\%(3.2\%)	&	9.5\%(4.8\%)	& 	3.6\%(0.5\%)	&	3.2\%(0.4\%)		&	9.5\%(3.5.\%)	\\	
					&	$K\!=\!9$	& 	1.9\%(0.2\%)	&	3.6\%(1.3\%)	& 	1.2\%(0.1\%)	&	0.9\%(0\%)		&	2.7\%(0.8\%)	\\						
	\tableline
	\end{tabular}
	\begin{tablenotes}[para]
 Ê Ê Ê Ê Ê	\item {Column description:}\\
 Ê Ê Ê Ê Ê	\item[a] fraction of simulations obtained with $K\!=\!\{7,8,9\}$ with number of extreme 
 			    pixels with density exceeding the 90-th percentile of the observed pixel 
			    residual distribution) for the entire sample of GCs.\\
		\item[b] as in column (a) for red GCs only.\\
		\item[c] as in column (a) for blue GCs only.\\
		\item[d] as in column (a) for high-L GCs only.\\
		\item[e] as in column (a) for low-L GCs only.\\
	\end{tablenotes}	
	\end{threeparttable}	
	}		
	\label{tab:galaxies_statistics}
\end{table*}

Based on the residual maps of the entire GC system evaluated for the 
optimal $K$ values 
of the six galaxies NGC4472, NGC4486, NGC4649, NGC4406, NGC4526 and NGC4621 and for 
$K\!=\!9$ for the remaining four galaxies, we have considered as ``over-density structures'' all 
the groups of contiguous over-dense pixels 
composed of at least one pixel with significance larger than $4\sigma$. The residual maps 
are shown in Figures~\ref{fig:ngc4472} through~\ref{fig:ngc4552}.
We calculate the total significance of a GC structure by taking into account the average 
significance of the enclosed pixels, its size and the geometry. The total significance of a GC structure expresses the 
probability of observing a similarly shaped GC structure with comparable size and average 
significance in a simulated GC density map. More details on the 
estimation of the total significance can be found in~\cite{dabrusco2013}.
We have classified the structures based on their total significance, size in pixels and 
morphology. 

The three classes of over-density structures are defined 
as follows: small structures: groups of 4 or more adjacent pixels with total significance 
$\geq\!4\sigma$, or groups of one, two or
three adjacent pixels with total significance $\geq\!5\sigma$; medium
structures: groups of 5 to 19 adjacent pixels with with total significance $\geq\!4\sigma$;
large structures, defined as groups of 20 or more adjacent pixels whose total 
significance is $\geq\!4\sigma$. All the remaining groups of over-dense pixels not meeting
the minimal requirements were discarded. We found 
24 medium and 14 ``large'' GC over-density structures, 
while the total number of structures is 229. In the following analysis, we will focus on the 
medium and large over-density structures because the small structures could be produced by 
stochastic fluctuations in the number of GCs in just one pixel. Table~3 
displays the number of all the over-density structures found in the residual maps for the 
$K$ values in Table~2, split according to their class, and the size (in number of 
pixels) for the medium and large structures. Table~3 
gives also the fractions of pixels of the residual maps contained in all the structures and 
only in the large ones. Table~4 contains the basic properties of the 38 
medium and large over-density structures, including the area of the structures in pixels, 
the total and excess number of GCs located within each structure. In the following, we will 
describe the main density features observed in each galaxy, listed in order of decreasing
brightness.

\begin{table*}
	\centering	
	\resizebox{\textwidth}{!}{	
	\begin{threeparttable}		
	\caption{Global properties of the over-density structures.}
	\begin{tabular}{lccccccc}
	\tableline
	Galaxy	&N$_{\mathrm{tot}}$	&N$_{\mathrm{medium}}$ &N$_{\mathrm{large}}$	&	\% observed area in 	& { \% observed area in}	&	{ \% GCs in}	&	{ \% excess GCs in}	\\
			&				&				  	&					&	{ medium + large}		& { large structures}		&    { all structures}&	{ all structures}		\\
			&	(a)			&		(b)			& (c)					&	{ structures} (d)		&		(e)				&	{ (f)}		&	{ (g)}				\\
	\tableline
	NGC4472 (M49)	& 	27	&	3				&	1				&	14.5\%				&	3.4\%				&	{ 11.7\%}	&	{ 4.9\%}			\\
	NGC4486 (M87)	& 	46	&	6				&	1				&	16.9\%				&	2.1\%				&	{ 17.2\%}	&	{ 6.5\%}					\\
	NGC4649 (M60)	& 	28	& 	0				& 	1				& 	16.6\%				&	8.5\%				&	{ 10.9\%}	&	{ 5.7\%}			\\
	NGC4406 (M86)	& 	12	&	3				&	1	  			& 	16.5\%				&	11.7\%				&	{ 26.4\%}	&	{ 13.2\%}			\\
	NGC4382 (M85)	& 	21	&	2				&	2				&	14.6\%				&	7.2\%				&	{ 22.3\%}	&	{ 9.9\%}			\\
	NGC4374 (M84)	& 	22	&	3				&	1	 			&	13.6\%				&	3.2\%				&	{ 14.4\%}	&	{ 6.8\%}			\\
	NGC4526			&	4	&	0				&	2				&	30.1\%				&	28.1\%				&	{ 49.2\%}	&	{ 28.6\%}			\\
	NGC4365			& 	38	&	4				&	1				&	15.2\%				&	2.8\%				&	{ 9.7\%}		&	{ 4.3\%}			\\
	NGC4621 (M59)	& 	10	&	0				&	2	 			&	18.4\%				&	16.2\%				&	{ 35.4\%}	&	{ 16.4\%}			\\
	NGC4552 (M89)	& 	21	&	3				&	2	 			& 	16.6\%				&	5.8\%				&	{ 21.3\%}	&	{ 9.8\%}			\\
	\tableline
	\end{tabular}
	\begin{tablenotes}[para]
 Ê Ê Ê Ê Ê	\item {Column description:}\\
 Ê Ê Ê Ê Ê	\item[a] number of all GC over-density structures.\\
		\item[b] number of medium GC over-density structures.\\
		\item[c] number of large GC over-density structures\\
		\item[d] fraction of observed area of the galaxy belonging to medium and large GC structures.\\
		\item[e] fraction of observed area of the galaxy belonging to the large GC over-density structures.\\
		\item[f] percentage of GCs located within GC structures over the total number of GCs\\
		\item[g] percentage of excess GCs located within GC structures over the total number of GCs\\
	\end{tablenotes}	
	\end{threeparttable}	
	}		
	\label{tab:galaxies_structures}
\end{table*}

\begin{table*}
	\footnotesize
	\centering
	\resizebox{\textwidth}{!}{	
	\begin{threeparttable}		
	\centering	
	\caption{Main properties of the ``medium'' and ``large'' GCs over-density structures.}
	\begin{tabular}{lccccccc}
	\tableline
	Galaxy			&	Structure		& {\bf Class}	&	Significance	&	Size [pixels]		&  Total GCs 		&	Excess GCs	&	Physical size [kpc]	\\
					&	(a)			&		(b)	&		(c)		&	(d)				&	(e)			&	(f)			&	(g) 				\\	
	\tableline
	NGC4472 (M49)	&	A1		& {\bf All}, {\bf red}, blue, high-L$^{*}$, low-L$^{*}$			&$\sim\!6.5\sigma$		&	22			&	25		&	8.1		& 14.4	\\
					&	A2		& {\bf All}, {\bf red}, blue, {\bf high-L},						&$\sim\!4.3\sigma$		&	9			&	46		&	23.2		& 5.3\\
					&	A3		& {\bf All}, {\bf red}, blue, high-L, low-L					&$\sim\!4.1\sigma$		&	10			&	47		&	16.6		&4.6	\\	
					&	A4		& {\bf All}, red, {\bf high-L}								&$\sim\!4\sigma$		&	7			&	6		&	2.1		&1.1	\\						
	NGC4486 (M87)	&	B1		& {\bf All}, red, blue, high-L, low-L						&$\sim\!5.8\sigma$		&	18			&	23		&	11.3		&5.7	\\
					&	B2		& {\bf All}, {\bf red}, blue, {\bf low-L}						&$\sim\!4.8\sigma$		&	20			&	31		&	12.1		& 1.2	\\
					&	B3		& {\bf All}, red, {\bf blue}, {\bf high-L}						&$\sim\!4.7\sigma$		&	9			&	15		&	6.8		& 1.4	\\
					&	B4		& All, red, blue, low-L								&$\sim\!4.5\sigma$		&	9			&	83		&	38.7		& 5.8	\\
					&	B5		& All, {\bf red}, {\bf low-L}								&$\sim\!4.5\sigma$		&	6			&	28		&	8.8		& 1.6	\\
					&	B6		& All, red, {\bf high-L}, low-L							&$\sim\!4.3\sigma$		&	7			&	25		&	12.8		& 4.6	\\
					&	B7		& All, red											&$\sim\!4.1\sigma$		&	18			&	18		&	5.6		& 1	\\						
	NGC4649 (M60)	&	C1		& {\bf All}, {\bf red}, high-L$^{*}$, low-L$^{*}$				& $\sim\!10\sigma$		&	60			&	160		&	80.3		& 29	\\
	NGC4406 (M86)	&	D1		& {\bf All}, {\bf red}, {\bf blue}, {\bf high-L}, {\bf low-L}			&$\sim\!15\sigma$		&	91			&	76		&	38.8		& 14.3	\\
					&	D2		&  All, blue, {\bf high-L}								& $\sim\!4.5\sigma$		&	7			&	4		&	1.2		& 2.4\\
					&	D3		&  All, {\bf red}, {\bf low-L}								& $\sim\!4.3\sigma$		&	8			&	5		&	2.4	 	& 4.6	\\
					&	D4		& All, red, high-L									&$\sim\!4\sigma$		&	8			&	4		&	2		& 5.1	\\			
	NGC4382 (M85)	&	E1		& {\bf All}, {\bf red}, blue, high-L, {\bf low-L}				&$\sim\!8.5\sigma$		&	61			&	36		&	18.6		& 7.9	\\
					&	E2		& {\bf All}, {\bf red}, blue, {\bf high-L}$^{*}$, low-L$^{*}$		&$\sim\!6.8\sigma$		&	22			&	26		&	11.8		& 6.4	\\	
					&	E3		& {\bf All}, red, {\bf high-L}								&$\sim\!6.3\sigma$		&	12			&	20		&	3.4		& 3.1	\\	
					&	E4		& {\bf All}, red, blue, high-L, low-L						&$\sim\!5.7\sigma$		&	19			&	19		&	9.1		&  6.2	\\									
	NGC4374 (M84)	&	F1		& {\bf All}, {\bf blue}, {\bf high-L}							& $\sim\!6.5\sigma$		&	30			&	18		&	9.2		& 6.4\\
					&	F2		& {\bf All}, {\bf red}, blue, {\bf high-L}, low-L				& $\sim\!6.3\sigma$		&	19			&	9		&	3.2		& 3.5	\\
					&	F3		& All, {\bf blue}, high-L 								& $\sim\!5.5\sigma$		&	17			&	23		&	6.6		& 3.6	\\
					&	F4		& All, red, {\bf high-L}, low-L 							& $\sim\!4.5\sigma$		&	11			&	7		&	1.4		& 2.7	\\
	NGC4526			&	G1		& {\bf All}, {\bf red}, {\bf blue}, {\bf high-L}, {\bf low-L} 			&$>\!20\sigma$			&	185			&	63		&	37.9		& 9.3	\\
					&	G2		& {\bf All}, {\bf red}, {\bf blue}, {\bf high-L}, {\bf low-L}			&$>\!20\sigma$			&	88			&	53		&	28.8		&16.7	\\
	NGC4365			&	H1		& {\bf All}, red$^{*}$, blue$^{*}$							&$\sim\!5.8\sigma$		&	22			&	72		&	29.5		& 23.2	\\
					&	H2		& All, red, blue, {\bf high-L}, low-L   						&$\sim\!5.1\sigma$		&	10			&	45		&	5.6		& 18.7	\\
					&	H3		& All, blue, high-L, low-L 								&$\sim\!4.6\sigma$		&	5			&	24		&	6.2		&11.2	\\
					&	H4		& {\bf All}, 	{\bf blue}, {\bf high-L}, {\bf low-L}				&$\sim\!4.5\sigma$		&	6			&	28		&	18.4		& 4.3\\
					&	H5		& All, red, blue, 	low-L								&$\sim\!4.1\sigma$		&	6			&	24		&	12.5		& 4.2	\\
	NGC4621 (M59)	&	I1		& {\bf All}, {\bf red}, {\bf blue}, {\bf high-L}, {\bf low-L}			& $>\!15\sigma$		&	71			&	42		&	23.4		& 10.9	\\
					&	I2		& {\bf All}, {\bf red}, {\bf blue}, {\bf high-L}, {\bf low-L}			& $>\!15\sigma$		&	85			&	56		&	21.3		& 7.7	\\
	NGC4552 (M89)	&	L1		& {\bf All}, {\bf blue}, high-L							&$\sim\!8.7\sigma$		&	30			&	7		&	4.1		& 5	\\
					&	L2		& {\bf All}, red, blue, low-L								& $\sim\!7.3\sigma$		&	22			&	11		&	6.4		& 4.2	\\
					&	L3		& All, red, blue, {\bf high-L}, low-L 						& $\sim\!6.4\sigma$		&	18			&	26		&	10.8		& 4.3	\\
					&	L4		& {\bf All}, 	red, {\bf blue}, {\bf high-L}, low-L				& $\sim\!5.7\sigma$		&	10			&	23		&	11.2		&  3.3	\\
					&	L5		& All, red, 	high-L$^{*}$, low-L$^{*}$						& $\sim\!4.8\sigma$		&	17			&	16		&	8.5		& 4.2	\\
	\tableline
	\end{tabular}
	\begin{tablenotes}[para]
 Ê Ê Ê Ê Ê	\item {Column description:}\\
 Ê Ê Ê Ê Ê	\item[a] unique code associated to the GCs structure\\
		\item[b] significance of each structure in the residual maps generated for all GCs, 
		the color and luminosity GC classes: boldface text is used for the GC classes 
		where the structure is clearly visible and has large statistical significance, plain 
		text is used for GC classes where the structure is still detected with lower 
		significance, while the GC classes where the structure is not detected - or detected in a 
		distinctly different shape and size - are not reported. Asterisks indicate cases 
		where the residual maps of either the color or luminosity GC classes suggest 
		the presence of spatial segregation\\
		\item[c] statistical significance of the structure\\
		\item[d] size of the structures measured in pixels\\
		\item[e] total number of GCs within the structures\\
		\item[f] number of GCs within the structures in excess to the smooth distribution of 
		GCs used to evaluate the residual maps\\
		\item[g] approximate maximum physical size of the GC structures [kpc]\\
	\end{tablenotes}	
	\end{threeparttable}	
	}		
	\label{tab:large_structures}
\end{table*}

\subsection{NGC4472}
\label{subsec:ngc4472}

\begin{figure*}[]
	\includegraphics[height=5.5cm,width=5.5cm,angle=0]{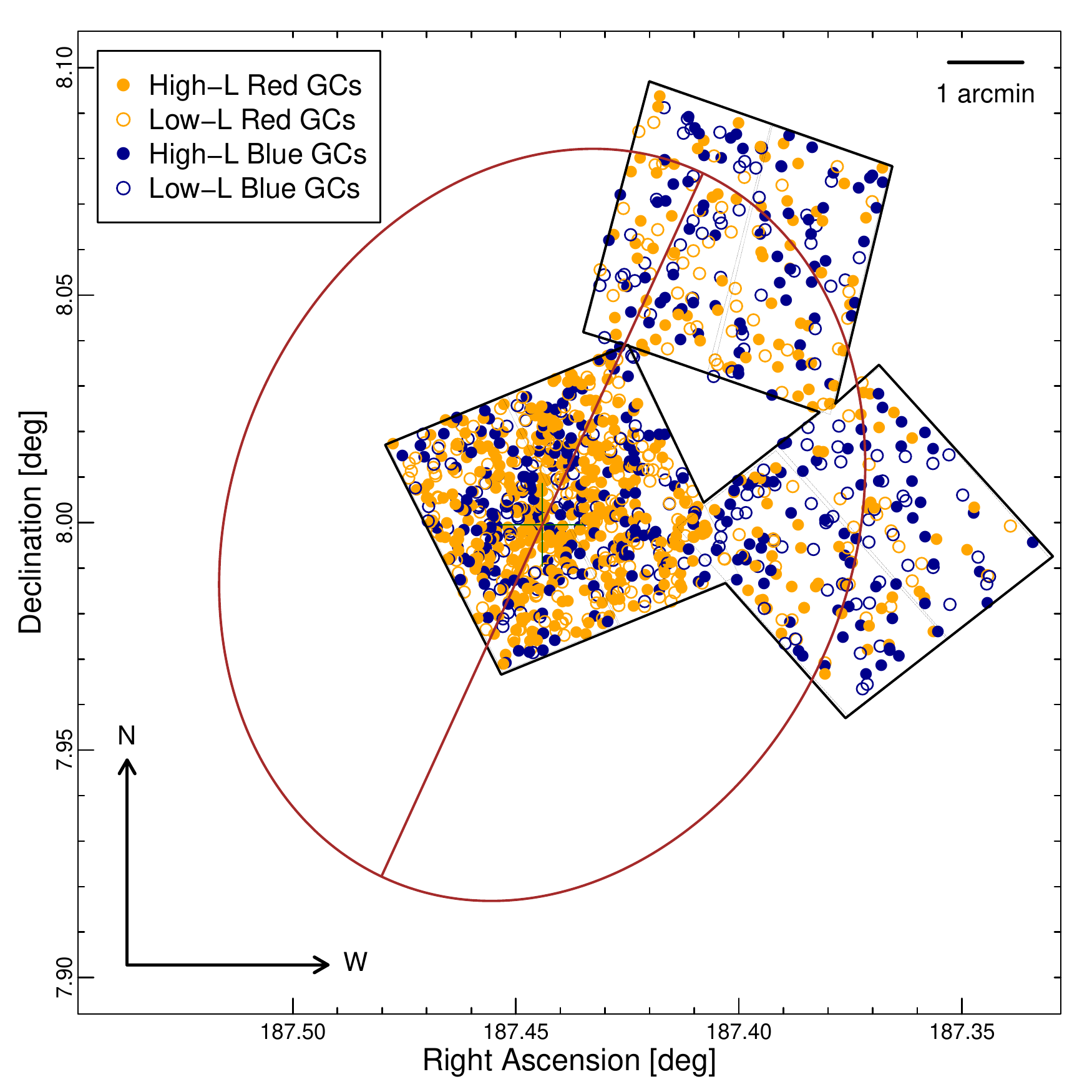}
	\includegraphics[height=5.5cm,width=5.5cm,angle=0]{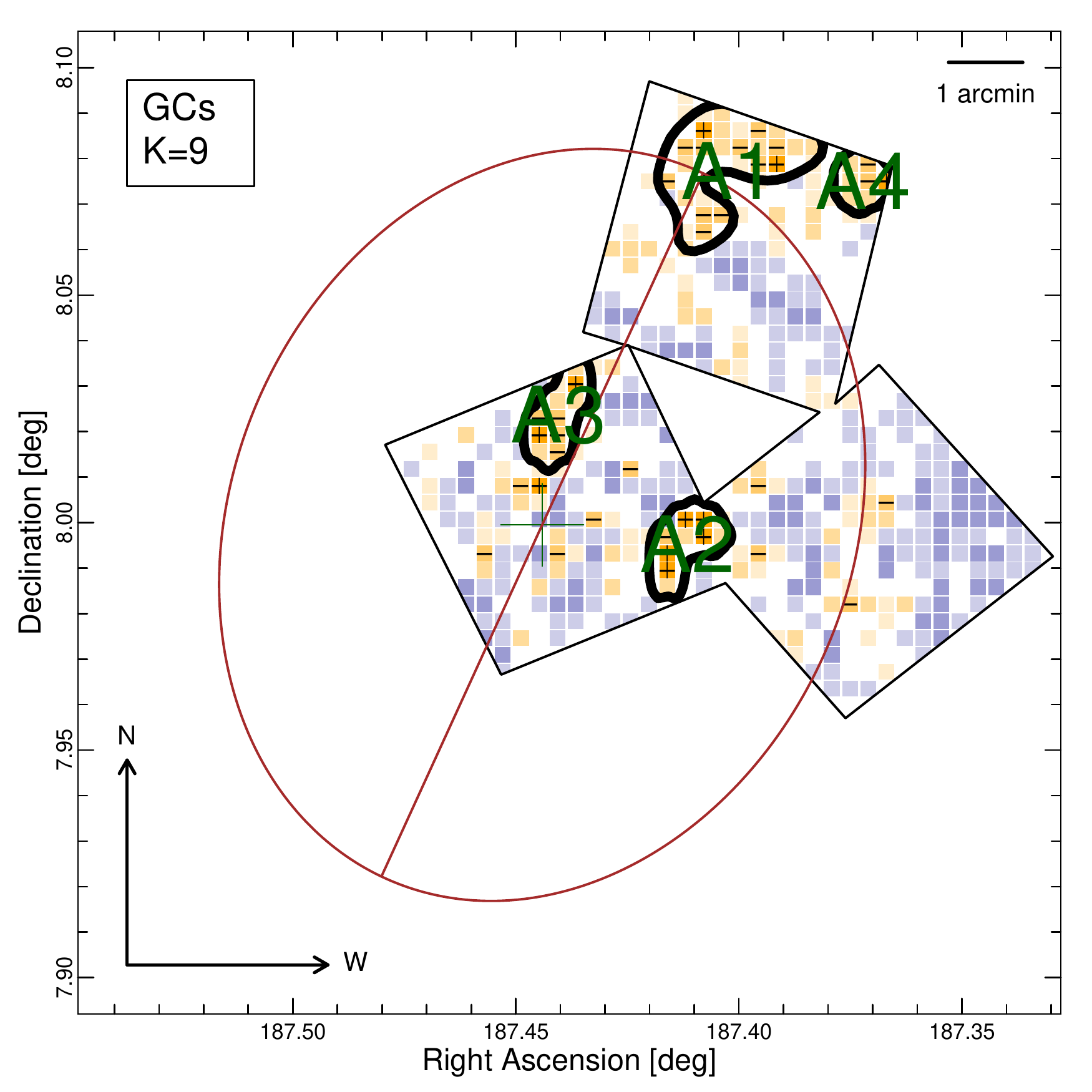}	
	\includegraphics[height=5.5cm,width=5.5cm,angle=0]{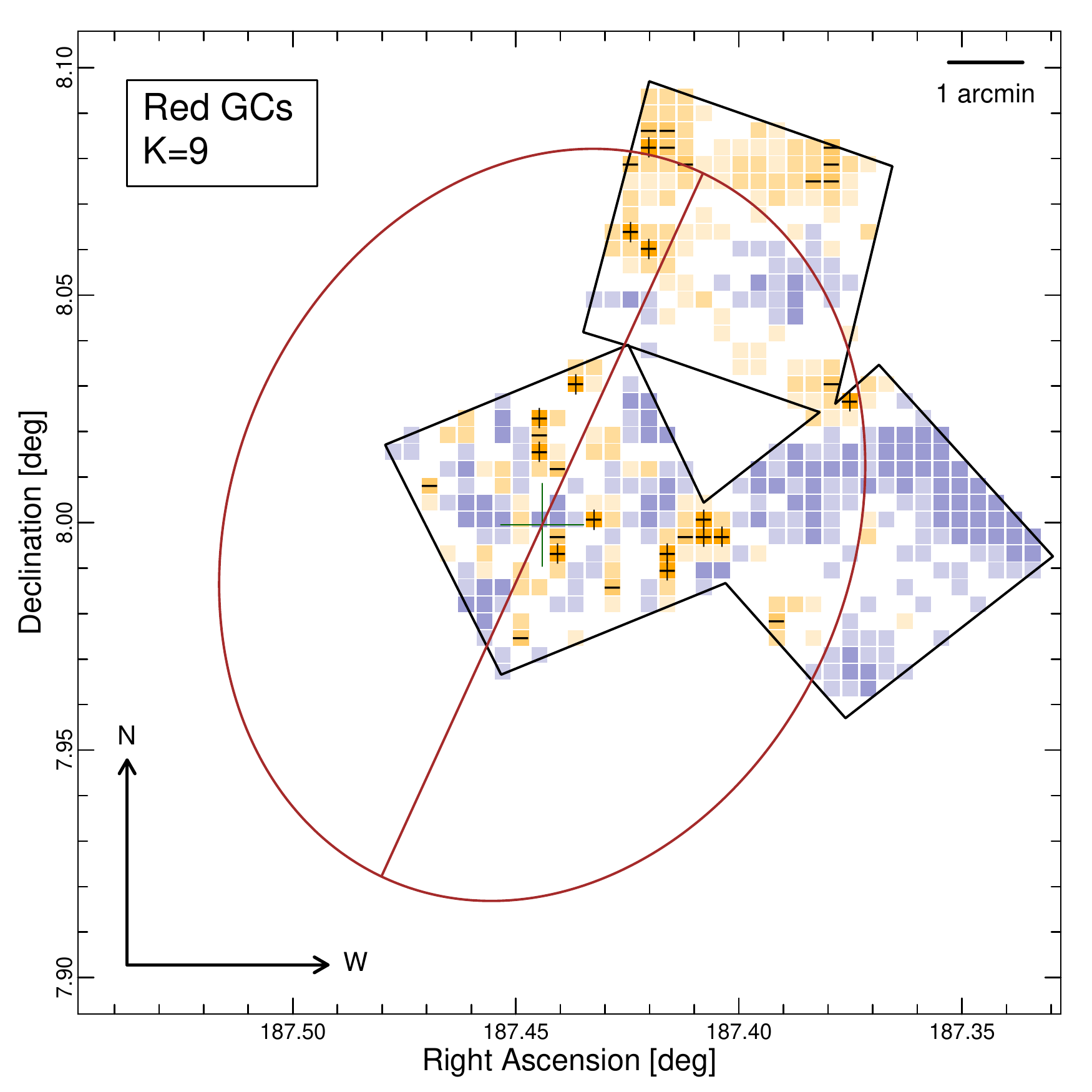}\\
	\includegraphics[height=5.5cm,width=5.5cm,angle=0]{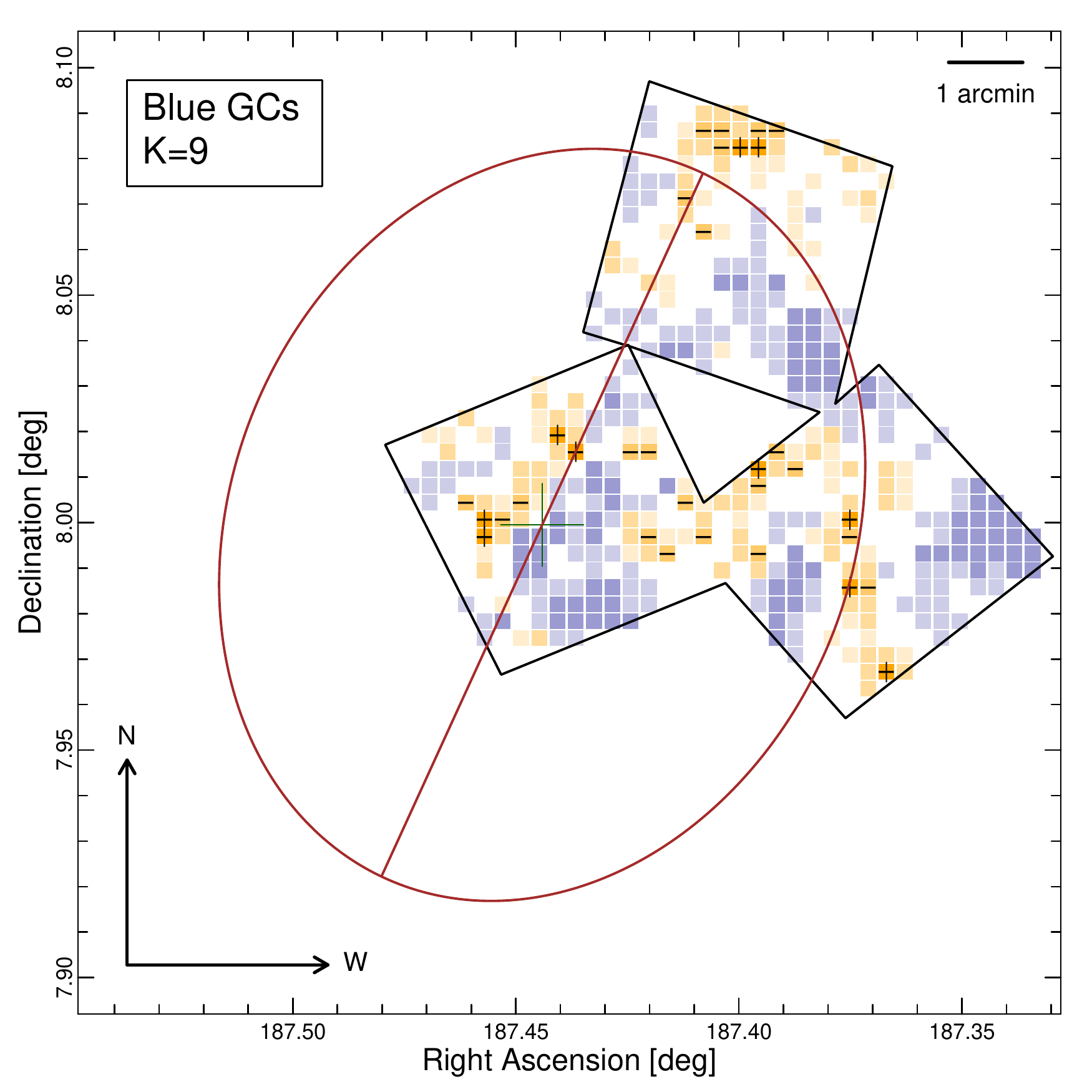}
	\includegraphics[height=5.5cm,width=5.5cm,angle=0]{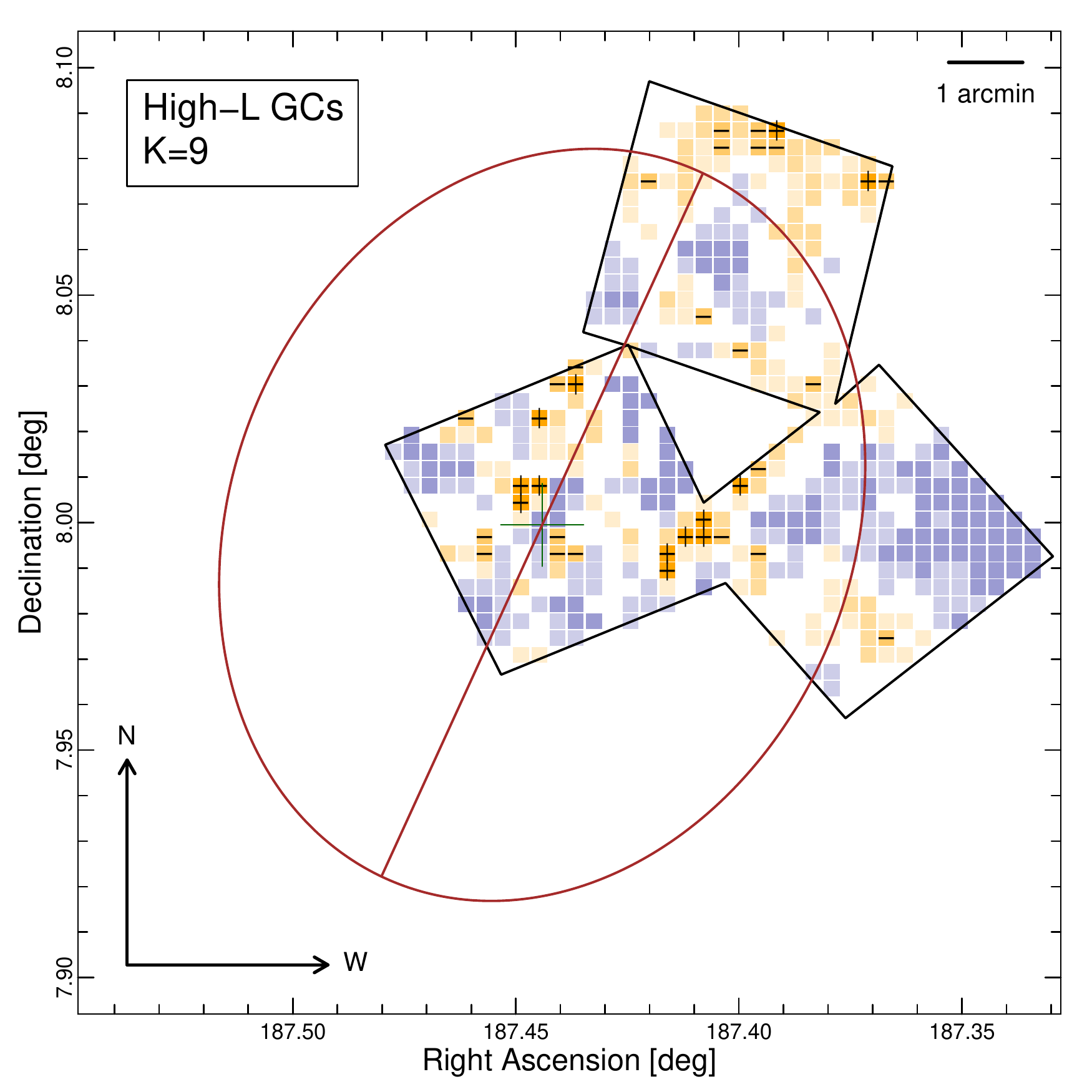}
	\includegraphics[height=5.5cm,width=5.5cm,angle=0]{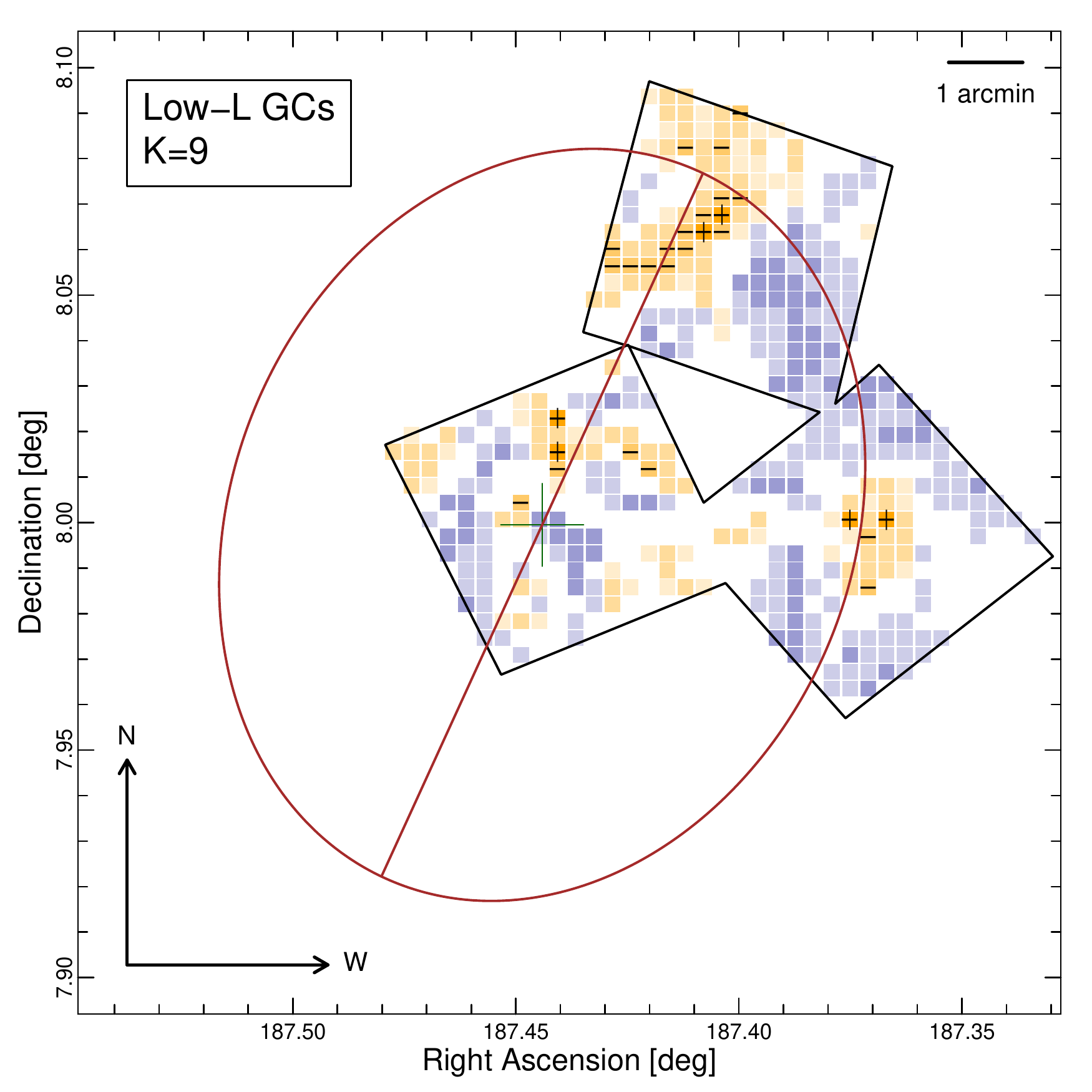}		
	\caption{(Upper left) Scatterplot of the positions of the GCs detected in NGC4472 and used
	in this paper. Orange and blue points correspond to red and blue GCs, while solid and open 
	symbols show high-luminosity and low-luminosity sources, respectively, defined using the 
	magnitude threshold discussed in Section~\ref{sec:data}. (Upper middle) Residual map obtained 
	for $K\!=\!9$ of the whole sample of GCs
	in NGC4472. Blue and orange pixels show negative and positive excess, i.e. under-density 
	and over-density residual. The intensity and symbols express the degree of significance of 
	each pixel: light blue/orange, dark blue/orange, horizontal bars and crosses indicate 
	over/under-density pixels with significance smaller or equal than 1, smaller or equal than 2, 
	smaller or equal than 3 and larger than 3, respectively. (Upper right) $K\!=\!9$ residual map 
	for red GCs. (Lower left) $K\!=\!9$ residual map for blue GCs. (Lower middle) $K\!=\!9$ residual map 
	for high-luminosity GCs. (Lower right) $K\!=\!9$ residual map for low-luminosity GCs. In all plots, 
	the ellipse shows the $D_{25}$ ($\sim\!2.9 r_{e}$) isophote of the galaxy 
	from~\cite{devaucouleurs1991}. The solid black line shows the footprint of the HST observations 
	used to extract the GCs.}
	\label{fig:ngc4472}
\end{figure*}

The $K\!=\!9$ residual map of the spatial distribution of all GCs in NGC4472 
(Figure~\ref{fig:ngc4472}, upper-mid panel) 
is characterized by a significant ($\sim\!6.5\sigma$) ``curly'' over-density structure located N of the center of 
the galaxy (structure A1). A1 lies almost entirely outside the $D_{25}$ { ($\sim\!2.9\ r_{e}$)} but it 
bends inward at its eastern end, radially following the major axis to one fourth of its length; it is
more prominent in the red GCs (Figure~\ref{fig:ngc4472}, upper right panel) 
than in the blue GCs (Figure~\ref{fig:ngc4472}, lower left panel). The high- and low-L residual maps
(Figure~\ref{fig:ngc4472}, lower mid and right panels) suggest spatial segregation, with 
low-L GCs mostly located in the radial section of the structure, and high luminosity GCs 
located outside of the $D_{25}$, not showing a clear radial dependence. We can rule out dynamical 
friction as cause of the observed spatial segregation since, in that case, we would have observed 
separation between the high-L, more massive GCs and the low-L, less massive GCs, 
with the former closer to center of the galaxy and the latter at larger galactocentric distances. 

The smaller distinct over-density structure A4 ($\sim\!4\sigma$) in the N-W corner could be an extension of A1.
The second and third more significant density features after A1, labeled as A3 ($\sim\!4.1\sigma$) and A2 
($\sim\!4.3\sigma$) in the upper-mid panel of Figure~\ref{fig:ngc4472}, are located N and W
of the center of the galaxy. They are visible also in the residual maps for the two color classes and the high-L
GCs. The first of these structures has a radial shape pointing N-W along the major axis. 

\subsection{NGC4486 (M87)}
\label{subsec:ngc4496}

\begin{figure*}[]
	\includegraphics[height=5.5cm,width=5.5cm,angle=0]{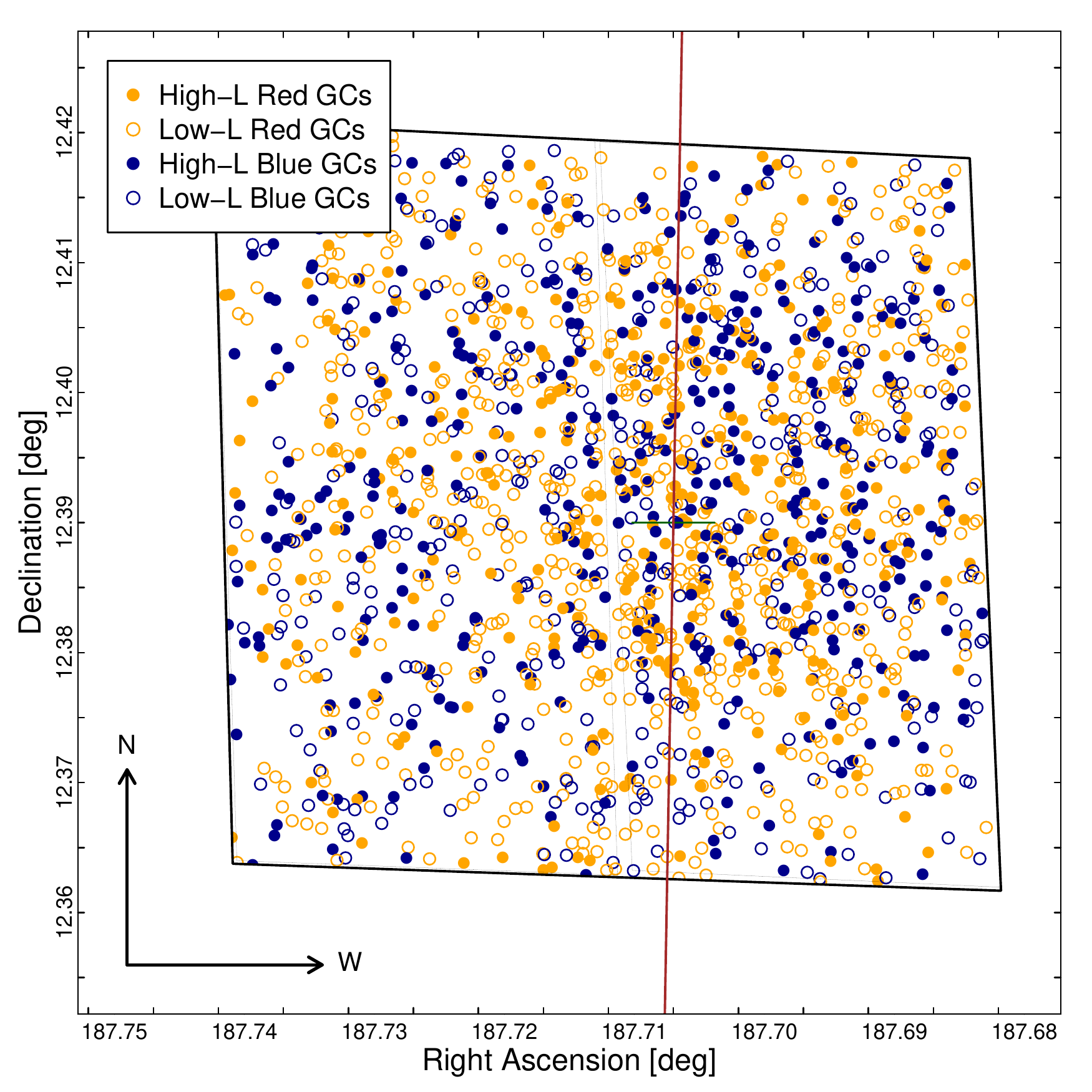}
	\includegraphics[height=5.5cm,width=5.5cm,angle=0]{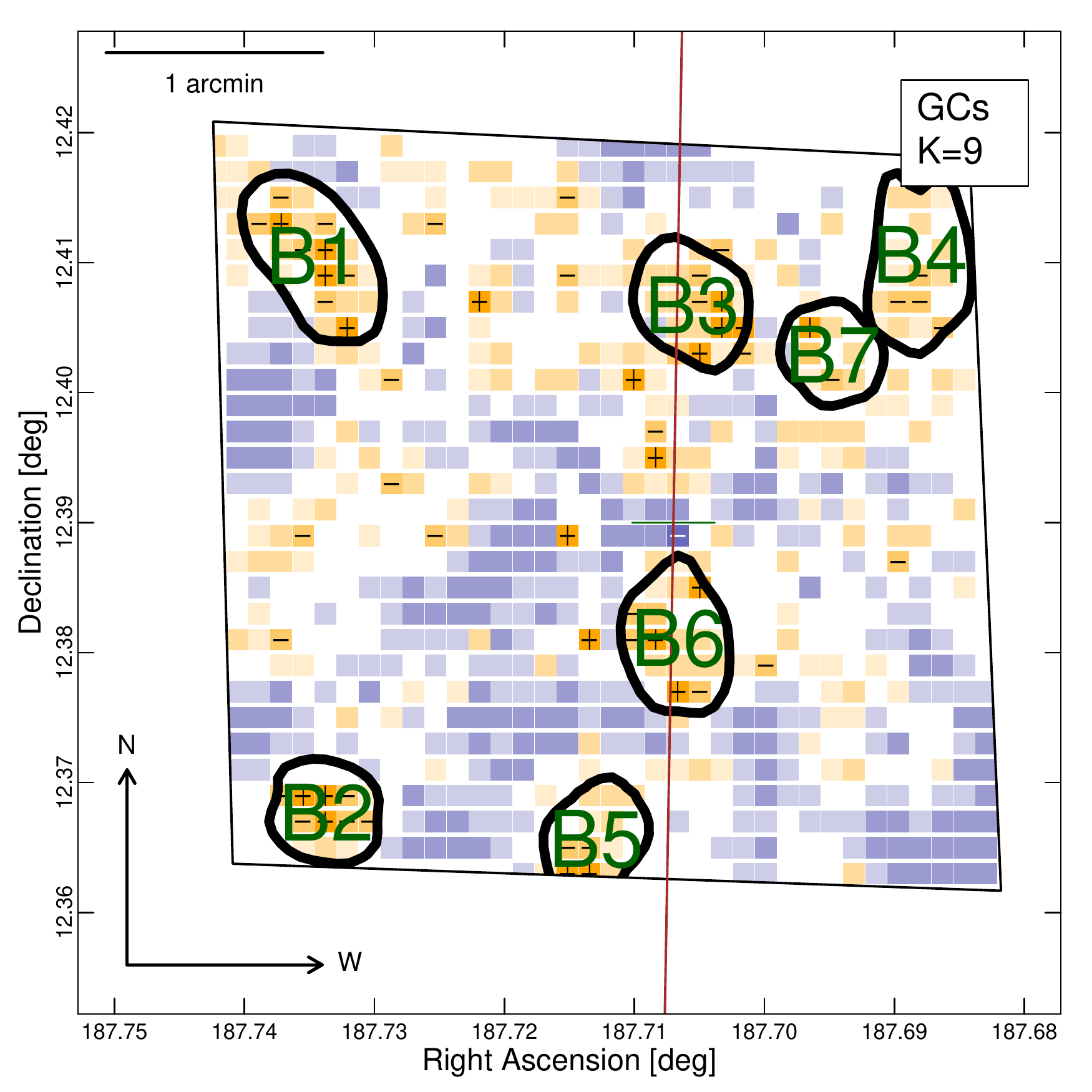}	
	\includegraphics[height=5.5cm,width=5.5cm,angle=0]{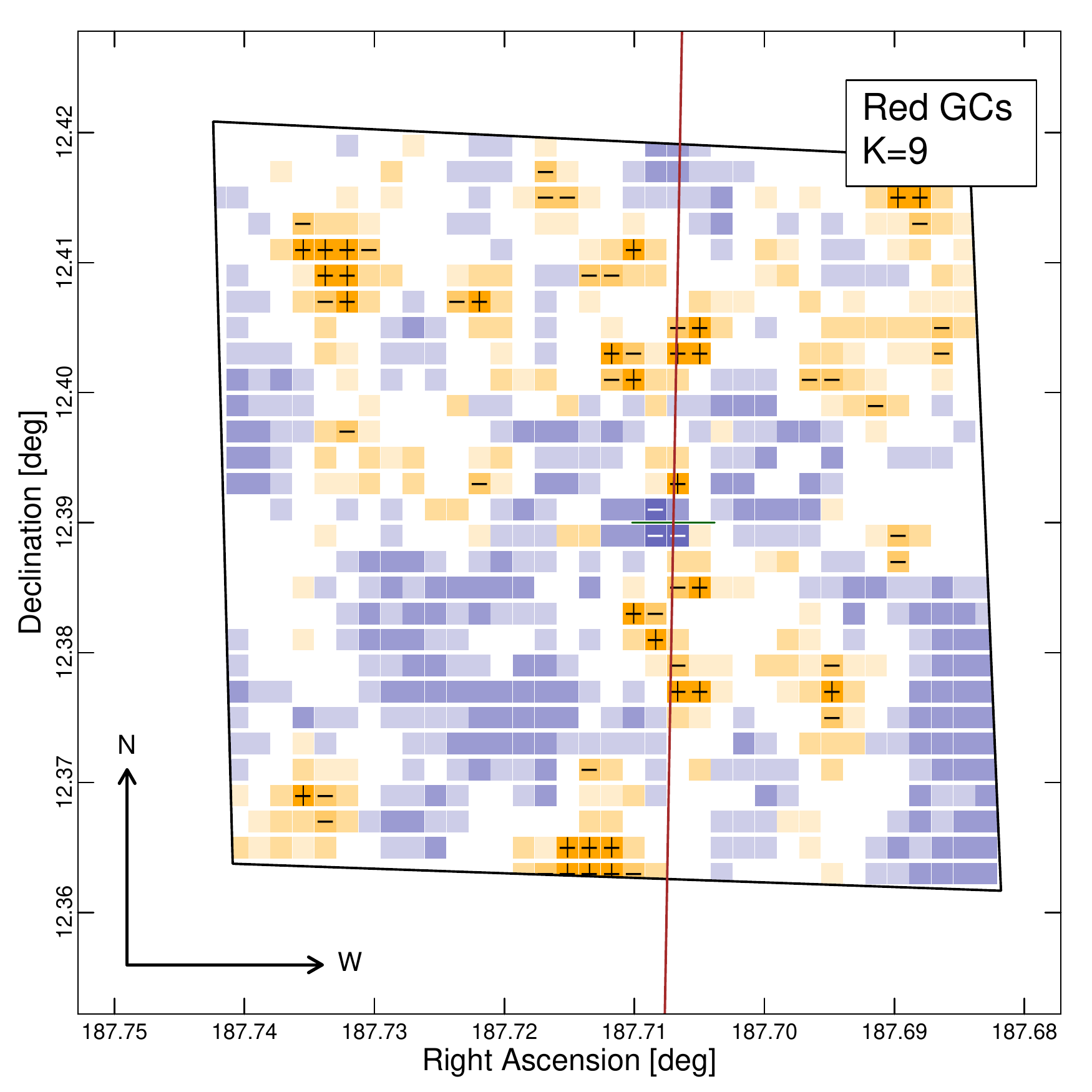}\\
	\includegraphics[height=5.5cm,width=5.5cm,angle=0]{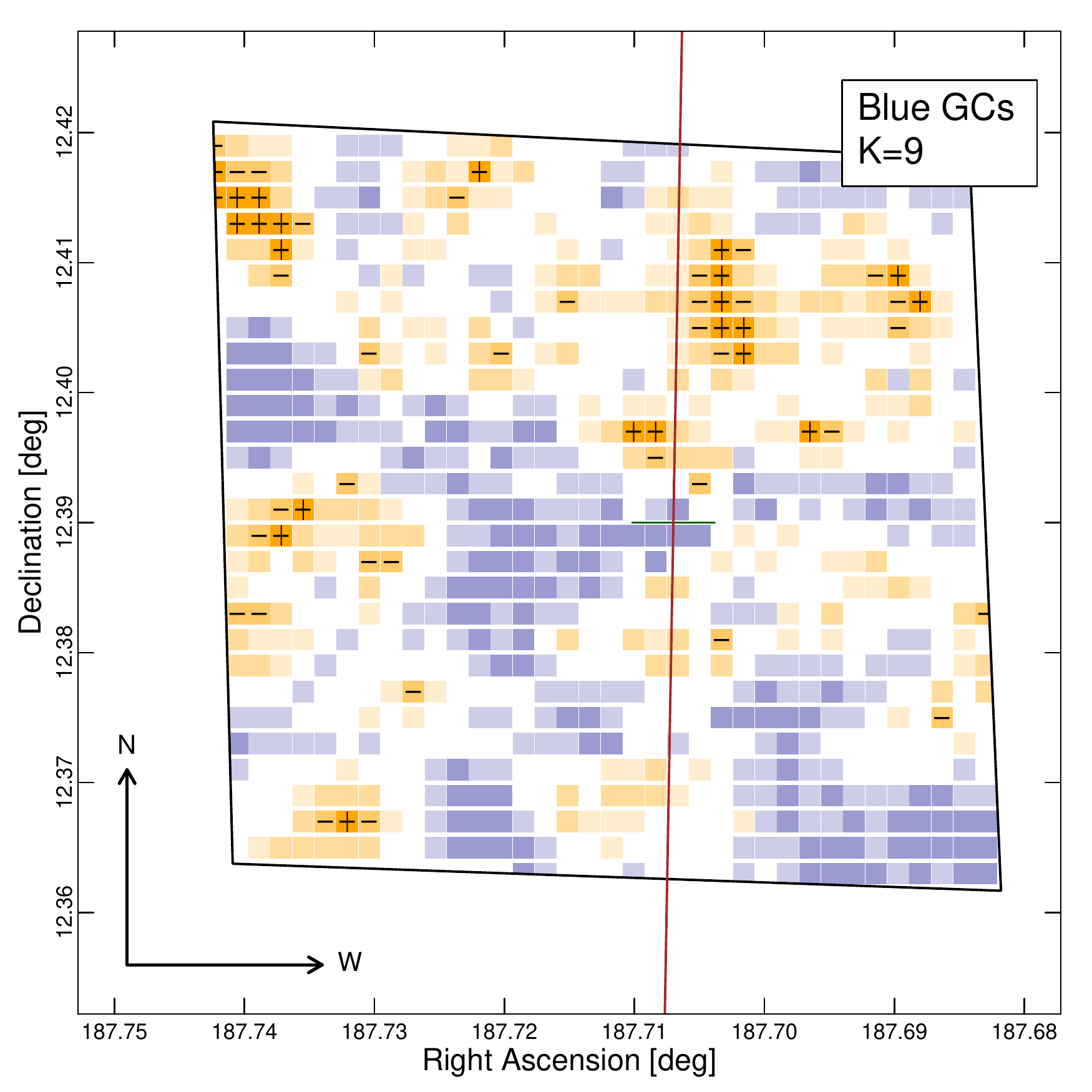}
	\includegraphics[height=5.5cm,width=5.5cm,angle=0]{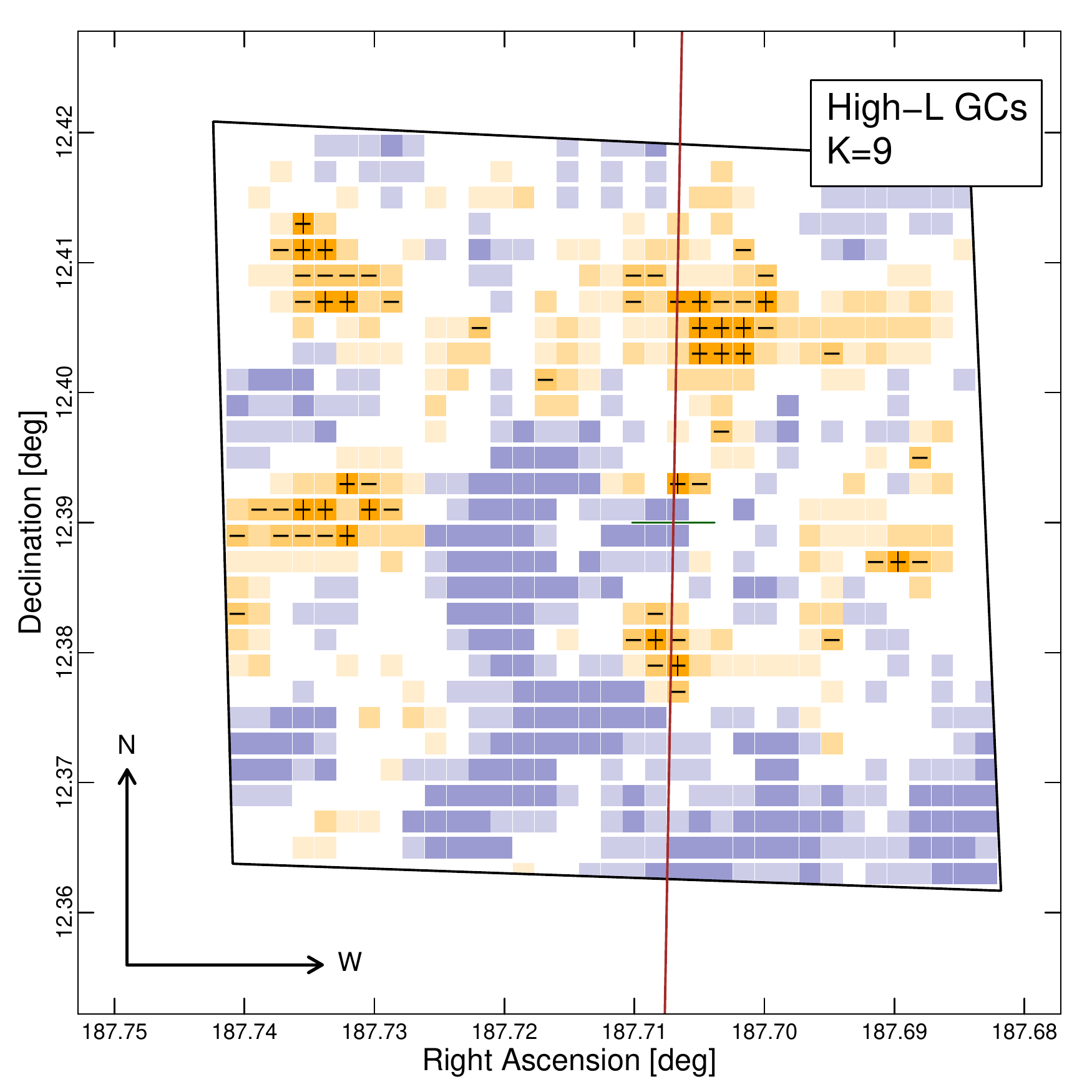}
	\includegraphics[height=5.5cm,width=5.5cm,angle=0]{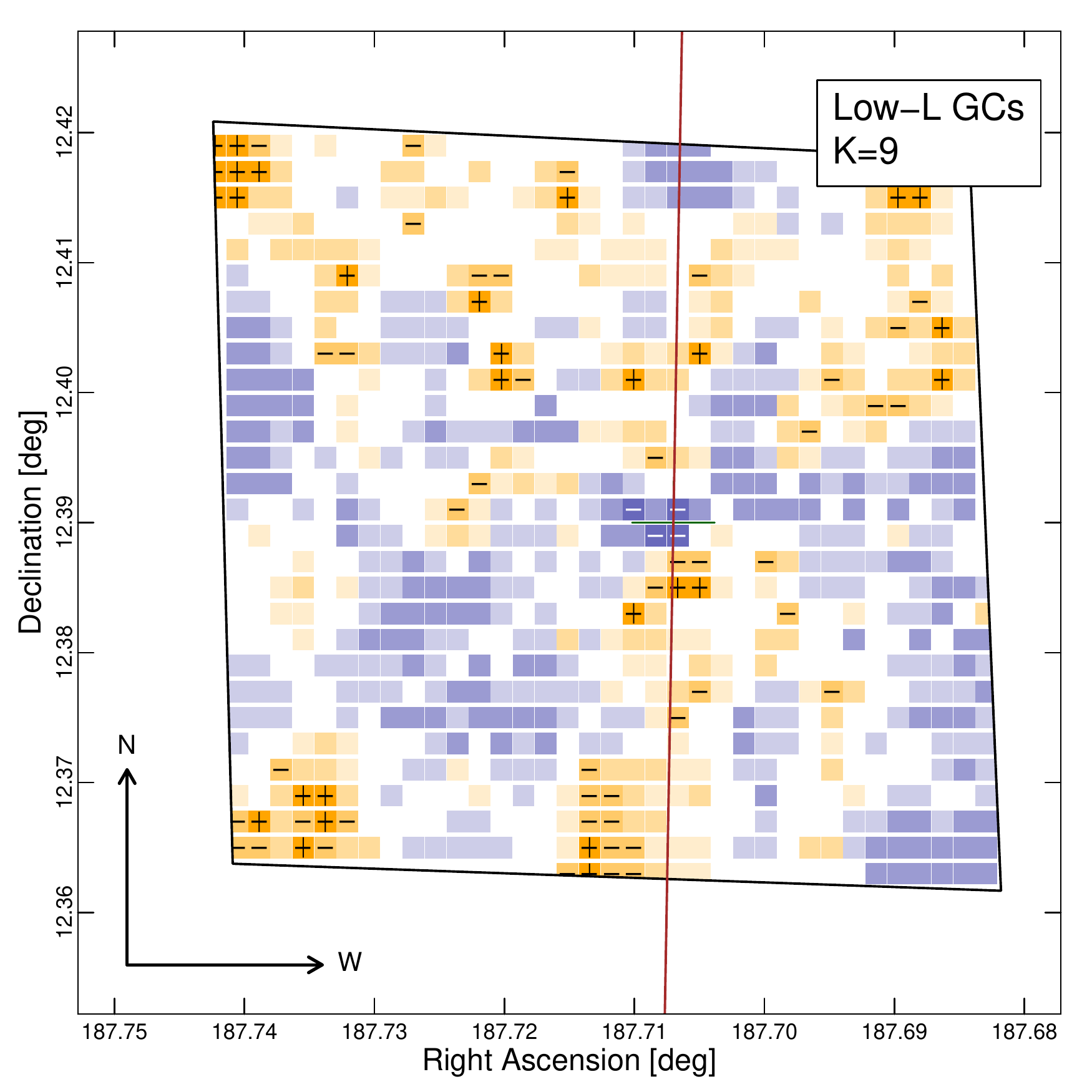}\\
	\includegraphics[height=5.5cm,width=5.5cm,angle=0]{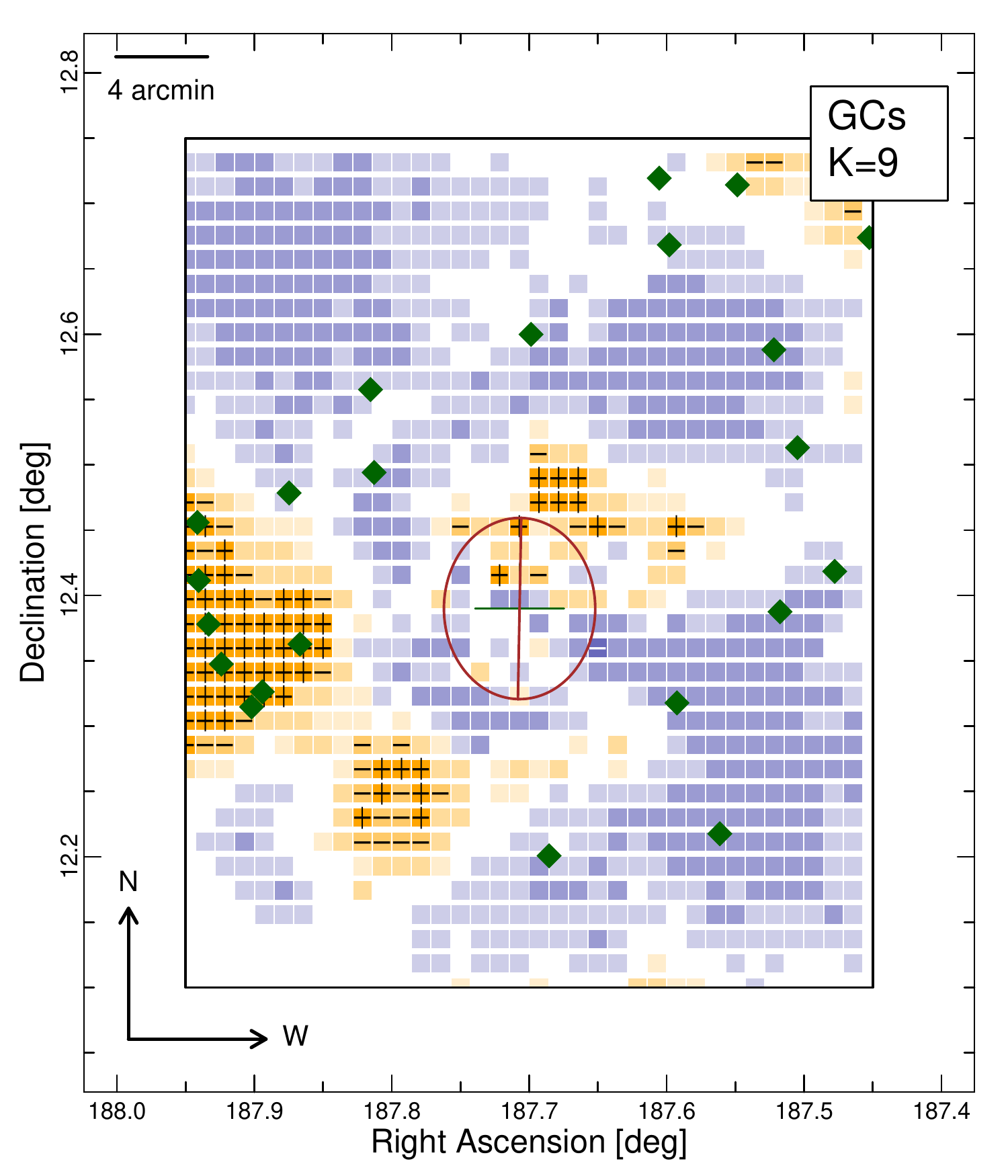}
	\includegraphics[height=5.5cm,width=5.5cm,angle=0]{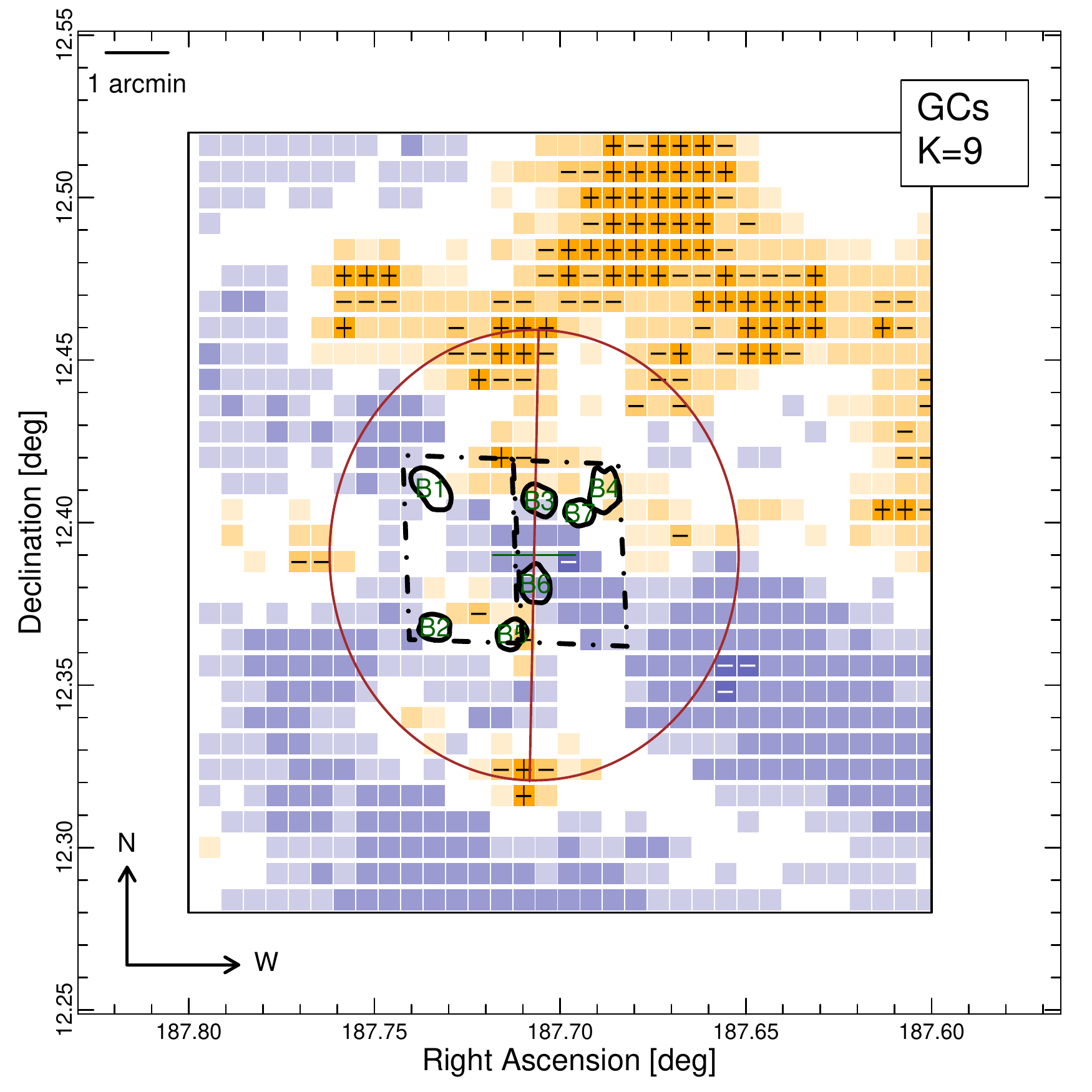}		
	\caption{Upper and mid panels: scatterplot of the position of the GCs in NGC4486 and 
	residual maps obtained for 
	$K\!=\!9$ for the whole sample of GCs and the two color and luminosity GC classes. 
	The $D_{25}$ elliptical isophote ($\sim\!2.9 r_{e}$) 
	is not shown because it falls outside of the plot borders.
	The major axis of the galaxy is shown by the vertical red line. Refer 
	to Figure~\ref{fig:ngc4472} for a description of the other panels. Lower
	panels: residual map for $K\!=\!9$ for the sample of GCs observed 
	spectroscopically~\citep{strader2011,romanowsky2012}. The GCs 
	belonging to the shell selected kinematically are shown as 
	green diamonds. On the right, a zoomed-in 
	version of the same map with the region covered by the ACSVCS data and 
	the GC structures discussed in Section~\ref{subsec:ngc4496} highlighted.}
	\label{fig:ngc4486}
\end{figure*}

The highly inhomogeneous 2D distribution of all GCs in NGC4486 (Table~4, Figure~\ref{fig:ngc4486}), 
is dominated by a large over-density structure (B1, $\sim\!5.8\sigma$), located in the 
N-E corner of the covered area (Figure~\ref{fig:ngc4486}, 
upper left and mid panels). B1 is also visible, with slightly smaller significance, in the 
color and luminosity GC classes (Figure~\ref{fig:ngc4486}, upper-right and 
lower panels). The second most significant over-density structure, in the S-E corner 
(B2, $\sim\!4.8\sigma$), is also observed in red and low-L GCs.
Multiple over-density structures (B5, B6, B3, B7 and B4 in Figure~\ref{fig:ngc4486}, upper mid panel) 
follow the major axis of NGC4486, turning to the N-W corner of the field, 
suggesting the presence of a large scale radial structure.
Four structures ($\geq\!4.2\sigma$) are also visible in high-L GCs (Figure~\ref{fig:ngc4486}, 
lower mid panel). Three of these, in the N-E corner of the field, correspond to B1, B3 and B6. 

~\cite{romanowsky2012} discussed two kinematically selected GC substructures in M87, 
one associated to the outer-halo
stellar filament N-W of the galaxy, and the other in the inner halo (the ``shell''). This second structure 
was interpreted as the signature of a disrupted infalling system. Both phase-space structures
are located outside of the region observed by ACSVCS. We have applied our method to the
sample of~\cite{romanowsky2012} (see also~\citealt{strader2011}) 
to assess our ability to detect these kinematically 
selected structures as over-densities in the GC spatial distribution (Figure~\ref{fig:ngc4486}, lower panels). 
We should keep in mind, however, 
that the ACSVCS and the~\cite{romanowsky2012} samples have different levels of completeness; 
in particular, the~\cite{romanowsky2012} sample is biased towards higher luminosity (mass) GCs.

At very large radii, we find a high-significance overdensity structure W
of NGC4486, located where seven GC members of the shell (green points in lower, left panel in 
Figure~\ref{fig:ngc4486}) can be found. While the other members of the shell are distributed at the
border of under-density areas (blue), they do not constitute a statistically significant overdensity.
A zoomed{-in} version of the residual map obtained from the~\cite{romanowsky2012}
GCs sample for $K\!=\!9$ (lower mid panel, Figure~\ref{fig:ngc4486}) shows that a very significant 
($\sigma\!\sim\!10$), extended
GC structure is located N and N-W of NGC4486, mostly outside the $D_{25}$ of the galaxy. A low-significance
bridge of overdensity pixels connects this large GC structure to the complex of GCs over-densities observed
in the ACSVCS residual map (B1, B3, B4 and B7). The average line-of-sight velocity of all~\cite{romanowsky2012} 
GCs enclosed within this large structure is consistent with the bulk velocity of the system within 2$\sigma$. 
The average velocities of the~\cite{romanowsky2012}
GCs located within the ACSVCS GC structures
are consistent with the bulk velocity of the system, but the numbers of GCs
are small (4~\citealt{romanowsky2012} GCs in B1, 3 in B2, 2 in B3, 8 in B4, 4 in B5, 3 in B6 and 0 in B7). 
For this reason, we cannot rule out a distinct kinematics, based on the~\cite{strader2011} sample only.
A complete photometric coverage of M87, at least including the area
within its $D_{25}$ and a deeper, more homogeneous spectroscopic campaign that can match the 
ACSVCS sample are 
needed to fully map its GC streamers and confirm their nature in the phase-space. 

NGC4486 shows dust filaments~\citep{ferrarese2006} and has been detected by~\cite{diserego2013} in the 
far infrared (FIR) images at 250 $\mu$m obtained by the Herschel Virgo Cluster Survey. The FIR emission 
indicates the presence of diffuse dust that may affect the detection 
of GCs and, in turn, our reconstruction of their spatial distribution. The location of 
the observed dust structures is not associated with significant under- or over-density regions in our 
residual GC maps, suggesting an overall negligible effect of the dust on the  
the reconstruction of the GC spatial distribution obtained with our method.

\subsection{NGC4649}
\label{subsec:ngc4649}

\begin{figure*}[]
	\includegraphics[height=5.5cm,width=5.5cm,angle=0]{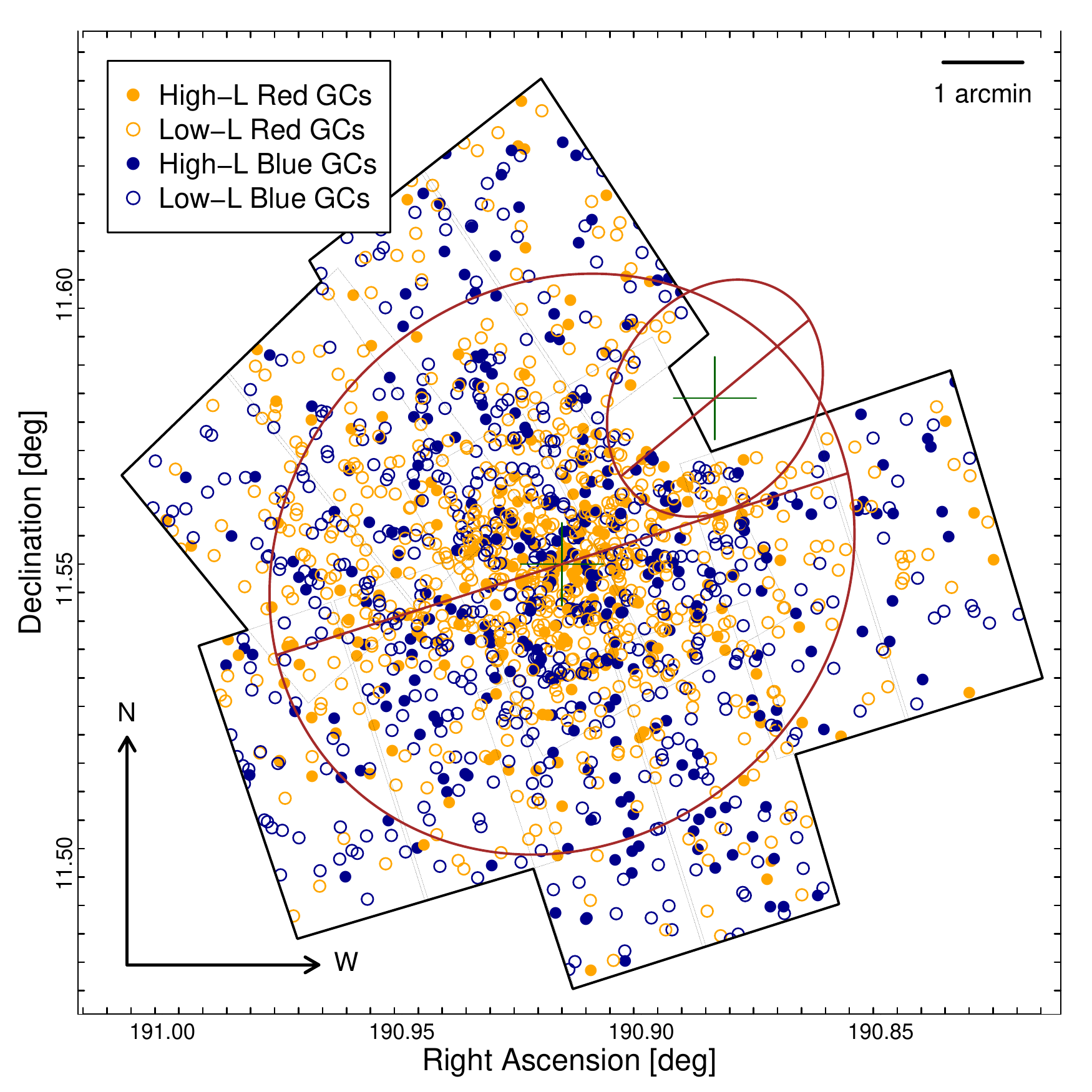}
	\includegraphics[height=5.5cm,width=5.5cm,angle=0]{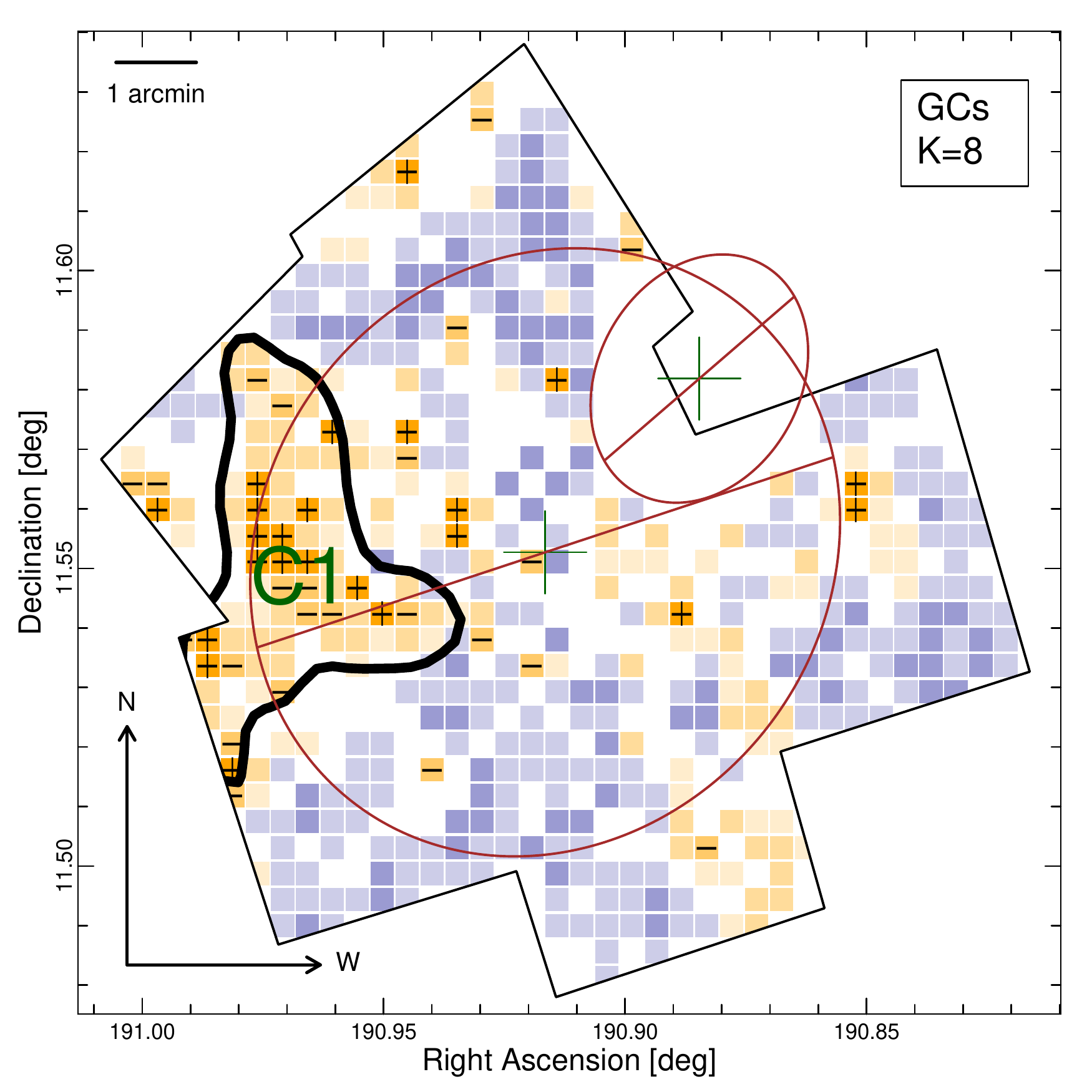}	
	\includegraphics[height=5.5cm,width=5.5cm,angle=0]{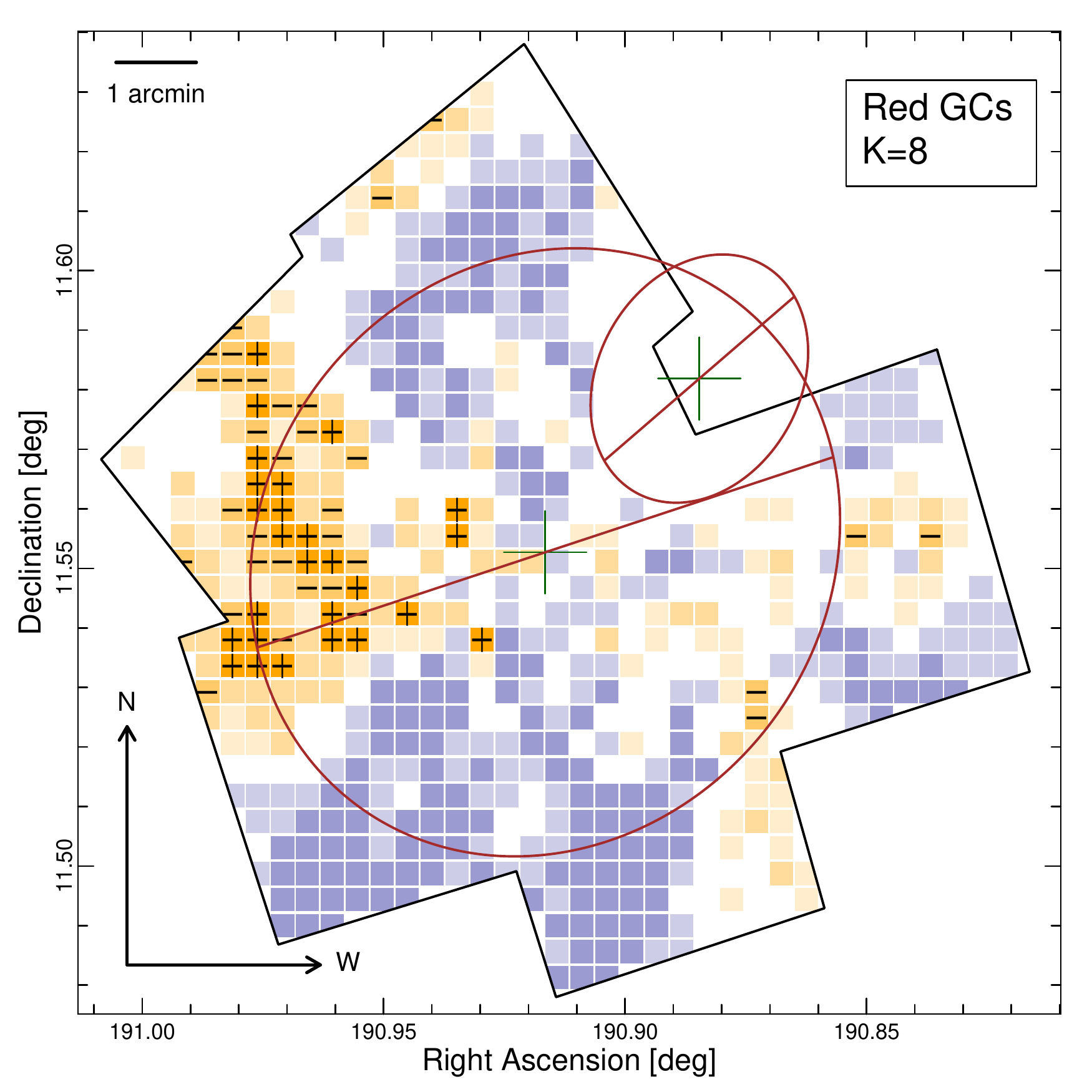}\\
	\includegraphics[height=5.5cm,width=5.5cm,angle=0]{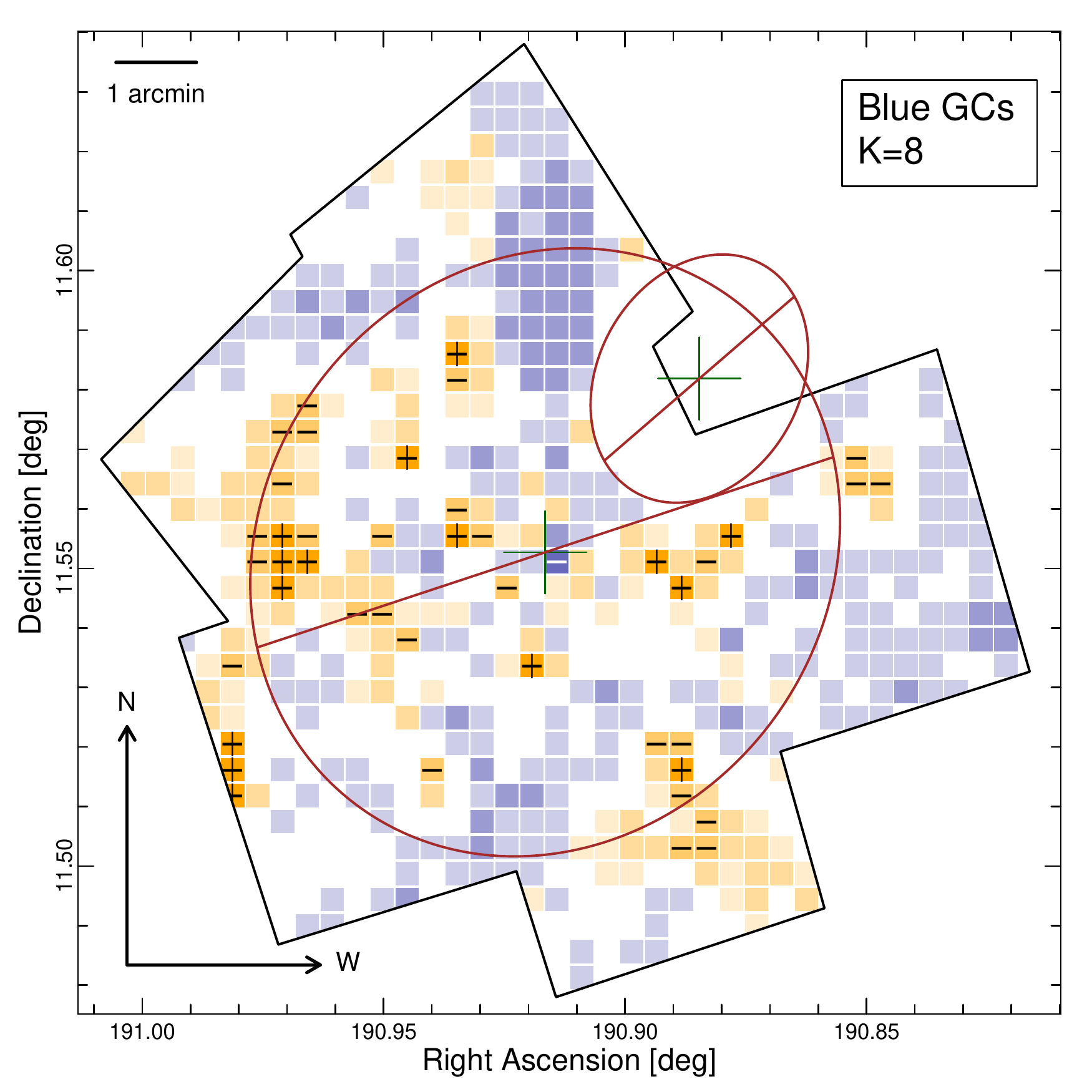}
	\includegraphics[height=5.5cm,width=5.5cm,angle=0]{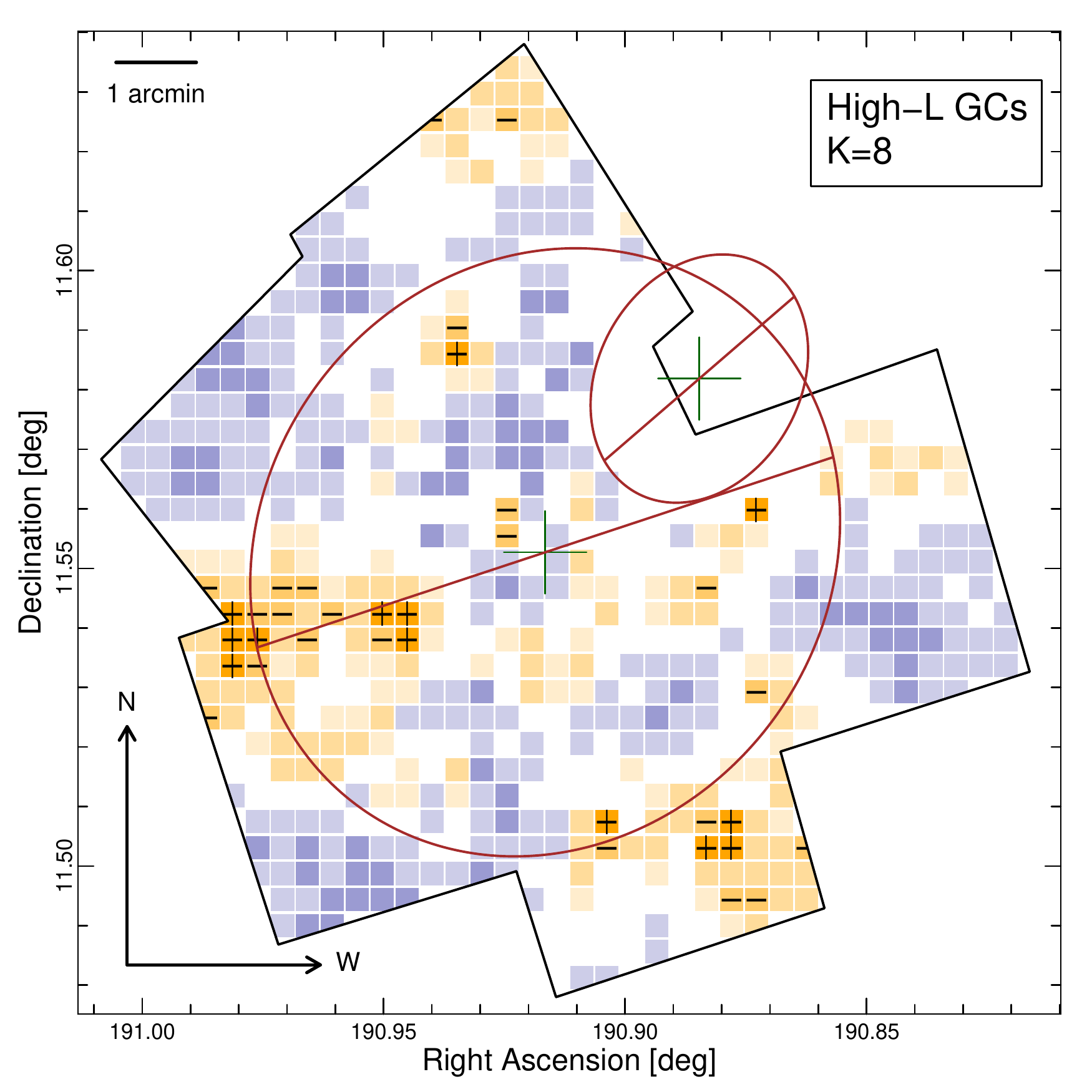}
	\includegraphics[height=5.5cm,width=5.5cm,angle=0]{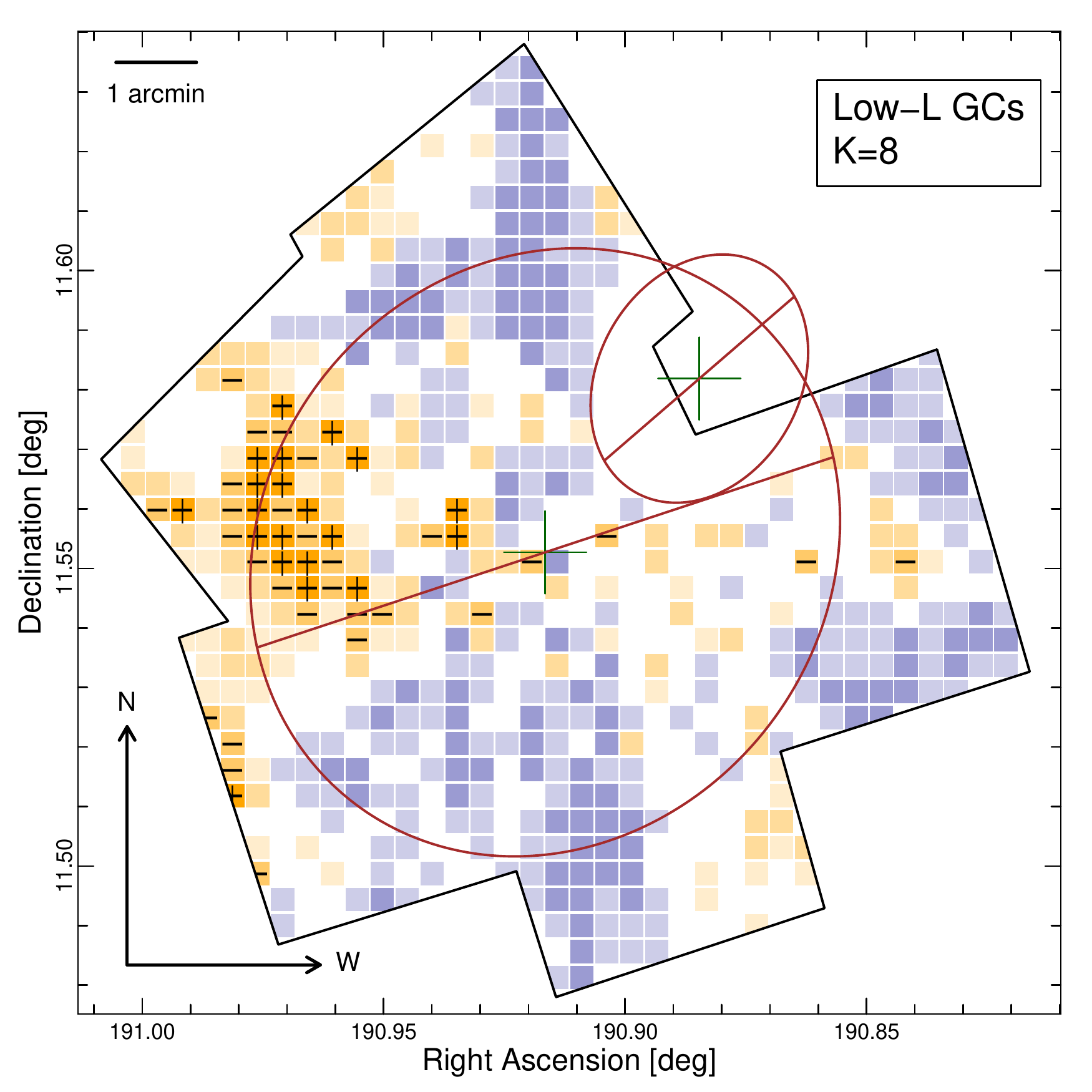}		
	\caption{Scatterplot of the position of the GCs in NGC4649 and residual maps obtained for 
	$K\!=\!8$ for the whole sample of GCs and the two color and luminosity classes. These plots
	are slightly modified versions of the plots shown by~\cite{dabrusco2014a}. Refer to 
	Figure~\ref{fig:ngc4472} for a description of each panel. The smaller ellipse represents
	the area occupied by NGC4647, a less massive spiral galaxy that is thought to be interacting with 
	NGC4649. This area was not observed with ACS~\citep{strader2012}.}
	\label{fig:ngc4649}
\end{figure*}

The residual map ($K\!=\!8$) of the spatial distribution of 
all GCs detected in NGC4649~\citep{dabrusco2014a} 
(Figure~\ref{fig:ngc4649}, upper mid panel) is dominated by a 
highly significant 
($\sim\!10\sigma$), spatially extended ($\sim\!50$ pixels) over-density structure (C1) in the E side of the 
galaxy. This feature, less prominent in red GCs, originates close to the major axis, following it out to the 
$D_{25}$ ($\sim\!4.5\ r_{e}$) before bending northward; high-L and low-L GCs appear spatially segregated in C1 
(Figure~\ref{fig:ngc4649}, lower mid and right panels), with the former 
mostly located along the major axis and the latter following the $D_{25}$ ellipse. 
The blue GCs residual map (Figure~\ref{fig:ngc4472}, lower left panel) 
also shows few more marginally significant over-density structures in the S and W sides of 
the galaxies. As discussed in~\cite{dabrusco2014a}, the GC over-density structure is not associated 
to any feature in the diffuse stellar light. Recent kinematic measurements~\citep{arnold2014} 
of the diffuse stellar light have revealed disk-like outer rotation, which suggests a major
dry merger undergone by a massive lenticular galaxy progenitor. No obvious correlation between
the GCs structure and the radial velocities of a sample of GCs observed spectroscopically 
by~\cite{lee2008} was found~\citep{dabrusco2014a}.

\subsection{NGC4406}
\label{subsec:ngc4406}

\begin{figure*}[]
	\includegraphics[height=5.5cm,width=5.5cm,angle=0]{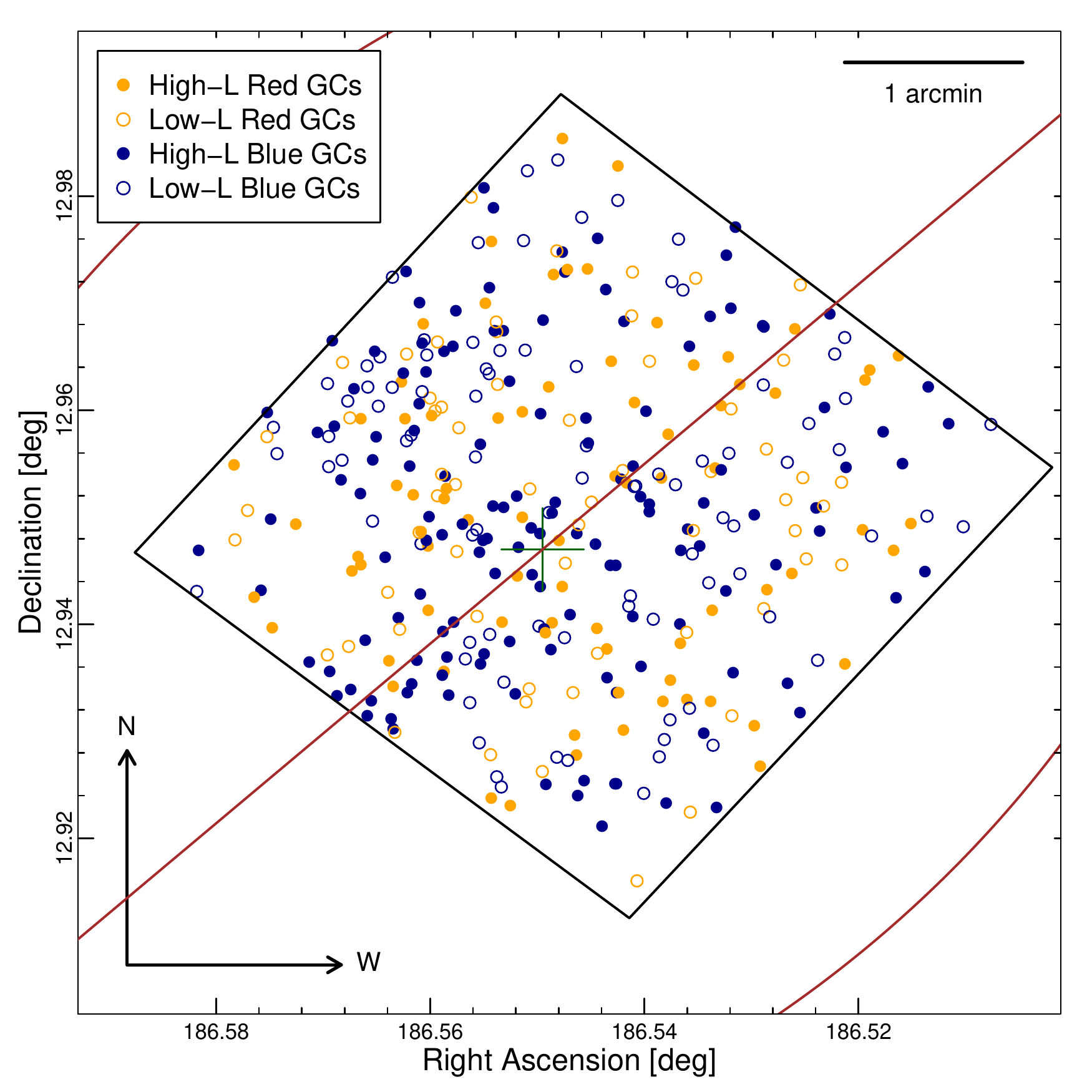}
	\includegraphics[height=5.5cm,width=5.5cm,angle=0]{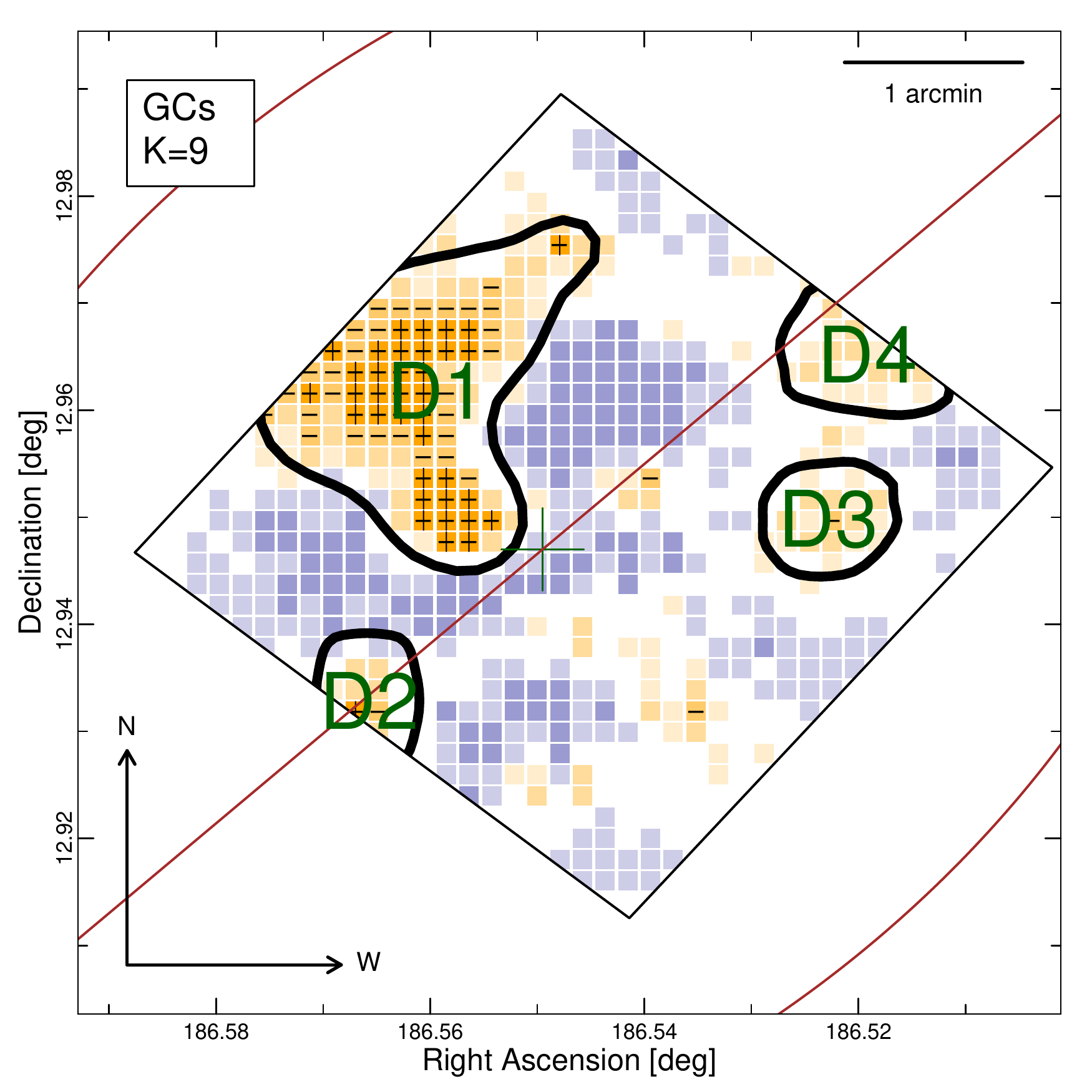}	
	\includegraphics[height=5.5cm,width=5.5cm,angle=0]{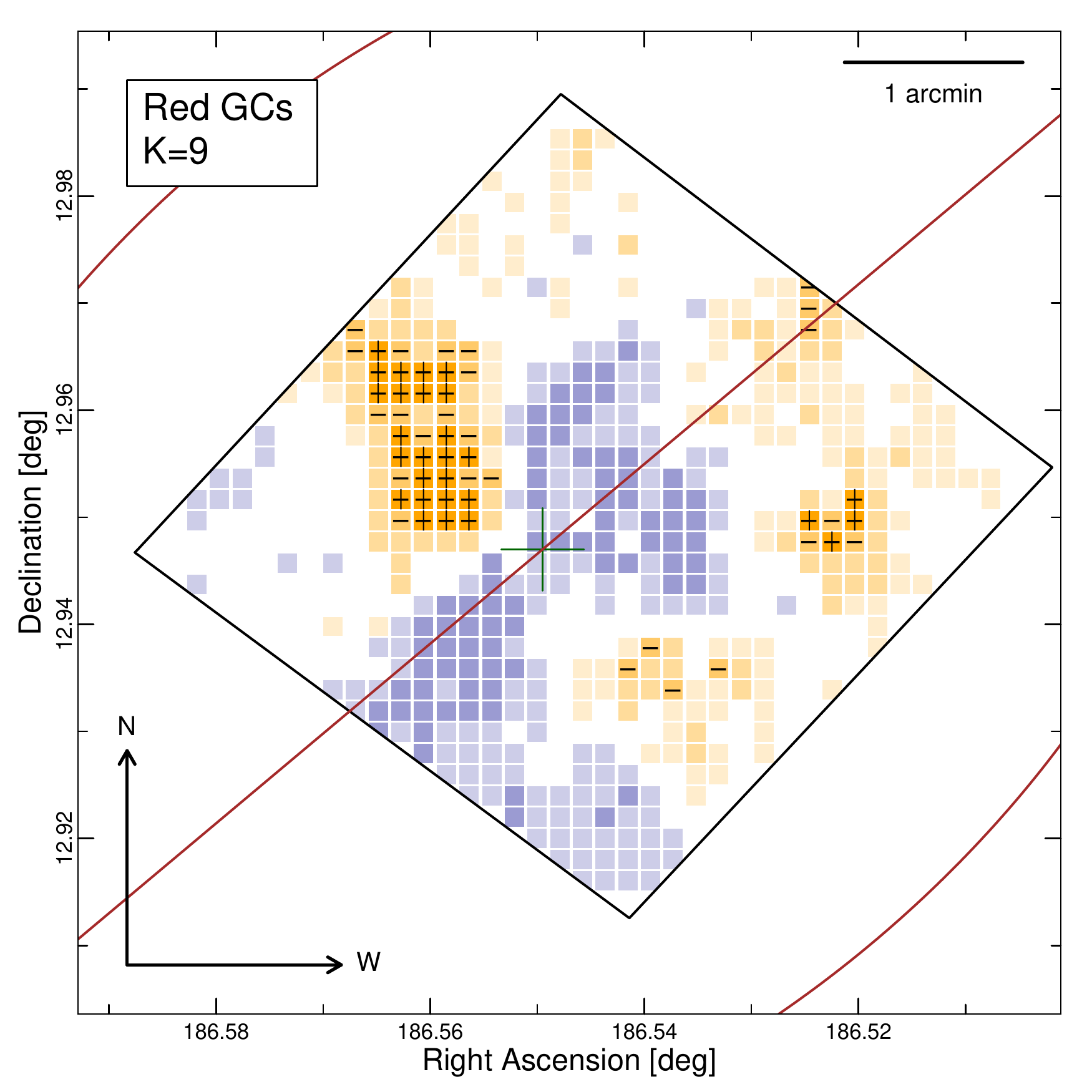}\\
	\includegraphics[height=5.5cm,width=5.5cm,angle=0]{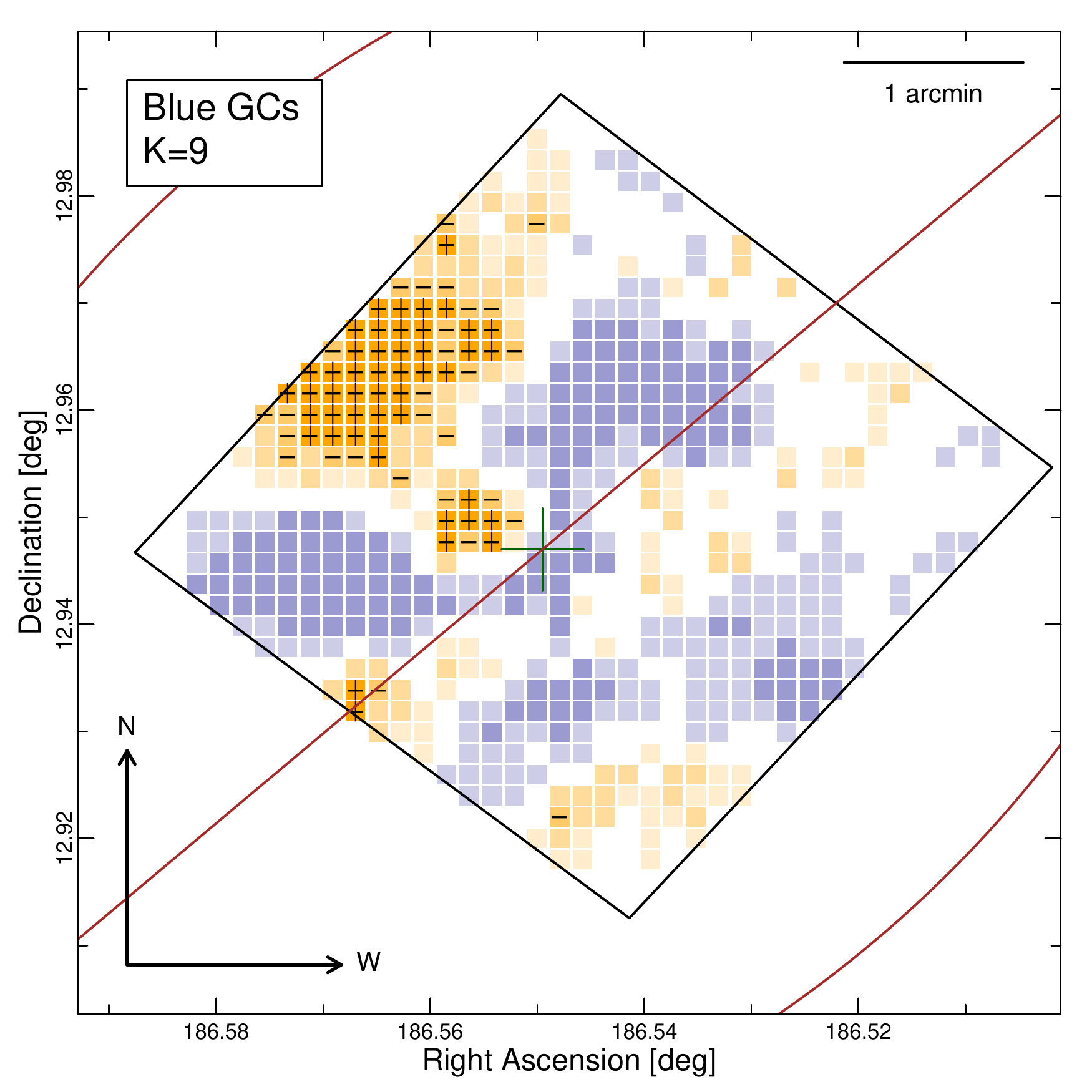}
	\includegraphics[height=5.5cm,width=5.5cm,angle=0]{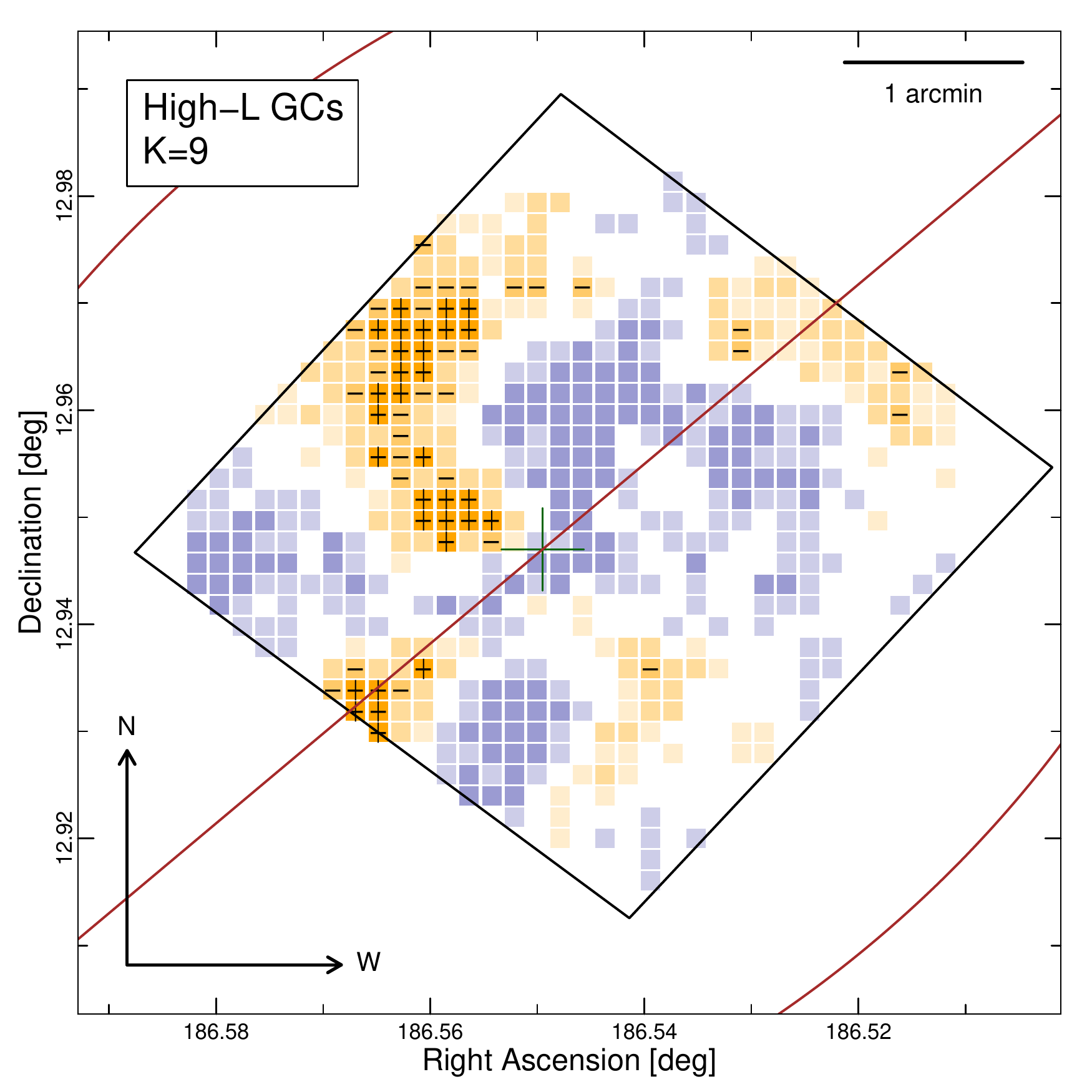}
	\includegraphics[height=5.5cm,width=5.5cm,angle=0]{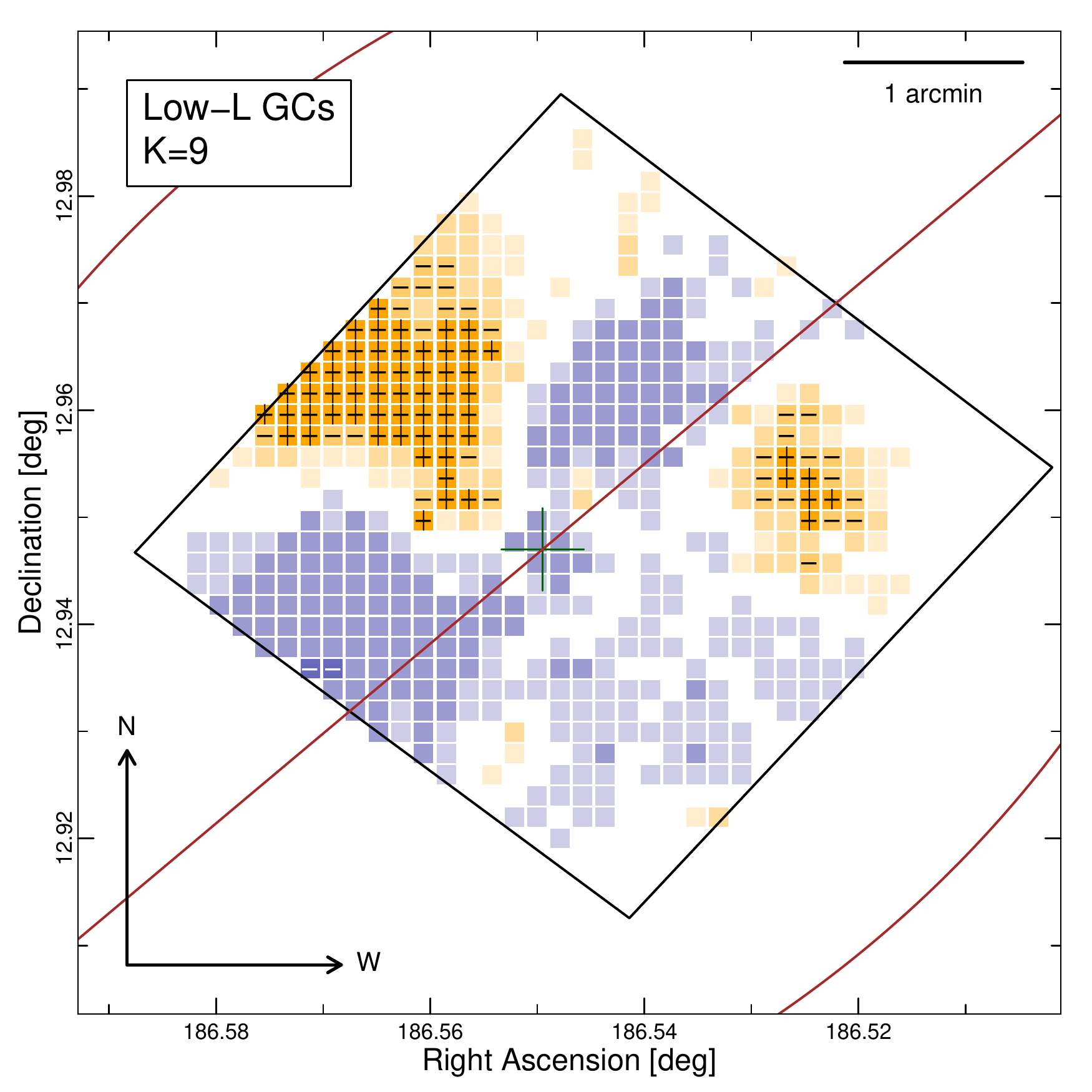}		
	\caption{Scatterplot of the position of the GCs in NGC4406 and residual maps obtained for 
	$K\!=\!9$ for the whole sample of GCs and the two color and luminosity GC classes. Refer 
	to Figure~\ref{fig:ngc4472} for a description of each panel.}
	\label{fig:ngc4406}
\end{figure*}

The $K\!=\!9$ residual map of all GCs in NGC4406 (Figure~\ref{fig:ngc4406}, upper mid panel) 
is dominated by a large, spatially coherent structure (D1, $\sim\!15\sigma$) located N-E of the center of 
the galaxy. At small galactocentric distances, D1 follows the radial direction, but it widens
as it reaches the boundary of the field. D1 is clearly visible in both color and luminosity GC classes 
(Figure~\ref{fig:ngc4406}, upper right and lower panels). While D1 in red GCs
is mostly located along the radial direction, in blue GCs it is more azimuthally extended. 
The second most significant residual structure (D2, $\sim\!4.5\sigma$) overlaps the 
S-E end of the major axis. D3 and D4 ($\sim\!4.3\sigma$ and $\sim\!4\sigma$ 
respectively) are visible in the whole sample of GCs, to the W of the major axis. 
D3 is more evident in the red (Figure~\ref{fig:ngc4406}, upper right panel) and low-L 
(Figure~\ref{fig:ngc4406}, lower left panel) GCs, indicating the presence of a cluster of 
metal-rich, low-luminosity GCs at galactocentric radius $\sim\!1.4\arcmin$.

A sub-arcsecond dusty disk has been observed in NGC4406~\citep{ferrarese2006,diserego2013}. 
Since the spatial scale of the structure that we detect in the GC distribution
is significantly larger that few arcseconds, the presence of the dust feature should not affect our results.

\subsection{NGC4382}
\label{subsec:ngc4382}

\begin{figure*}[]
	\includegraphics[height=5.5cm,width=5.5cm,angle=0]{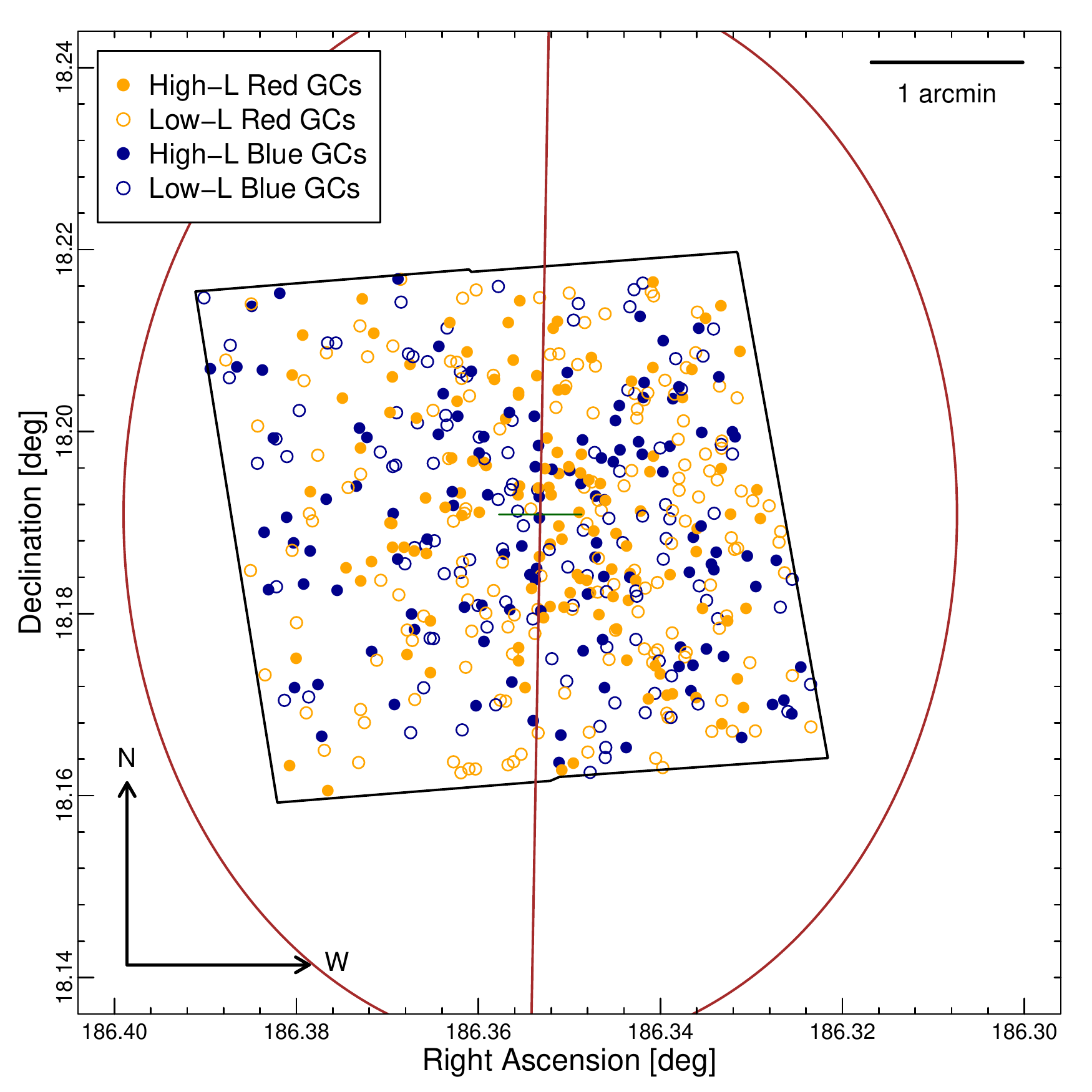}
	\includegraphics[height=5.5cm,width=5.5cm,angle=0]{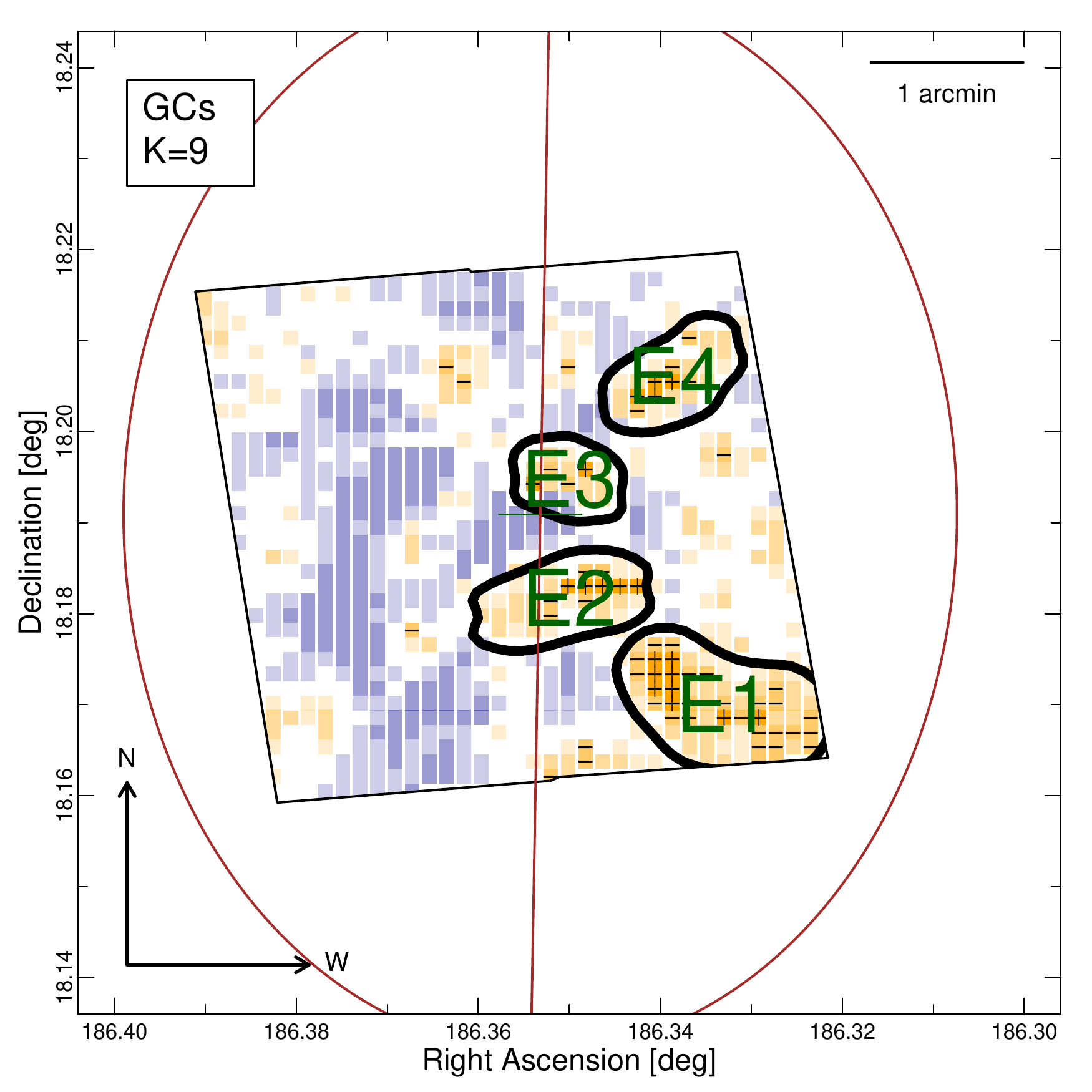}	
	\includegraphics[height=5.5cm,width=5.5cm,angle=0]{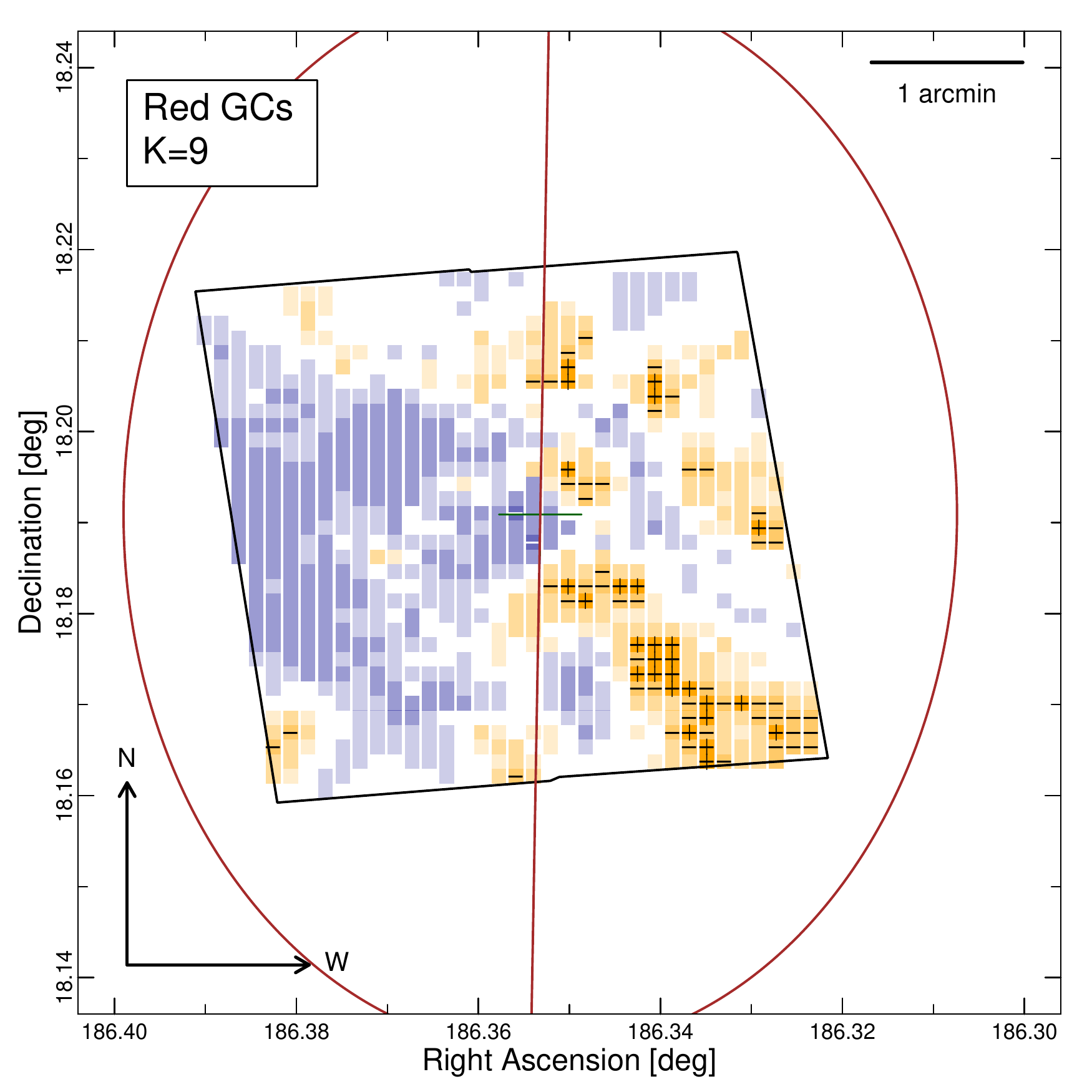}\\
	\includegraphics[height=5.5cm,width=5.5cm,angle=0]{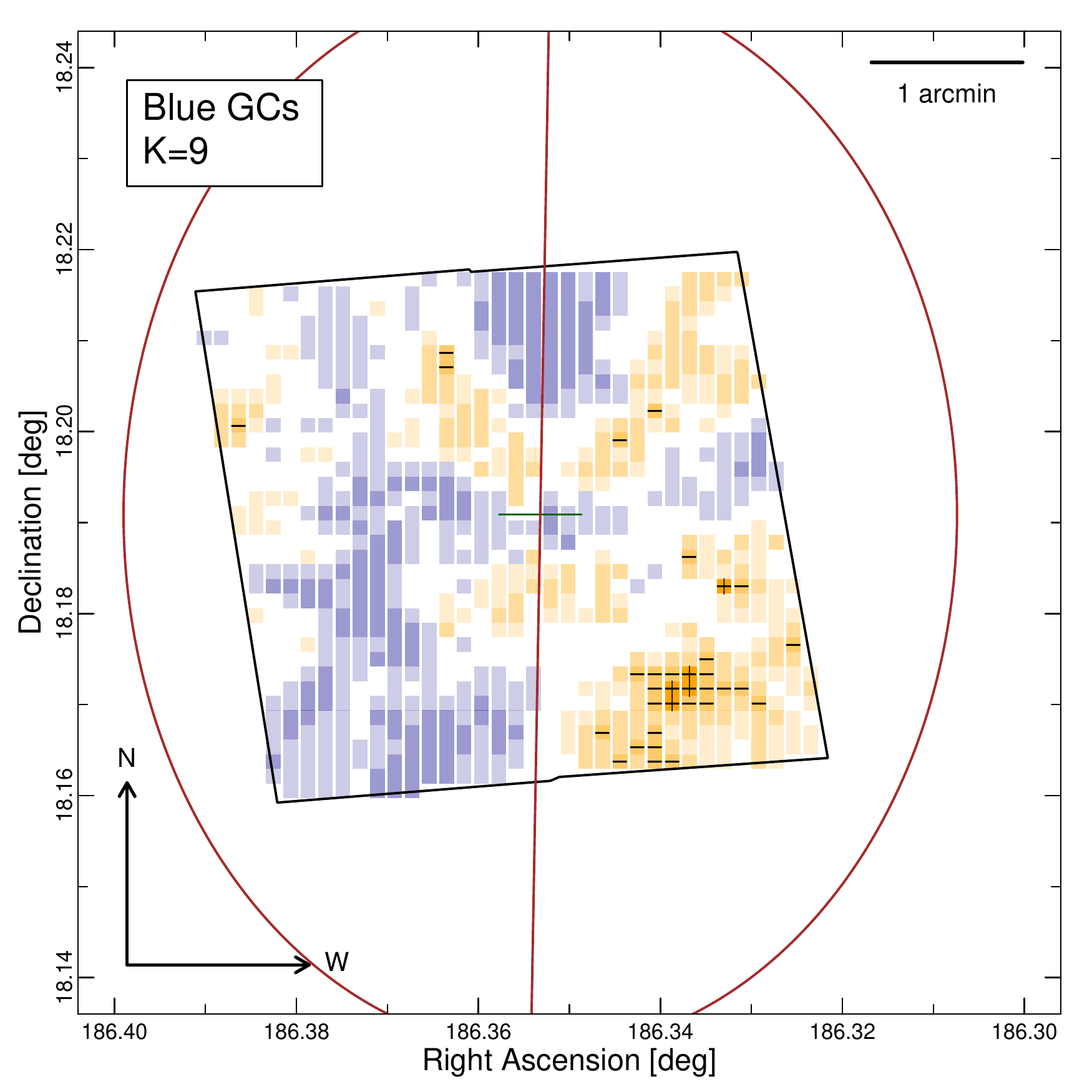}
	\includegraphics[height=5.5cm,width=5.5cm,angle=0]{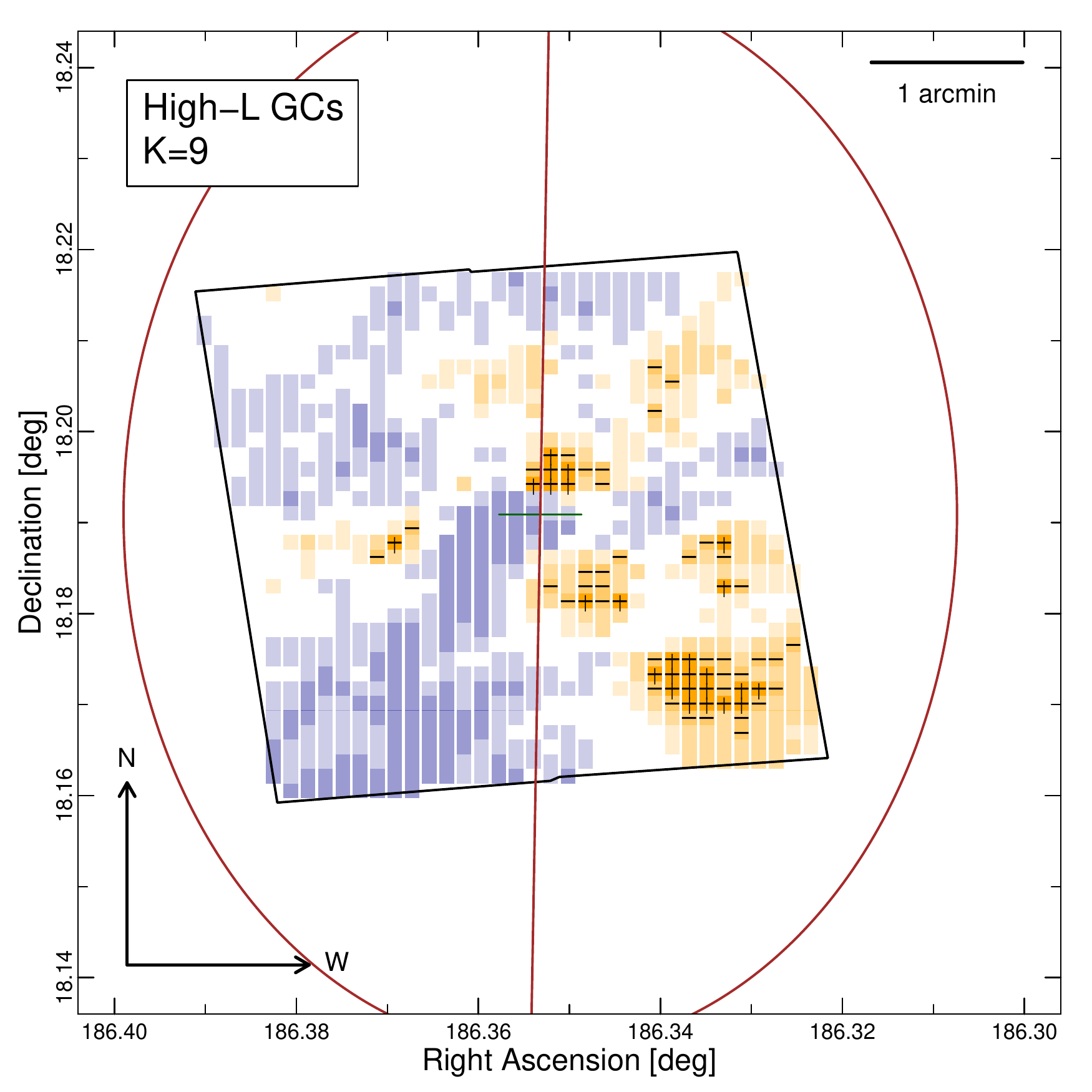}
	\includegraphics[height=5.5cm,width=5.5cm,angle=0]{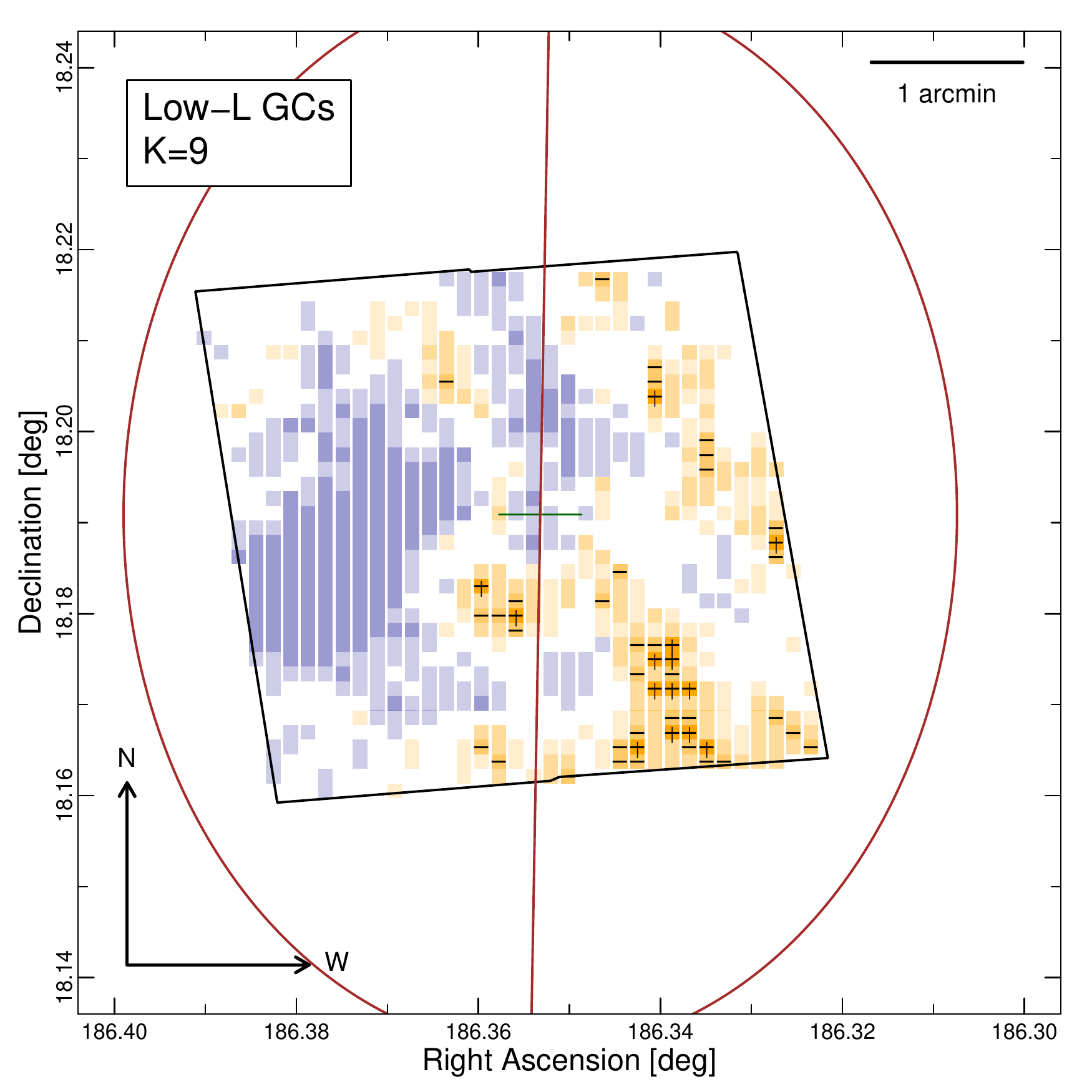}		
	\caption{Scatterplot of the position of the GCs in NGC4382 and residual maps obtained for 
	$K\!=\!9$ for the whole sample of GCs and the two color and luminosity GC classes. Refer 
	to Figure~\ref{fig:ngc4472} for a description of each panel.}
	\label{fig:ngc4382}
\end{figure*}

The $K\!=\!9$ spatial distribution of all GCs in NGC4382, 
(Figure~\ref{fig:ngc4382}, upper mid panel), is characterized by the presence of two large, significant 
over-density structures. E1 ($>\!10\sigma$), located in the S-W corner of the 
HST field, is radial in the entire GC sample and is visible in all GC color classes (Figure~\ref{fig:ngc4382}, 
upper right and lower left panels). E2 ($\sim\!6.8\sigma$), S of the center of the galaxy, crosses the major 
axis at $\sim\!0.8\arcmin$ from the center. E2 and E1 
could be distinct sections of a single elongated structure. A second group 
of significant and spatially coherent structures is located N-W; these are clearly 
visibile in the residual maps of all and red GCs: E3 ($\sim\!6.3\sigma$) 
and E4 ($\sim\!5.7\sigma$). E4 looks similar to 
E1, and extends radially from the center to the N-W corner of the field 
(Figure~\ref{fig:ngc4382}, upper mid panel). Another property of the spatial distribution of all GCs in 
NGC4382 is the very significant ($\sim\!6.5\sigma$) E-W anisotropy, with $\sim\!68\%$ of 
all GCs located E of the major axis of the galaxy. Since the ACSVCS 
image is not centered on NGC4382 center, it is not surprising tat the asymmetry between the E and W 
equal area halves of the ACSVCS field is slightly more significant, with $\sim\!72\%$ of the total GCs located in the 
western half of the field.

\subsection{NGC4374}
\label{subsec:ngc4374}

\begin{figure*}[]
	\includegraphics[height=5.5cm,width=5.5cm,angle=0]{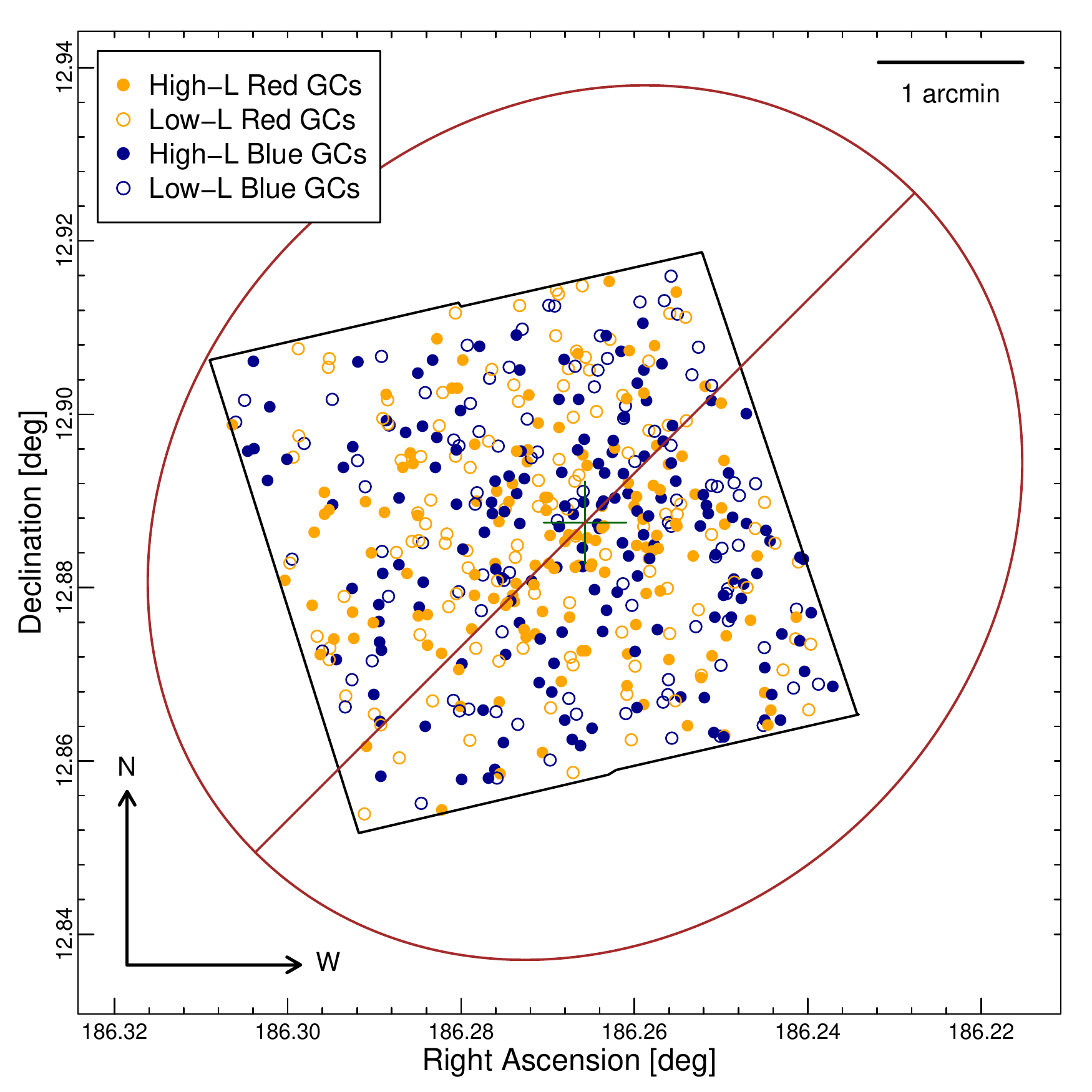}
	\includegraphics[height=5.5cm,width=5.5cm,angle=0]{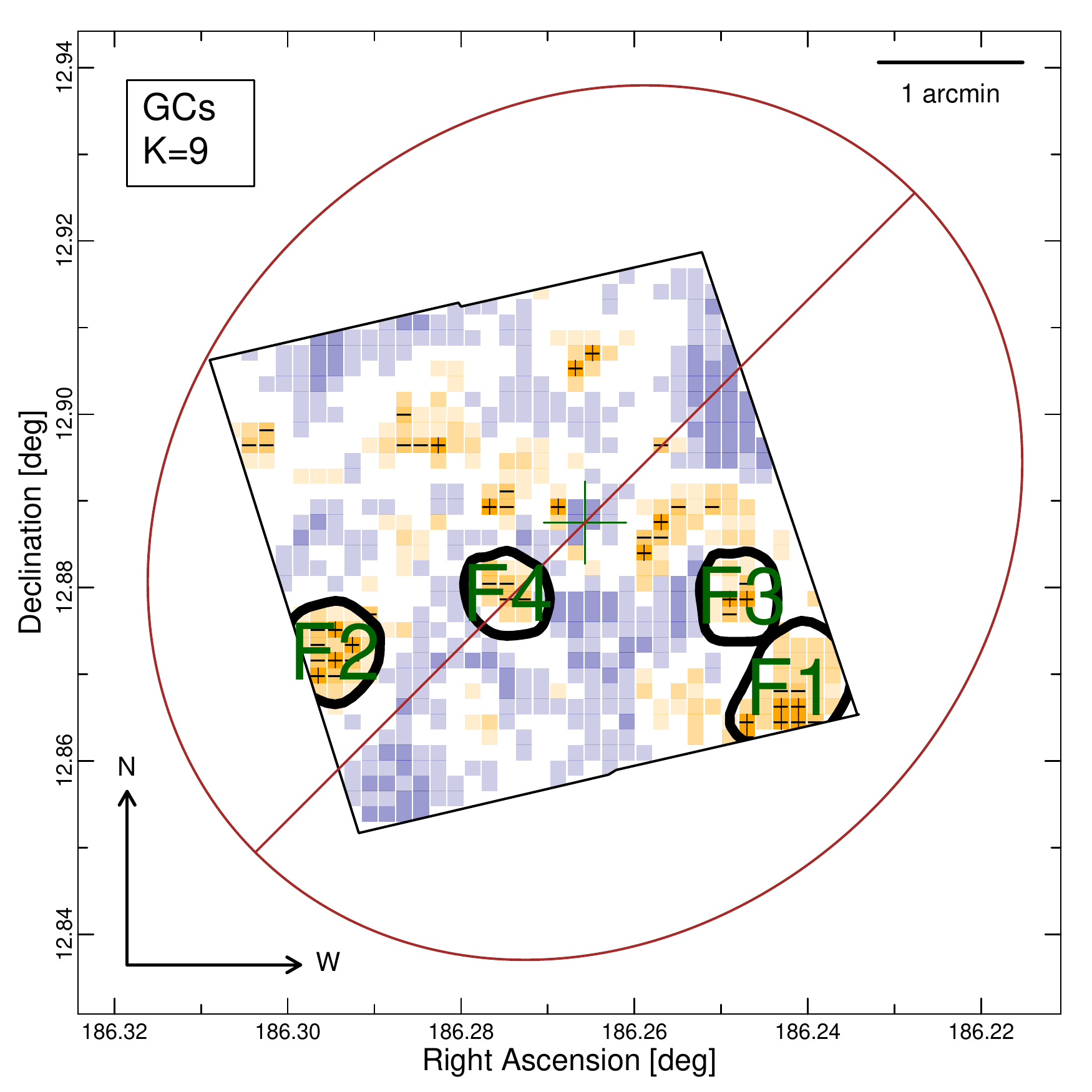}	
	\includegraphics[height=5.5cm,width=5.5cm,angle=0]{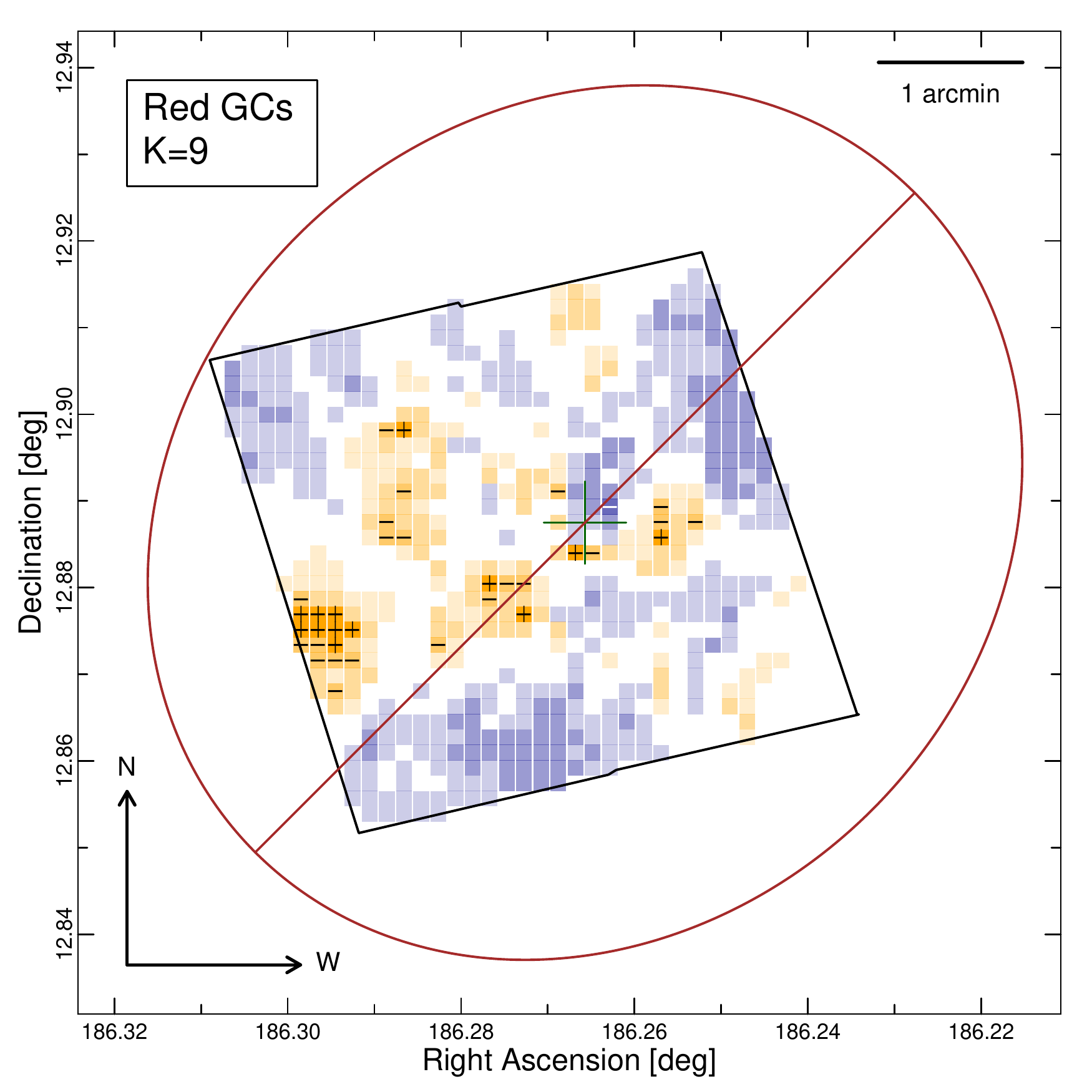}\\
	\includegraphics[height=5.5cm,width=5.5cm,angle=0]{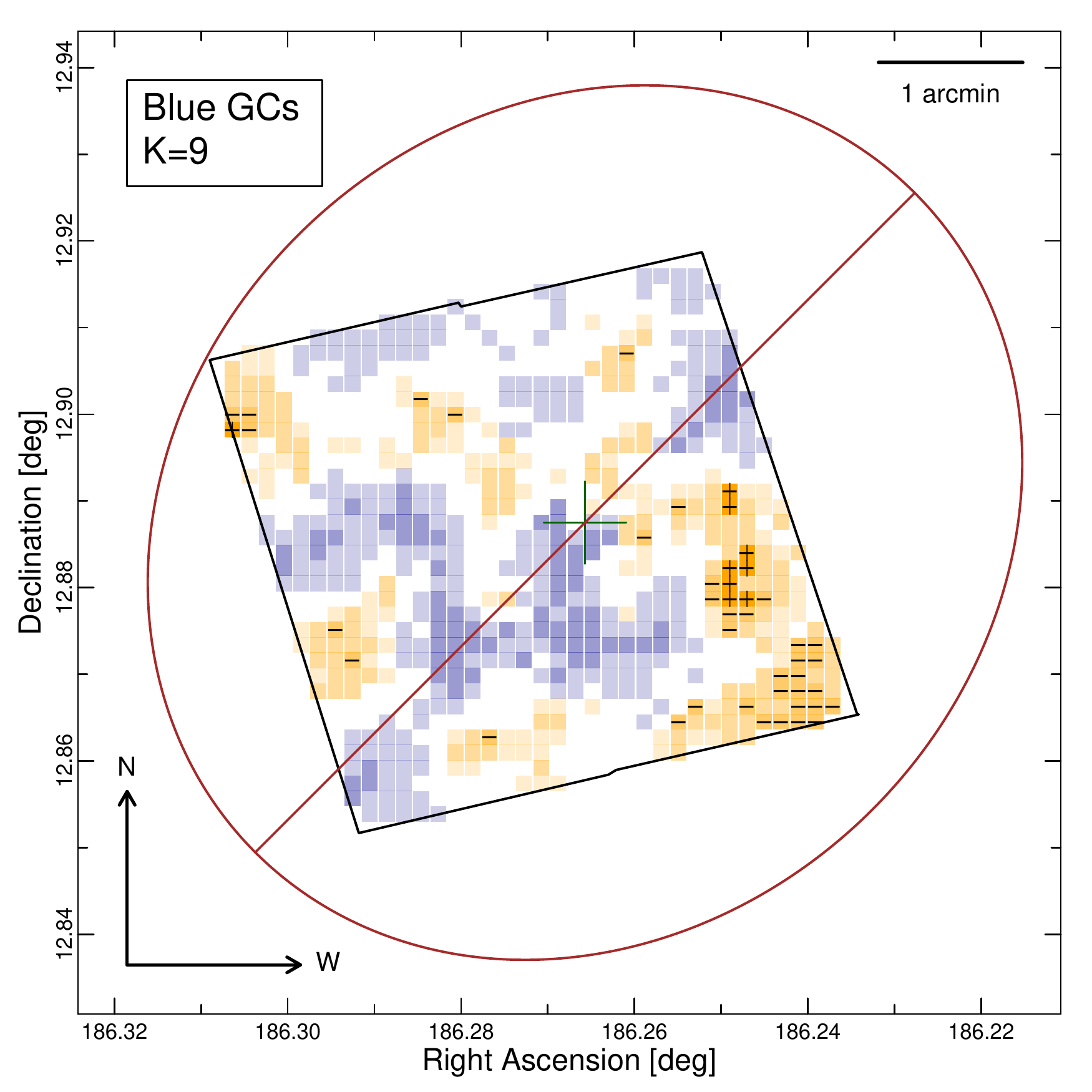}
	\includegraphics[height=5.5cm,width=5.5cm,angle=0]{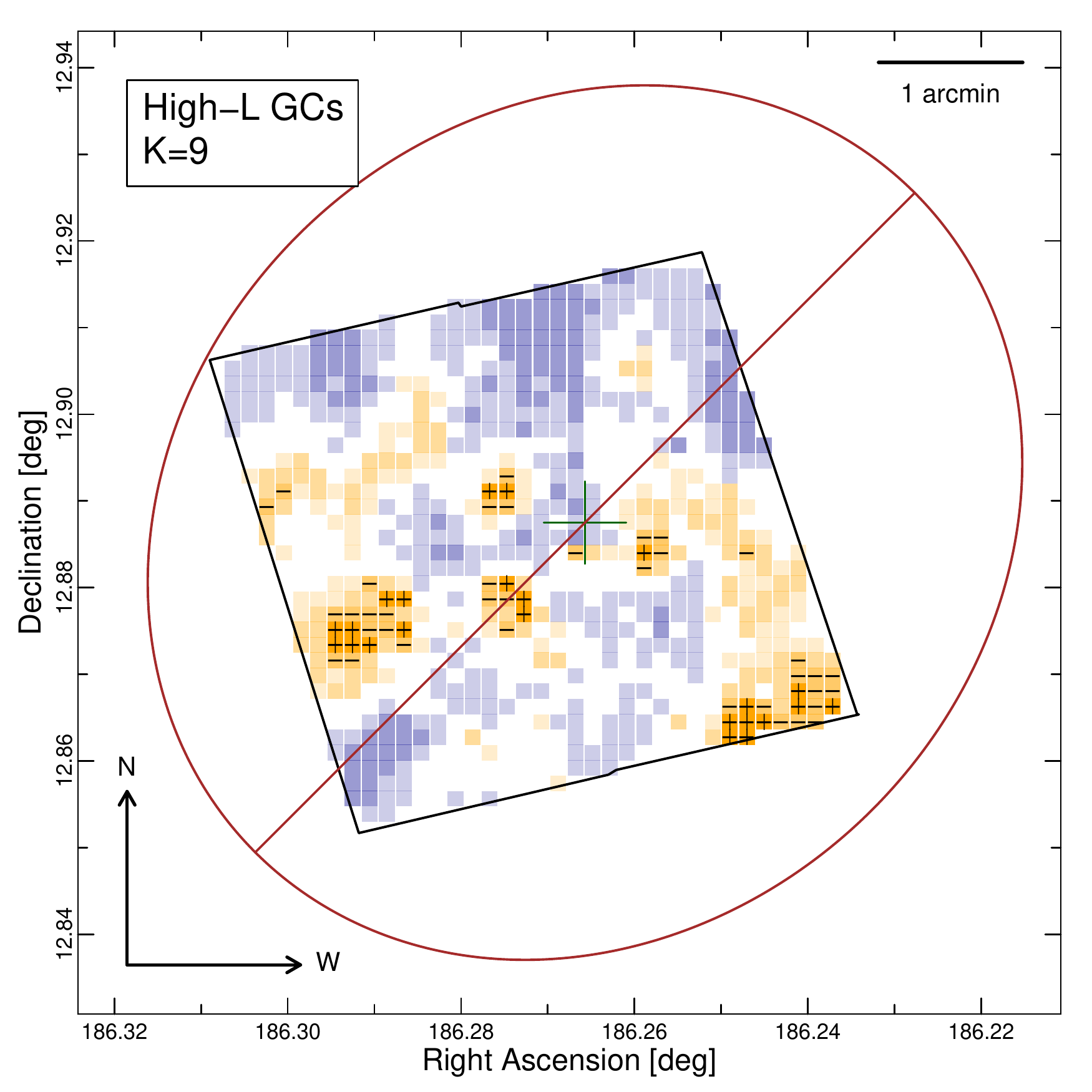}
	\includegraphics[height=5.5cm,width=5.5cm,angle=0]{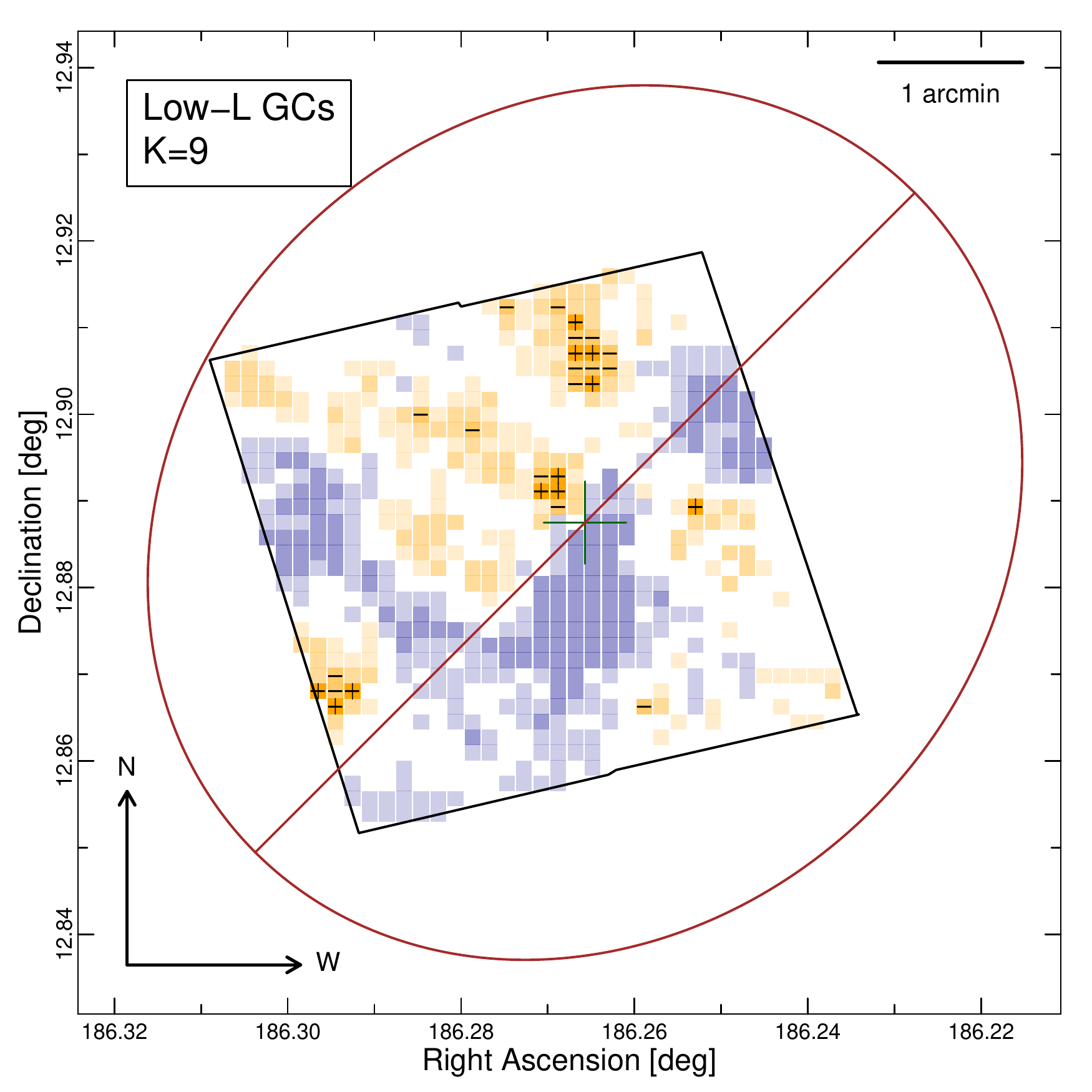}		
	\caption{Scatterplot of the position of the GCs in NGC4374 and residual maps obtained for 
	$K\!=\!9$ for the whole sample of GCs and the two color and luminosity GC classes. Refer 
	to Figure~\ref{fig:ngc4472} for a description of each panel.}
	\label{fig:ngc4374}
\end{figure*}

This galaxy has only one large over-density structure, F1 ($\sim\!6.5\sigma$), located in the S-W corner of 
the ACSVCS data field (Figure~\ref{fig:ngc4374}, upper mid panel). F1 is not visible in the red and 
low-luminosity GC classes (Figure~\ref{fig:ngc4374}, upper right 
and lower mid panels). The smaller over-density structures F2 ($\sim\!6.3\sigma$), F3 
($\sim\!5.5\sigma$) and F4 ($\sim\!4.5\sigma$) are roughly located along the major axis. 
F3 lies nearby F1, suggesting that they may be physically connected.

NGC4374 is known for the presence of abundant dust~\citep{diserego2013}. In particular, two parallel dust lanes,
located close to the center of the galaxy and slightly N of it along the E-W direction~\citep{ferrarese2006}, extend
for $\sim\!5\arcsec$ and $\sim\!14\arcsec$, respectively. While the presence of the dust lane may 
have affected the detection of GCs close to the position of F4, the GC structures F1, F2 and F3 are located 
too far from the dust to be significantly affected.

\subsection{NGC4526}
\label{subsec:ngc4526}

\begin{figure*}[]
	\includegraphics[height=5.5cm,width=5.5cm,angle=0]{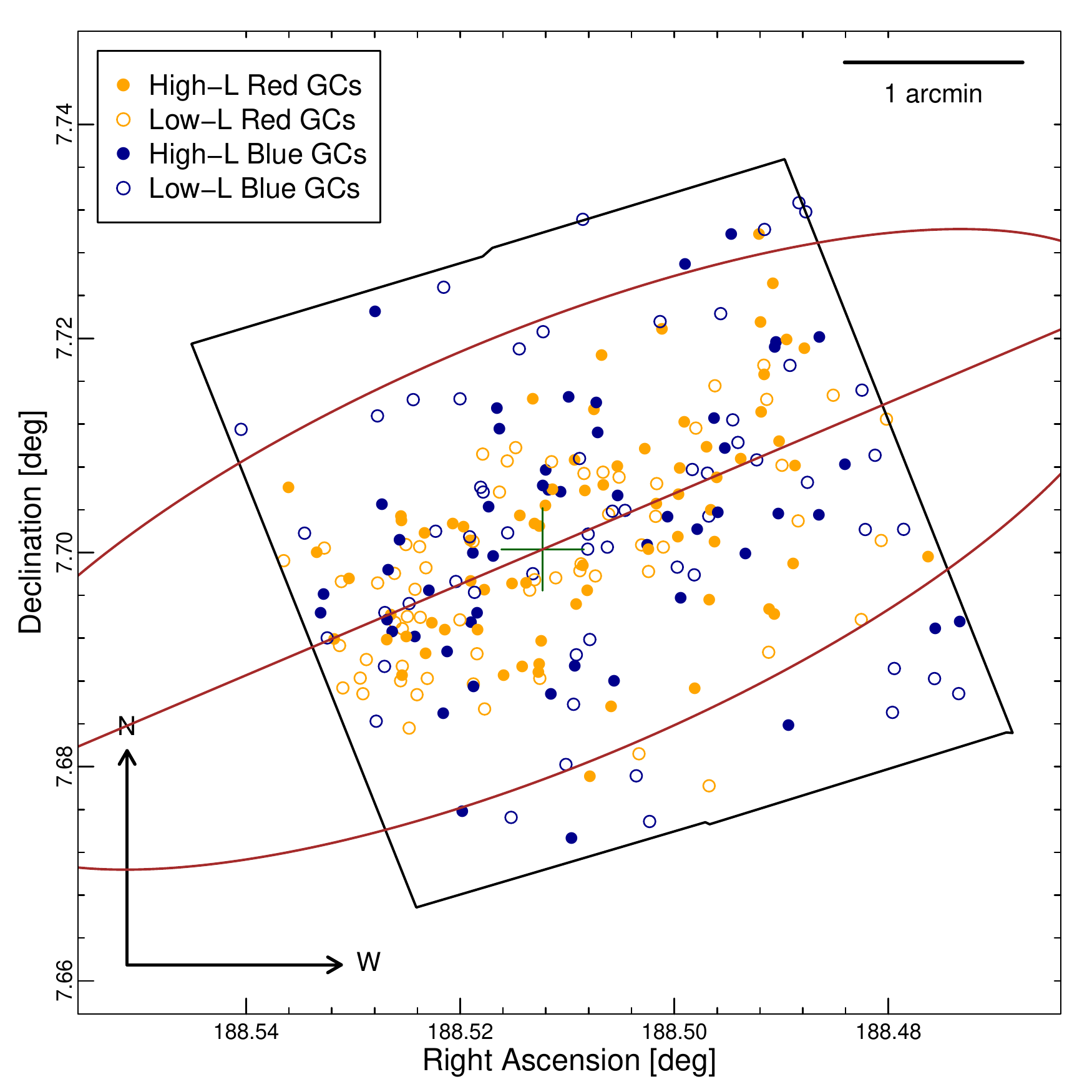}
	\includegraphics[height=5.5cm,width=5.5cm,angle=0]{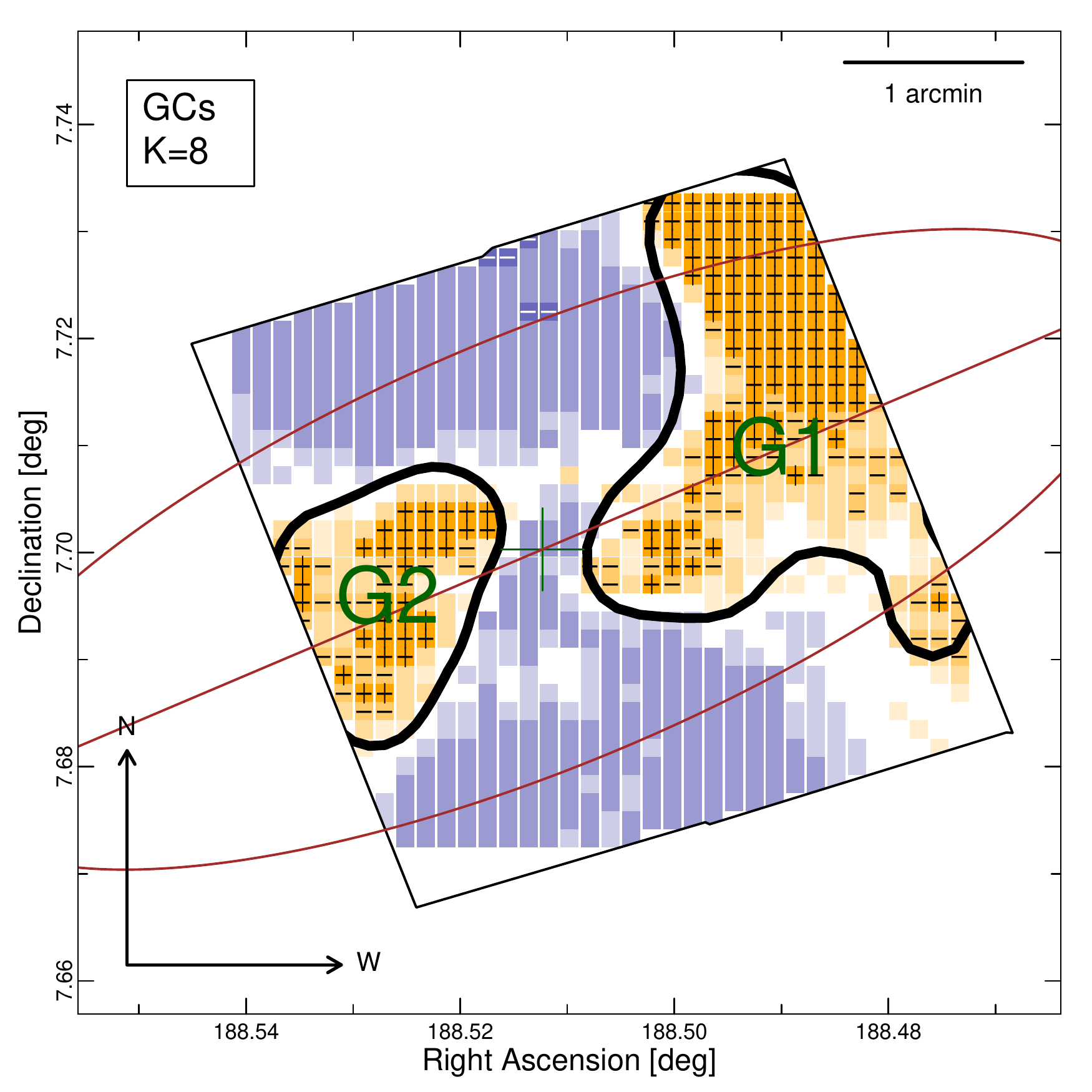}	
	\includegraphics[height=5.5cm,width=5.5cm,angle=0]{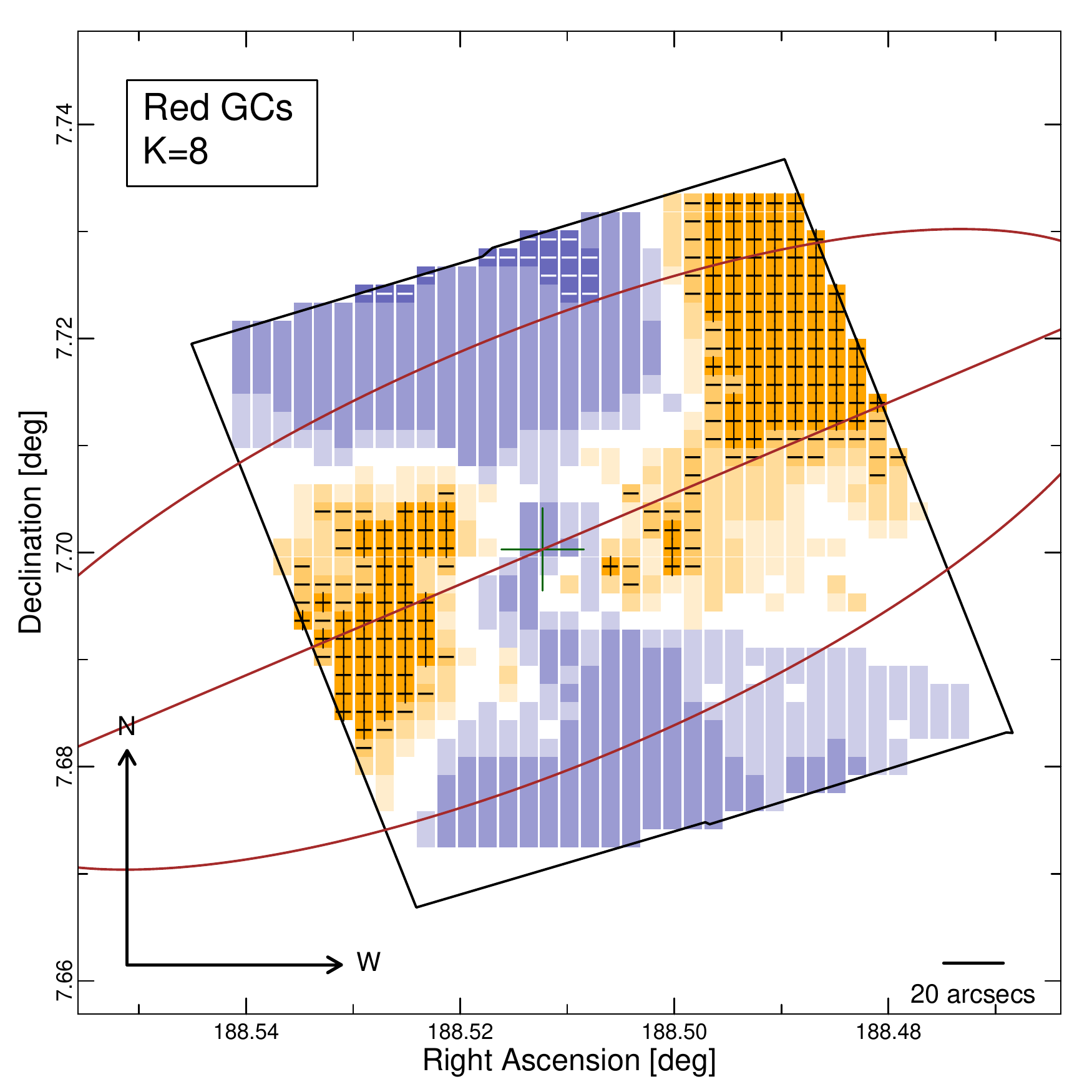}\\
	\includegraphics[height=5.5cm,width=5.5cm,angle=0]{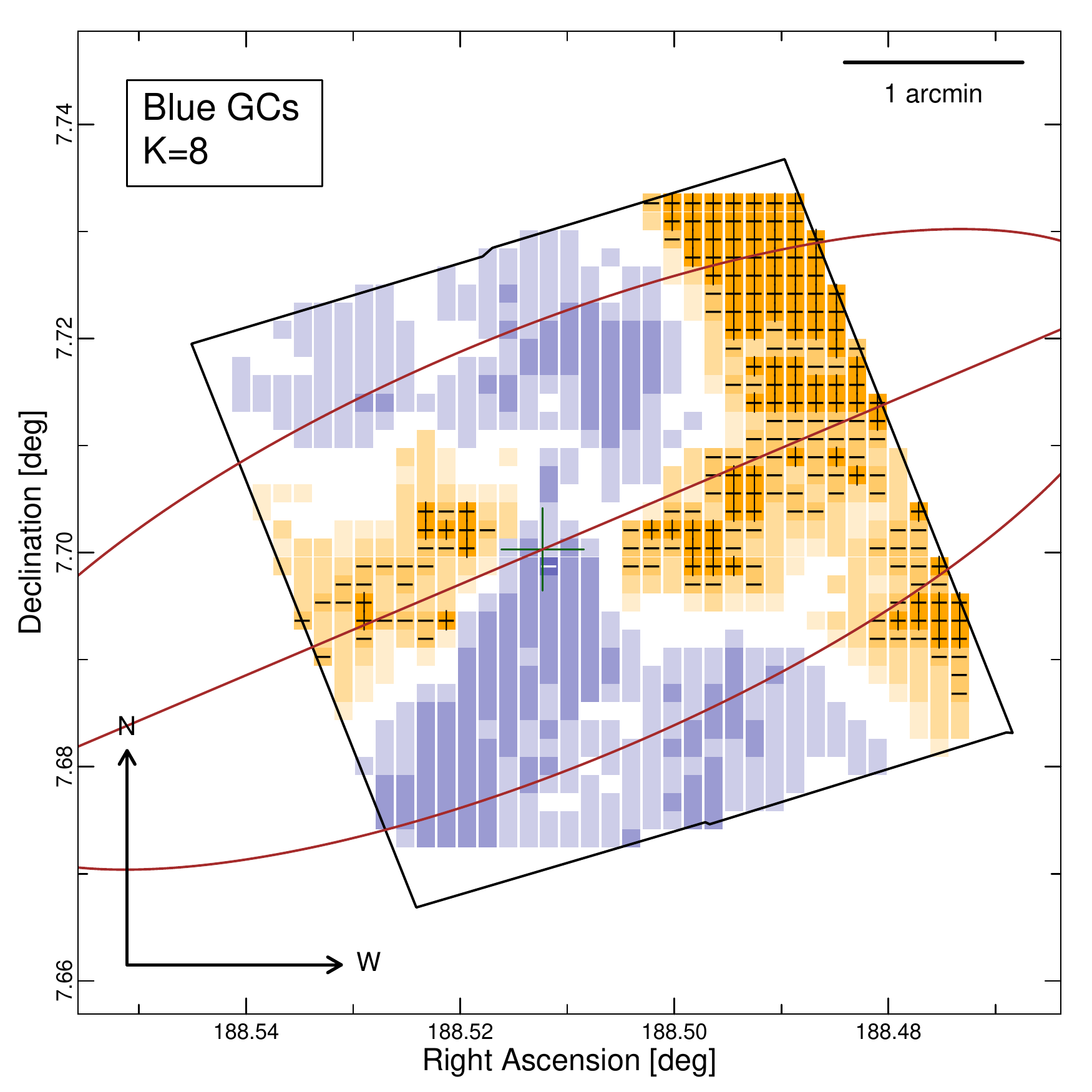}
	\includegraphics[height=5.5cm,width=5.5cm,angle=0]{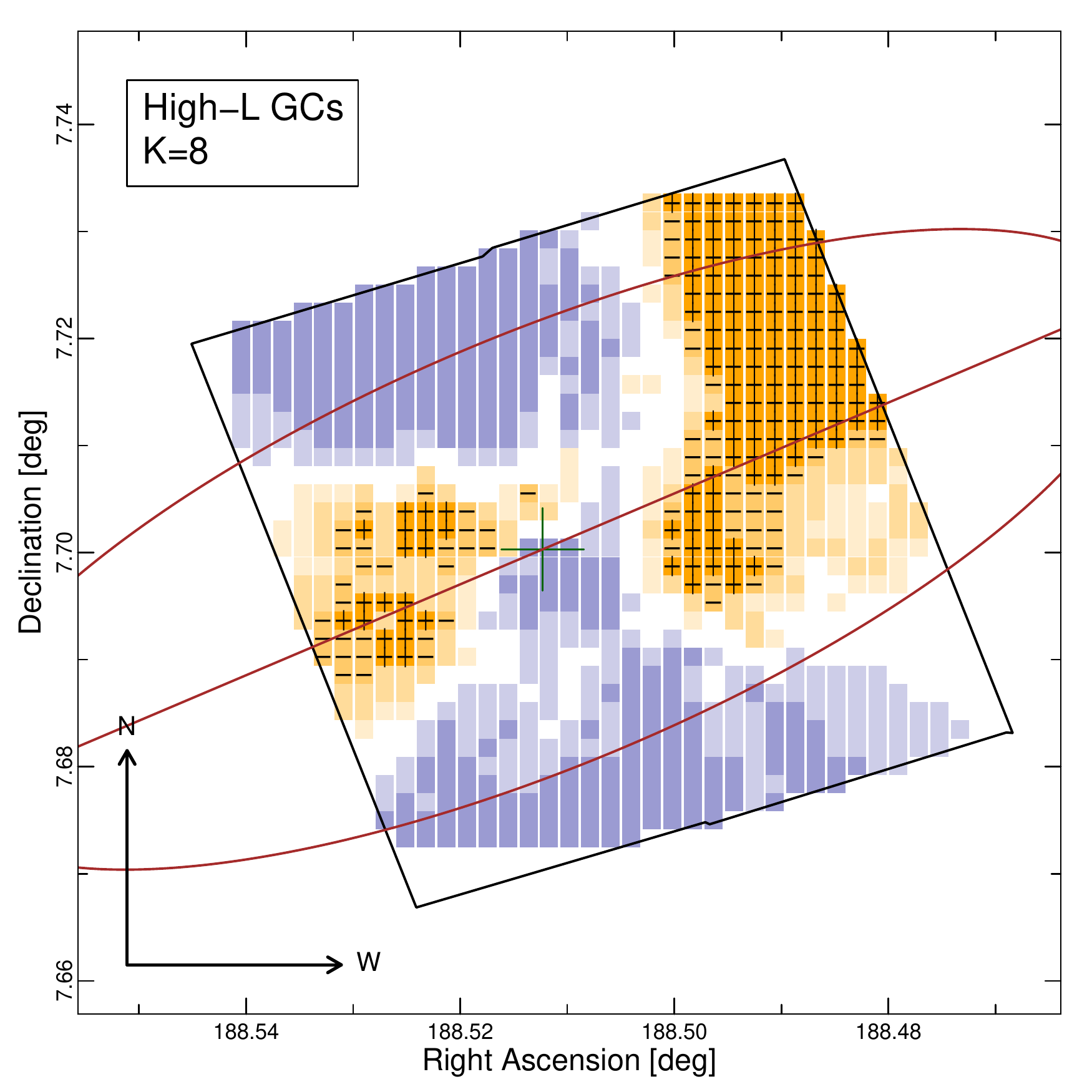}
	\includegraphics[height=5.5cm,width=5.5cm,angle=0]{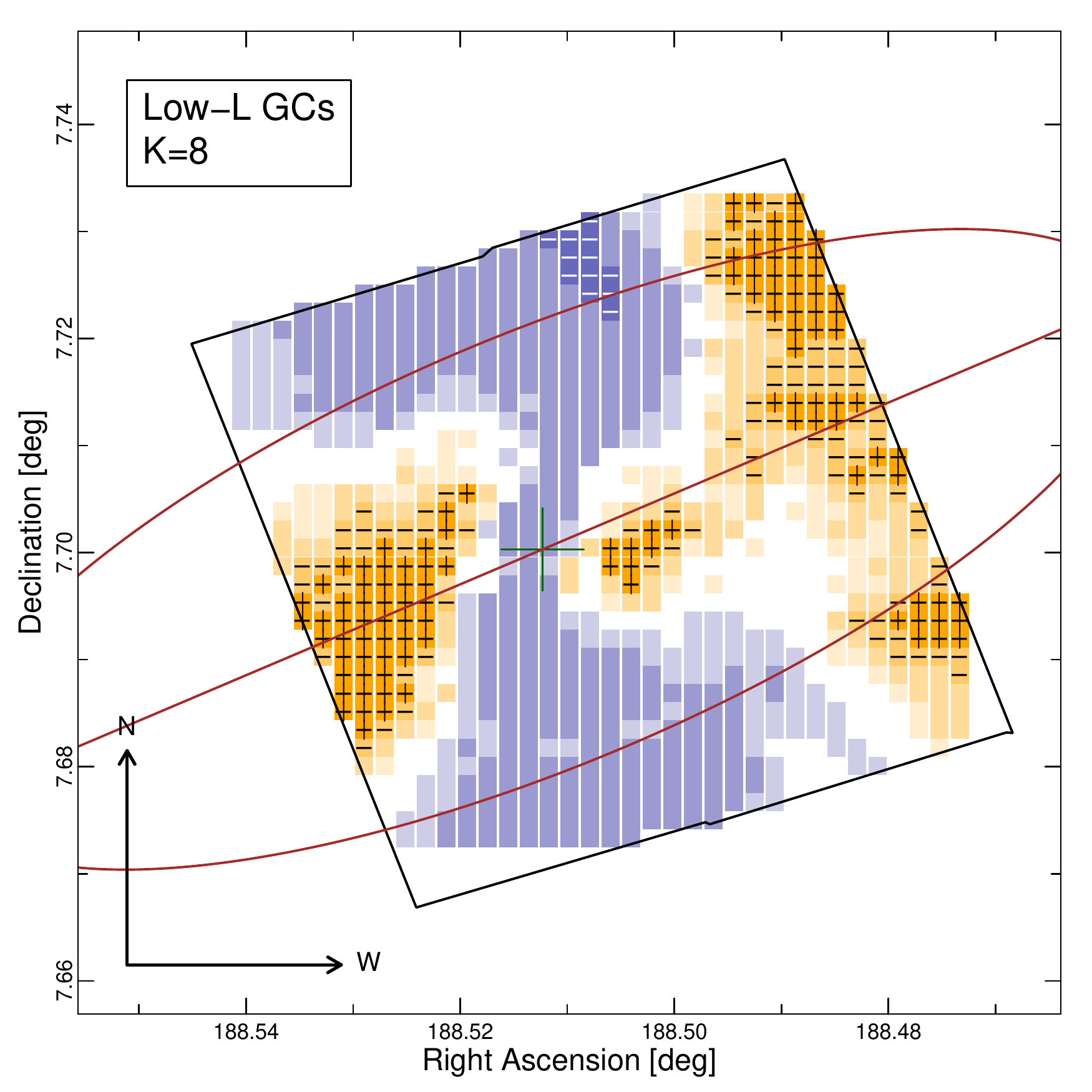}		
	\caption{Scatterplot of the position of the GCs in NGC4526 and residual maps obtained for 
	$K\!=\!9$ for the whole sample of GCs and the two color and luminosity GC classes. Refer 
	to Figure~\ref{fig:ngc4472} for a description of each panel.}
	\label{fig:ngc4526}
\end{figure*}

We detected (Figure~\ref{fig:ngc4526}, upper mid panel) two strong spatially extended over-density 
structures located W (G1) and E (G2) of the center of the galaxy, both $\geq\!10\sigma$ significant and 
clearly visible in all the classes (Figure~\ref{fig:ngc4526}, upper right and lower panels). G1, 
at small galactocentric distances ($<\!1\arcmin$), roughly follows the W section of the major axis
of the galaxy, and it extends towards N as it approaches the boundary of the field covered 
by the HST data, at galactocentric distance between $\sim\!1\arcmin$ and $\sim\!1.6\arcmin$. G2
overlaps the E section of the major axis. 

Since NGC4526 is a lenticular (S0\_3\_(6)) galaxy~\citep{ferrarese2006}, 
the two structures G1 and G2 could be enhanced by projection effects, assuming a smooth three-dimensional 
distribution of GCs that follows the galaxy light. In order to check whether projection effect may bias our results, 
we have produced additional residual maps
using a modified model for the simulated spatial distribution of GCs.
We add to the smooth background component with isotropic azimuthal distribution (the halo) 
an azimuthally anisotropic component aligned along the major axis, spanning  
$15\deg$ above and below the major axis measured at the $D_{25}$, to mimic the presence of a population of 
GCs associated to the disk. We have 
performed different experiments where the fractions of simulated GCs associated to either component 
varies between 30\% and 70\%. Both G1 and G2 are still visible, although with decreasing significance as the 
fraction of simulated GCs in the disk component grows. In the worst case scenario, the significance of G1 is 
$\sim\!7\sigma$ and the significance of G2 is $\sim3.5\sigma$.Based on the results, we can rule 
out that both G1 and G2 structures are entirely spurious.

Moreover, NGC4526 contains abundant dust~\citep{ferrarese2006,diserego2013}, mostly located in 
a large-scale ($15.6\arcsec$) disk structure aligned along the major axis of the galaxy. 
The non-negligible dust obscuration may have affected the detection efficiency of GCs and, in turn, 
the characterization of their spatial structure. Spectroscopic measurements could confirm the nature 
of G1 and G2 as physically distinct groups of GCs by detecting them also in the phase-space.

\subsection{NGC4365}
\label{subsec:ngc4365}

\begin{figure*}[]
	\includegraphics[height=5.5cm,width=5.5cm,angle=0]{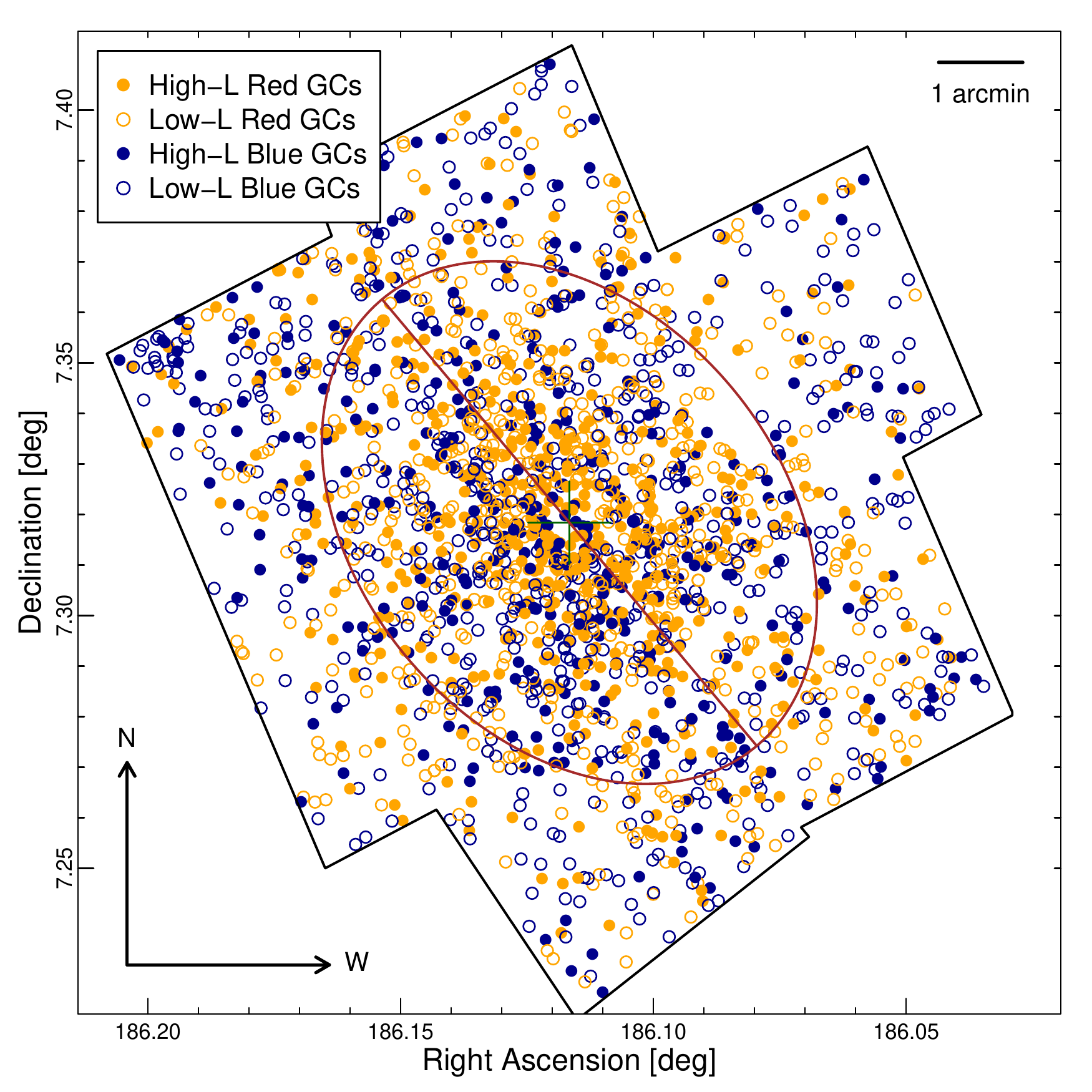}
	\includegraphics[height=5.5cm,width=5.5cm,angle=0]{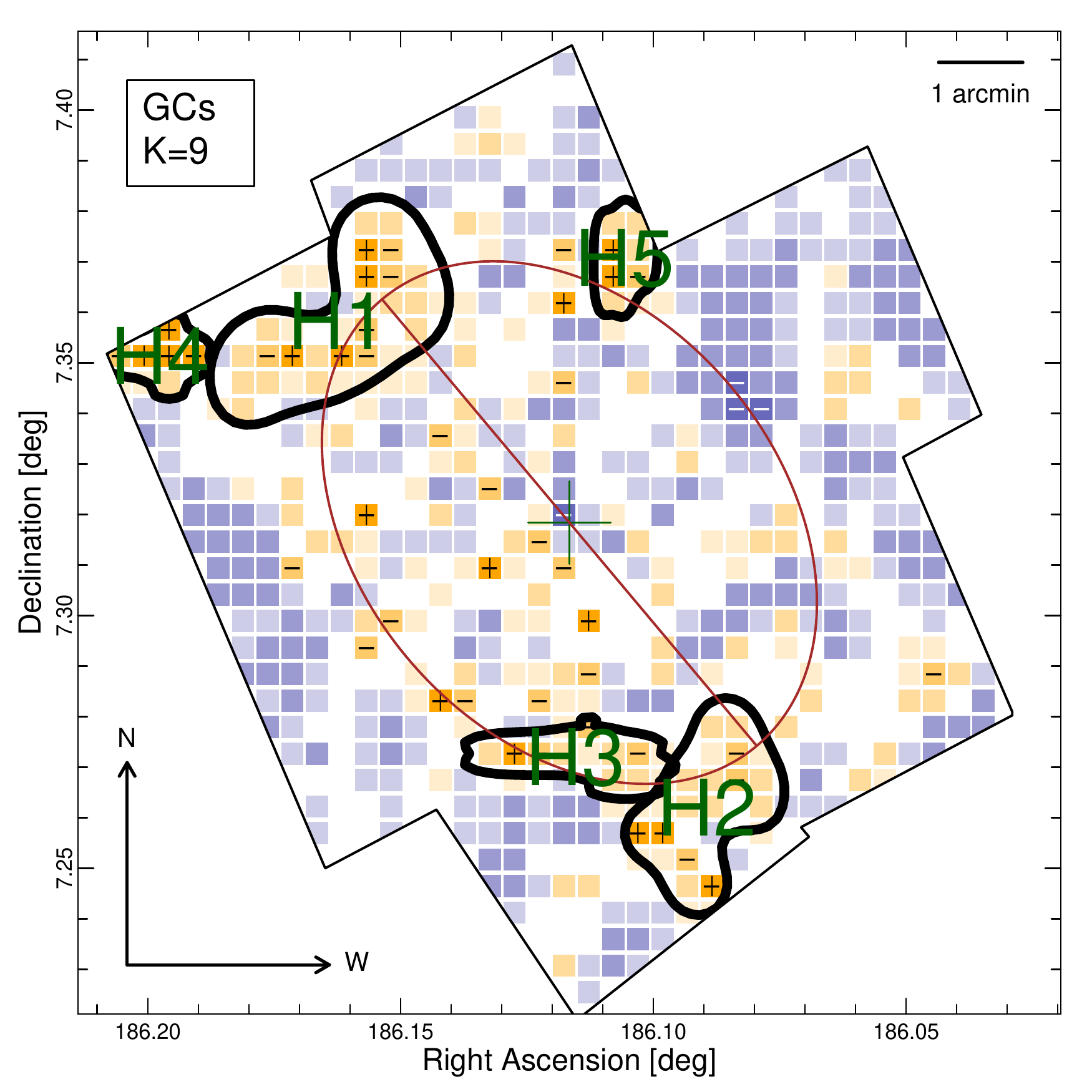}	
	\includegraphics[height=5.5cm,width=5.5cm,angle=0]{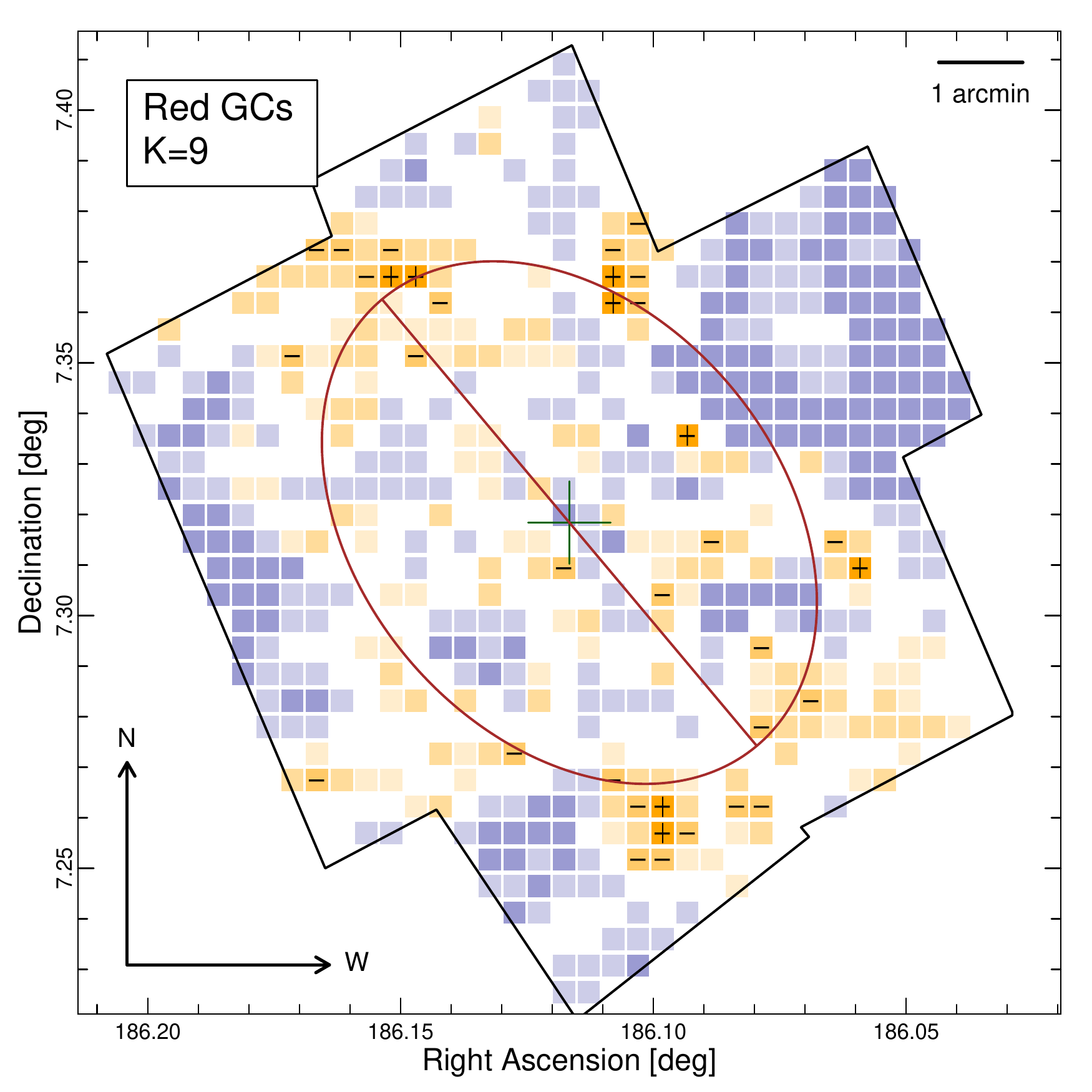}\\
	\includegraphics[height=5.5cm,width=5.5cm,angle=0]{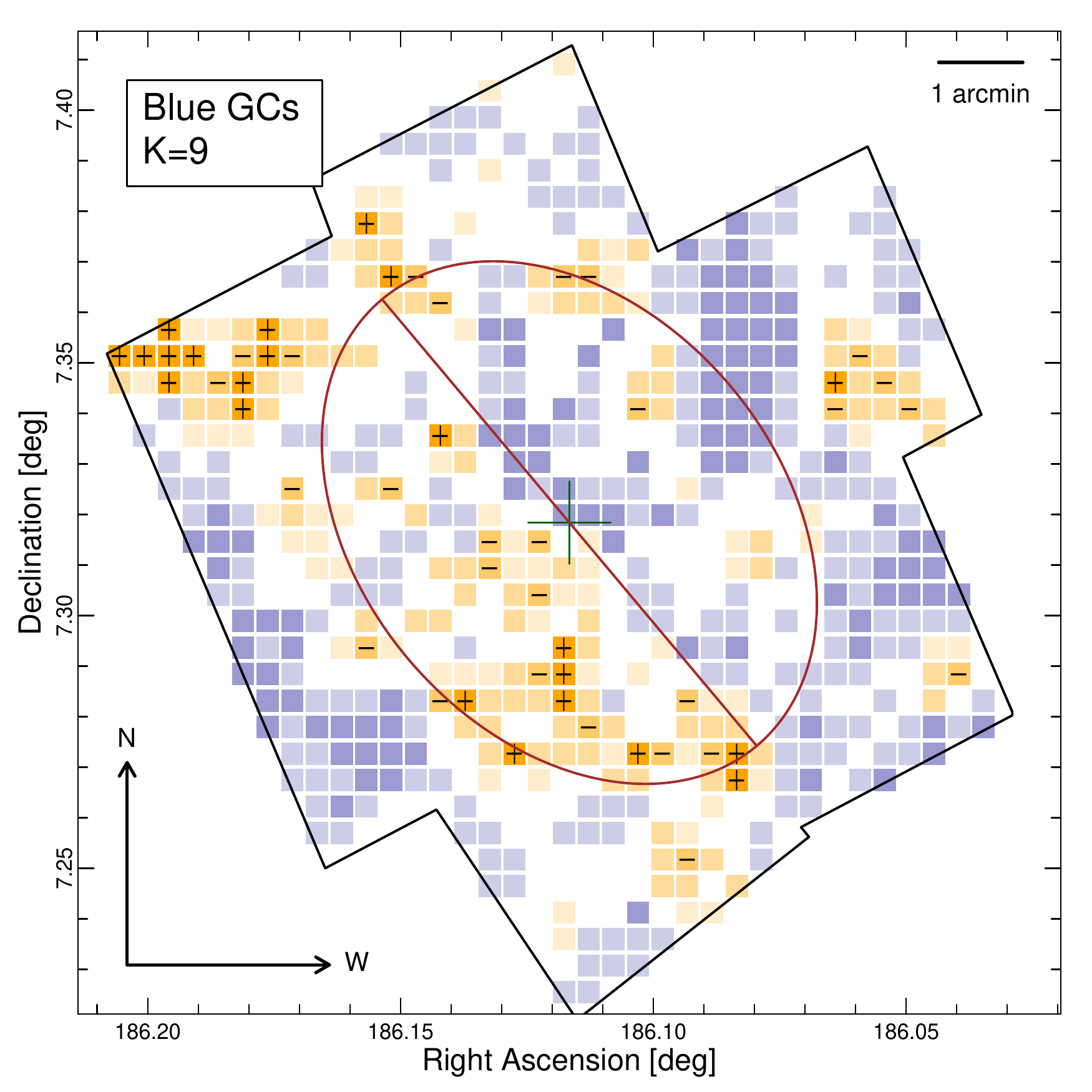}
	\includegraphics[height=5.5cm,width=5.5cm,angle=0]{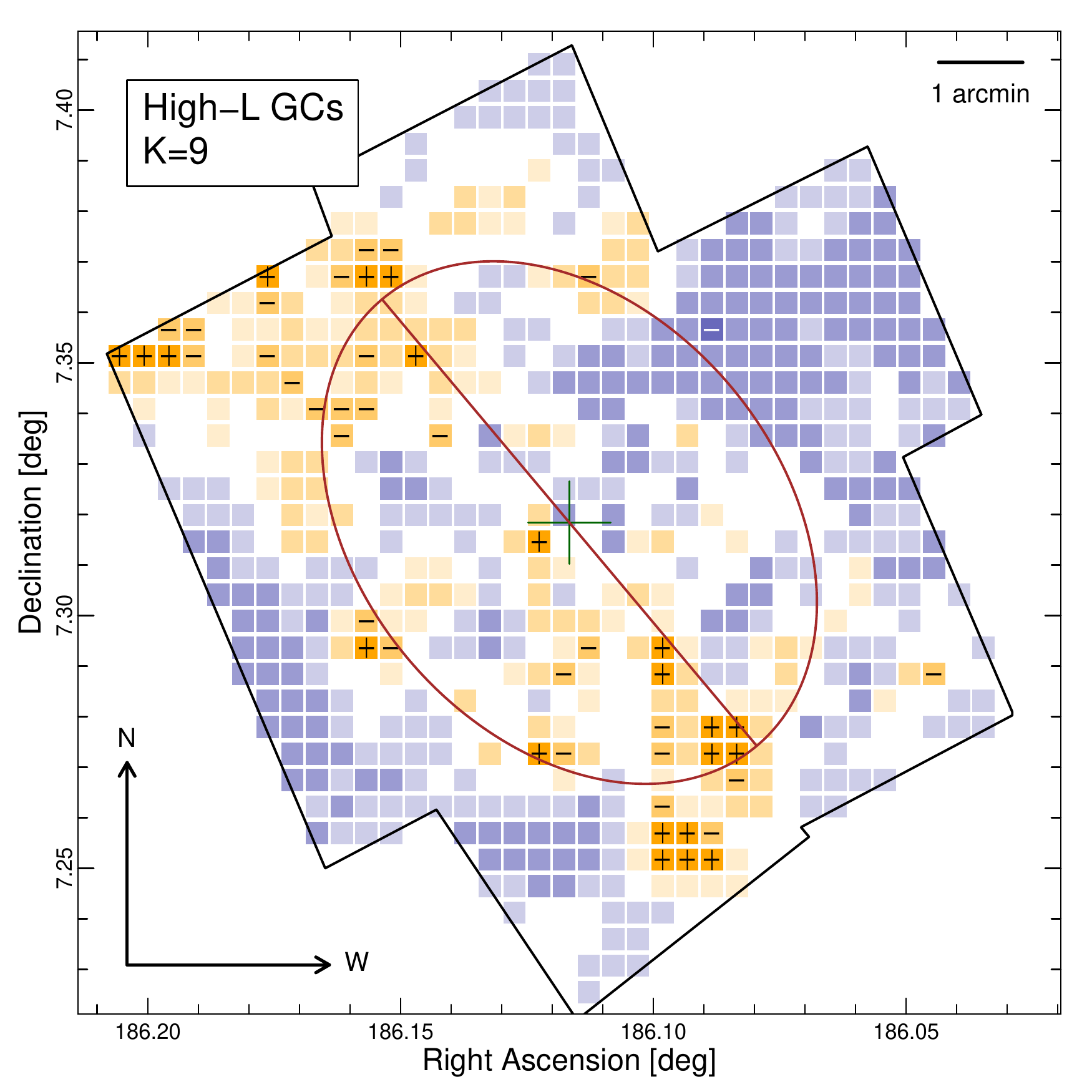}
	\includegraphics[height=5.5cm,width=5.5cm,angle=0]{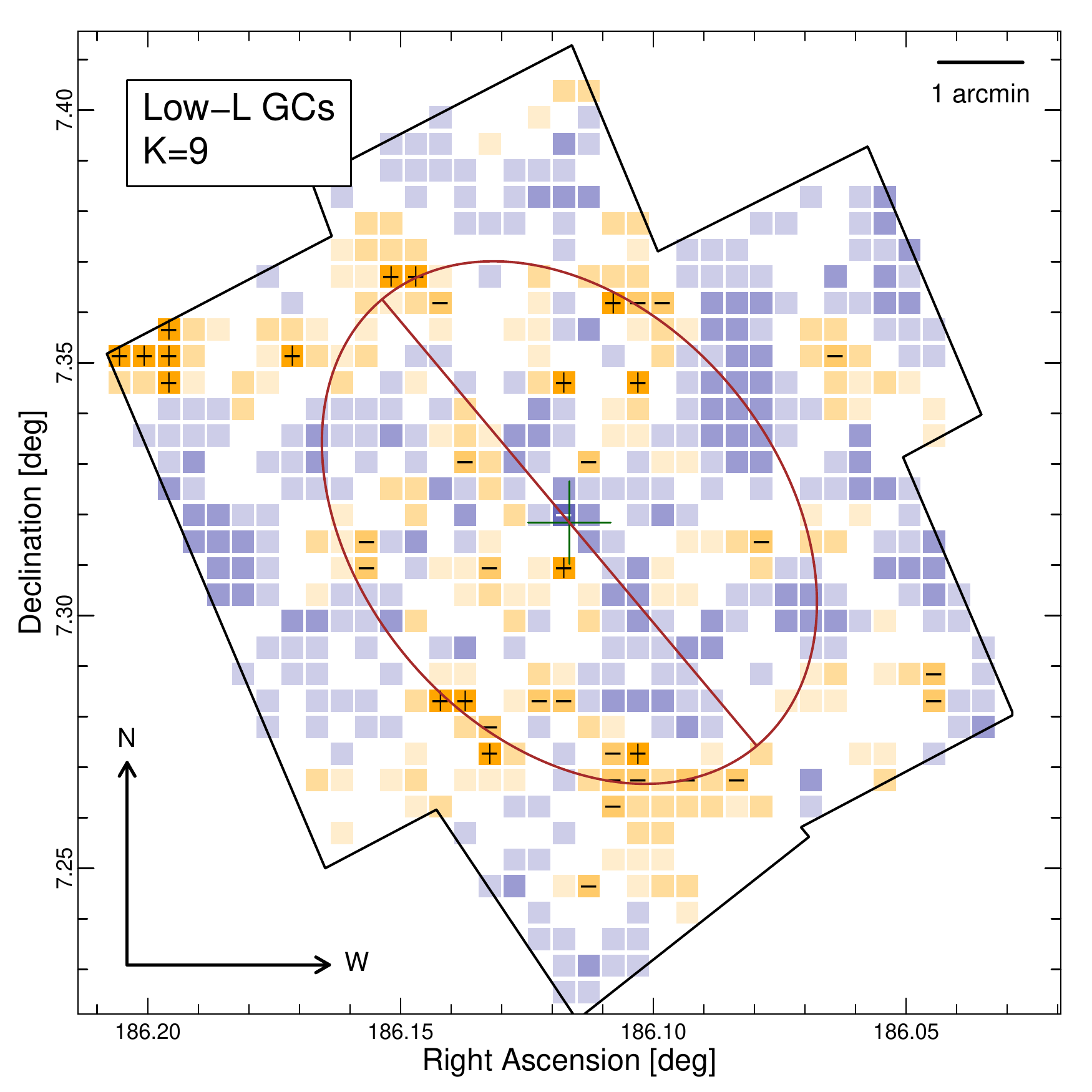}		
	\caption{Scatterplot of the position of the GCs in NGC4365 and residual maps obtained for 
	$K\!=\!9$ for the whole sample of GCs and the two color and luminosity GC classes. Refer 
	to Figure~\ref{fig:ngc4472} for a description of each panel.}
	\label{fig:ngc4365}
\end{figure*}

The spatial distribution of all GCs detected in NGC4365 (Figure~\ref{fig:ngc4365}, upper mid panel) 
is characterized by the absence of large structures and by a marginally significant ($2.8\sigma$)
asymmetry in the distribution of GCs, which are more numerous to the E of the major axis of the 
galaxy. The most significant feature H1 ($\sim\!5.8\sigma$), 
is close to the intersection of the $D_{25}$ ($\sim\!3.8\!r_{e}$) ellipse with the major axis. 
H1 extends towards the N-E corner of 
the field, where H4 ($\sim\!4.5\sigma$) can be 
observed. H1 is visible in the residual maps for both red and blue GC classes 
(Figure~\ref{fig:ngc4365}, upper right and 
lower left panels) with a strong indication of spatial segregation: red GCs 
in the W of the L structure and blue GCs in 
the E. H2 ($\sim\!5.1\sigma$), slightly E 
of the intersection between the $D_{25}$ and the S end of the major axis, is
more prominent in the high-luminosity and blue GCs (Figure~\ref{fig:ngc4365}, 
lower right and left panels). H3 ($\sim\!4.6\sigma$) is located E of H2. 

Both H3 and H2
lie along the S-W direction, where the companion galaxy NGC4341 is 
located.~\cite{bogdan2012} reported that the centers of NGC4365 and NGC4342 are connected
by a stellar-light stream. This stream is an indication of tidal interaction 
between NGC4365 and NGC4342 and corresponds to a stream of GCs 
detected by~\cite{blom2014} with kinematics consistent with the 
NGC4342 GC system~\cite[see Figures~10 and~11 from][]{blom2014}. 
Using data from~\cite{blom2014}, we find a 
marginally significant ($\sim\!2.2\sigma$) GC overdensity consistent with 
the GC. H2 and H3 are roughly located at the E end of the GC stream. However, 
only one GC of the sample of 31 selected to have line-of-sight 
velocities compatible with the motion of the NGC4342 GC system~\citep[Figure~10 in][]{blom2014} is 
located within the region occupied by H2. The average line-of-sight velocity of the
spectroscopic GCs from the list presented by~\cite{pota2013} located within 
H2 and H3 is compatible with the systemic velocity of the GC system 
of NGC4365. H5 ($\sim\!4.1\sigma$), N of the center of the galaxy, across the 
$D_{25}$ ellipse is also visible (with varying significance) in the other GC color and 
luminosity classes.

\subsection{NGC4621}
\label{subsec:ngc4621}

\begin{figure*}[]
	\includegraphics[height=5.5cm,width=5.5cm,angle=0]{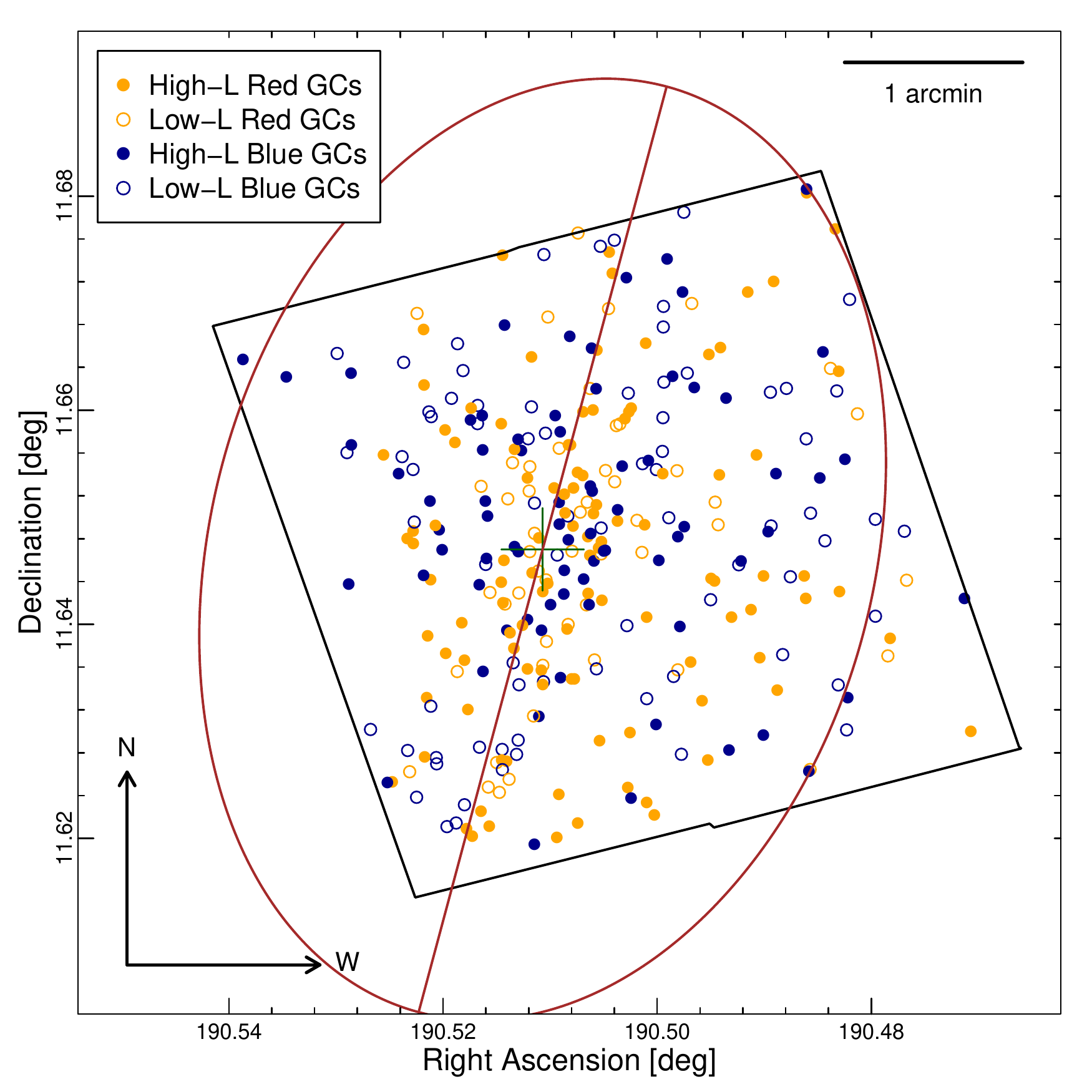}
	\includegraphics[height=5.5cm,width=5.5cm,angle=0]{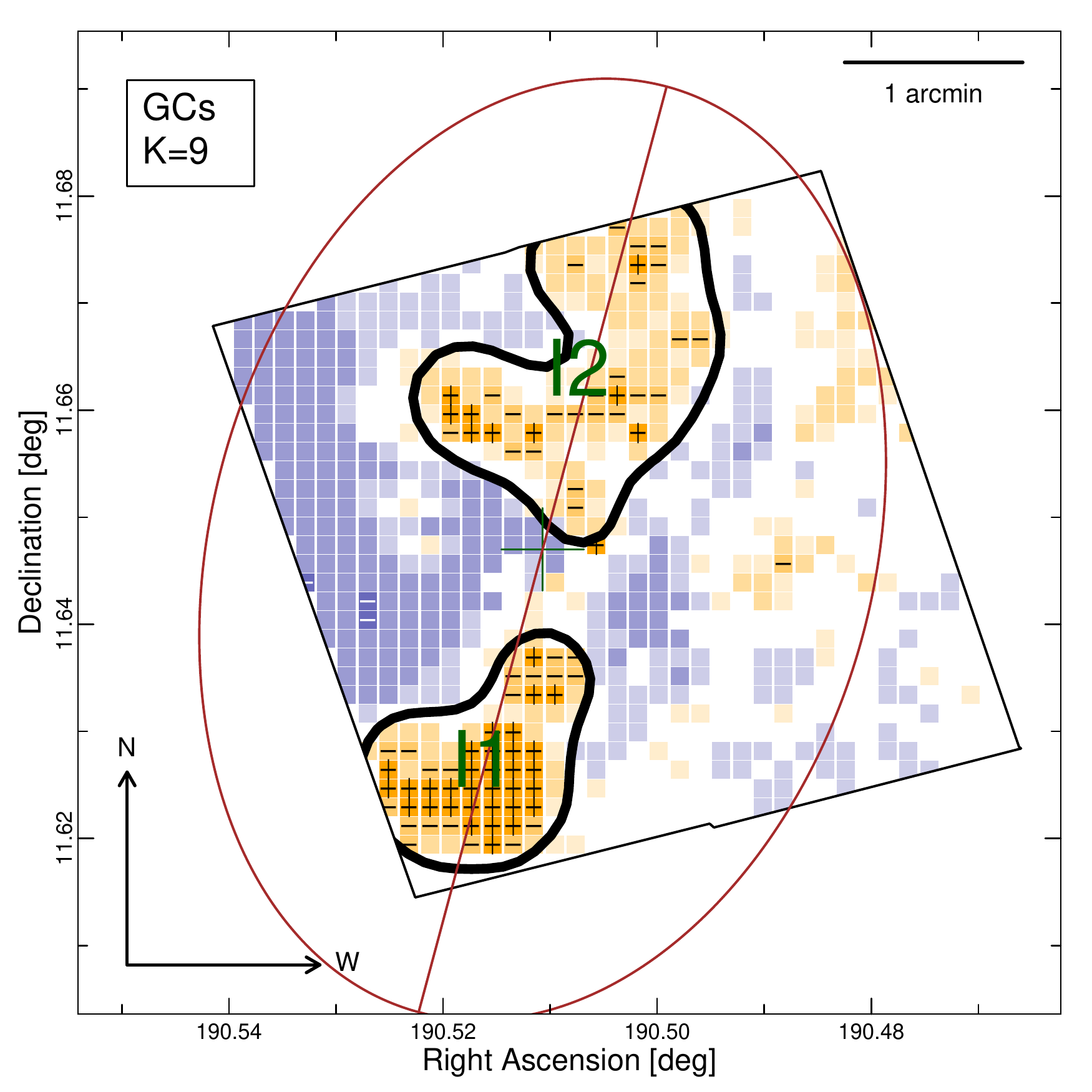}	
	\includegraphics[height=5.5cm,width=5.5cm,angle=0]{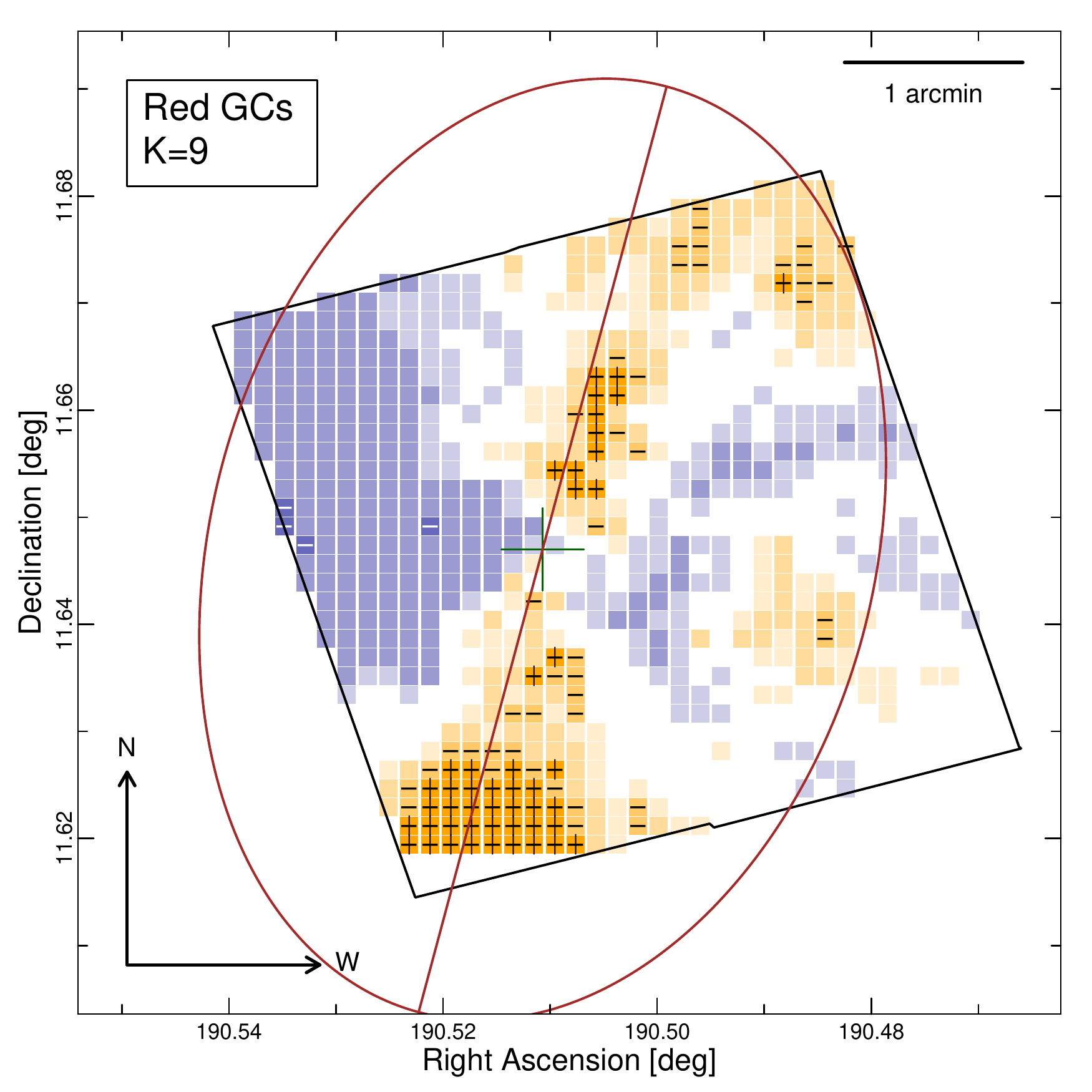}\\
	\includegraphics[height=5.5cm,width=5.5cm,angle=0]{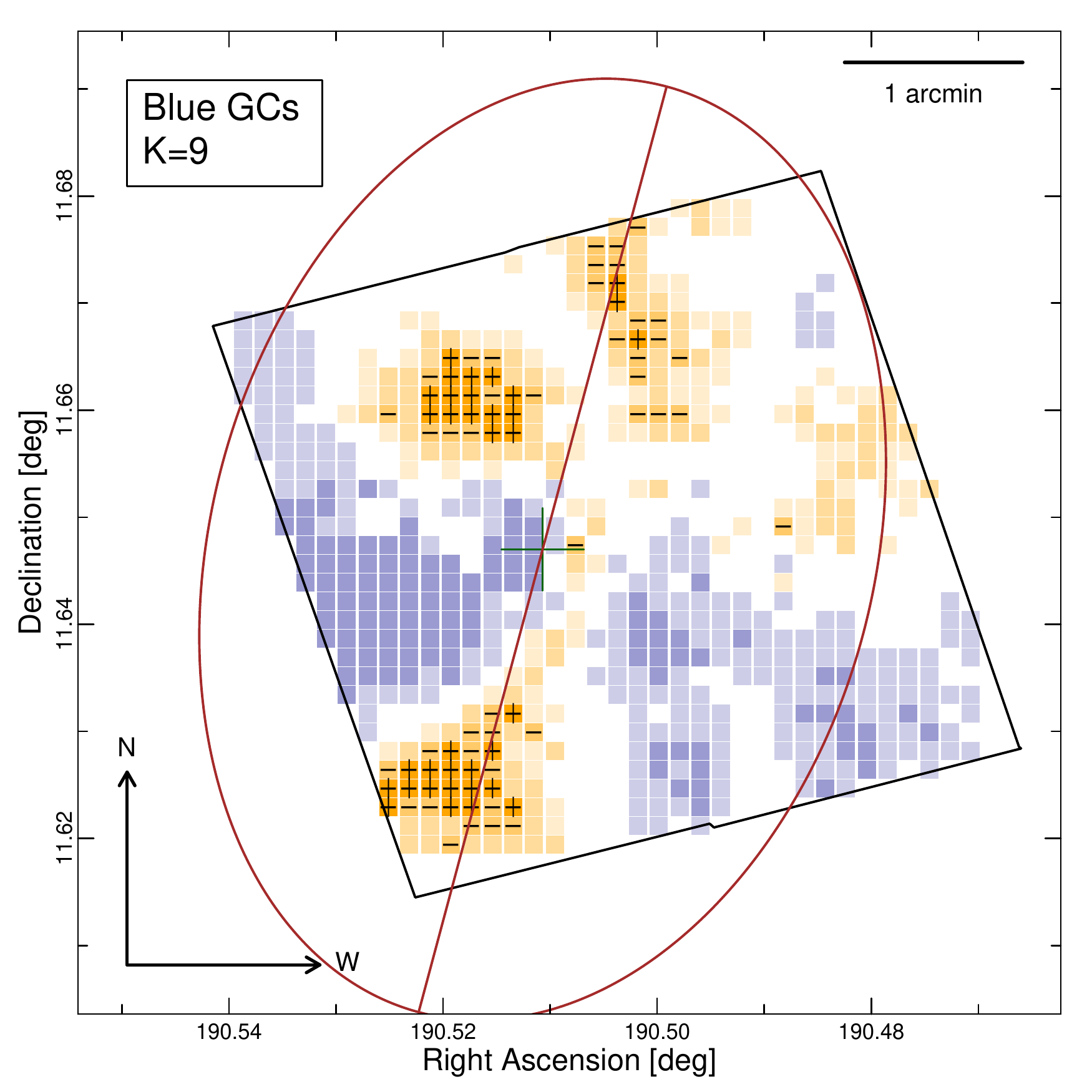}
	\includegraphics[height=5.5cm,width=5.5cm,angle=0]{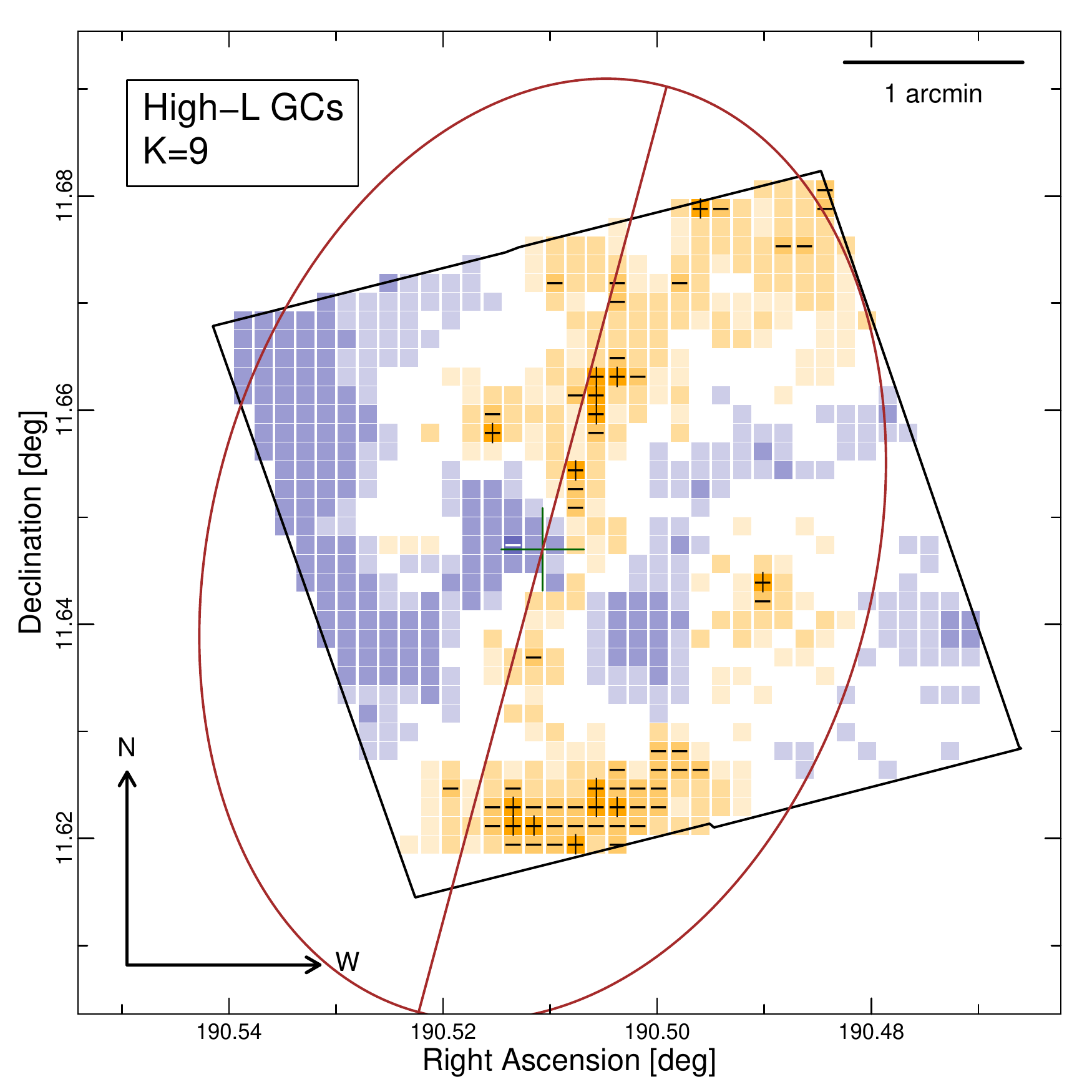}
	\includegraphics[height=5.5cm,width=5.5cm,angle=0]{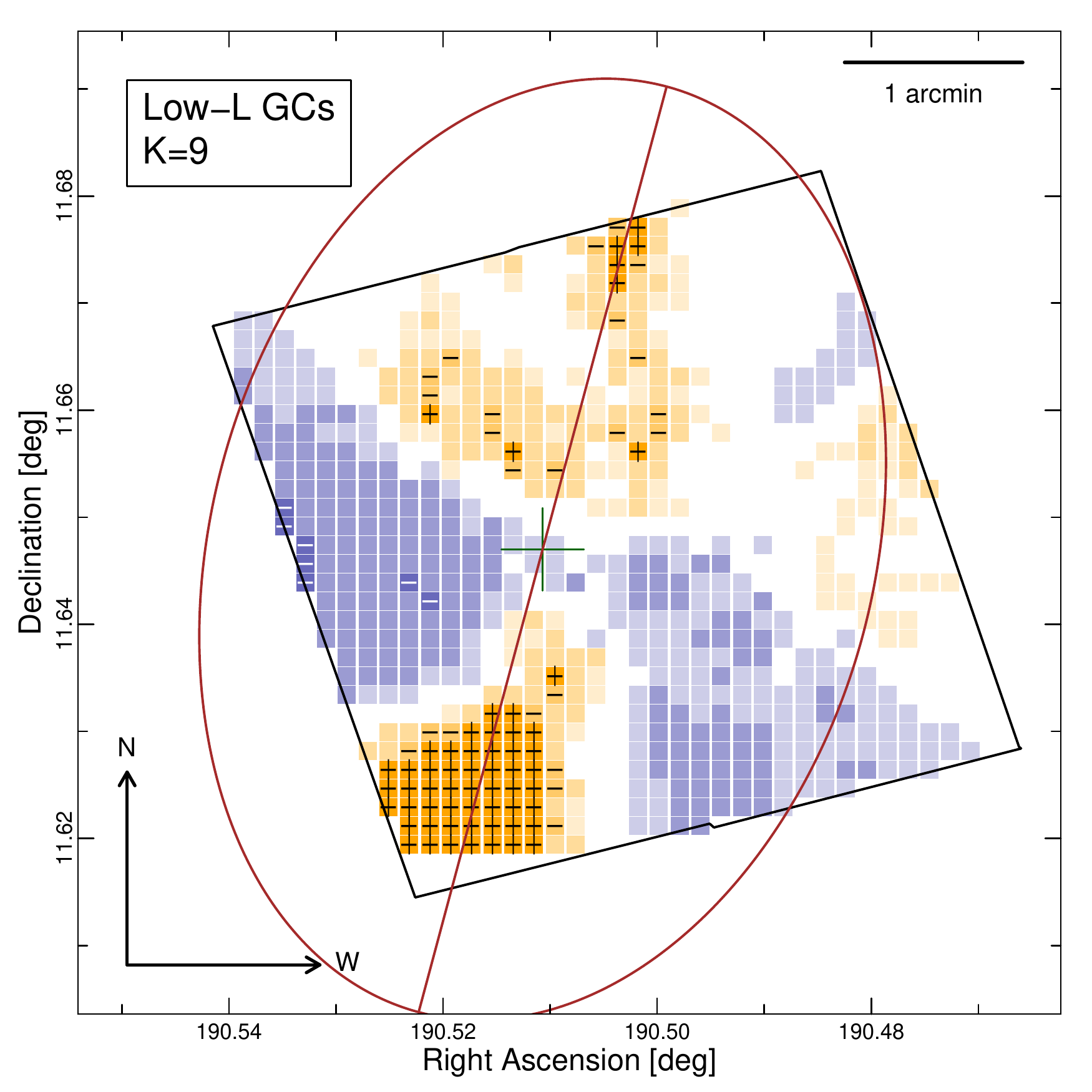}		
	\caption{Scatterplot of the position of the GCs in NGC4621 and residual maps obtained for 
	$K\!=\!9$ for the whole sample of GCs and the two color and luminosity GC classes. Refer 
	to Figure~\ref{fig:ngc4472} for a description of each panel.}
	\label{fig:ngc4621}
\end{figure*}

Similar to NGC4526, the spatial distribution of GCs 
in NGC4261 (Figure~\ref{fig:ngc4621}, upper mid panel) 
shows two large and strong structures along the major
axis. I1 ($>\!12\sigma$), 
in the S corner of the field observed, has similar significance and shape
in the color and low-L classes (Figure~\ref{fig:ngc4621}, 
upper right and lower panels), but it appears positionally shifted in the high-L GCs 
(Figure~\ref{fig:ngc4621}, lower mid panel), I2 is similarly significant in all the classes, 
but its shape and size change considerably. For example, in 
the red (Figure~\ref{fig:ngc4621}, upper right panel) and high-L GCs (Figure~\ref{fig:ngc4621}, 
lower mid panel), the section of I2 that extends from the major axis towards N-E is 
barely visible, while in the blue (Figure~\ref{fig:ngc4621}, lower left panel) and 
the low-L GCs (Figure~\ref{fig:ngc4621}, lower right panel), it splits in three distinct 
components that occupy most of the area of the HST field N of the center of the galaxy.

NGC4621 has large ellipticity~\citep[E4][]{ferrarese2006}, and displays a small stellar 
disk extending $\sim\!10\arcsec$ from the center along the major axis. Its dust content
is negligible~\citep{ferrarese2006,diserego2013}. In order to estimate the influence of the projection of an 
elongated three-dimensional distribution of 
GCs on the detection of the GC structures I1 and I2, we have used the same 
approach described in Section~\ref{subsec:ngc4526}. The worst case scenario significance of the 
structure I1 is $\sim5.8\sigma$, while I2 disappears as a single structure when the fraction 
of simulated GCs belonging to the component aligned with the major axis or the galaxy
becomes larger than 55\%. Based on these results, we can assume that I1 is not 
entirely produced by projection effects while the real nature of I2 can only be determined by
a more detailed modeling of the GC system, which is not possible with the data currently 
available. 

\subsection{NGC4552}
\label{subsec:ngc4552}

\begin{figure*}[]
	\includegraphics[height=5.5cm,width=5.5cm,angle=0]{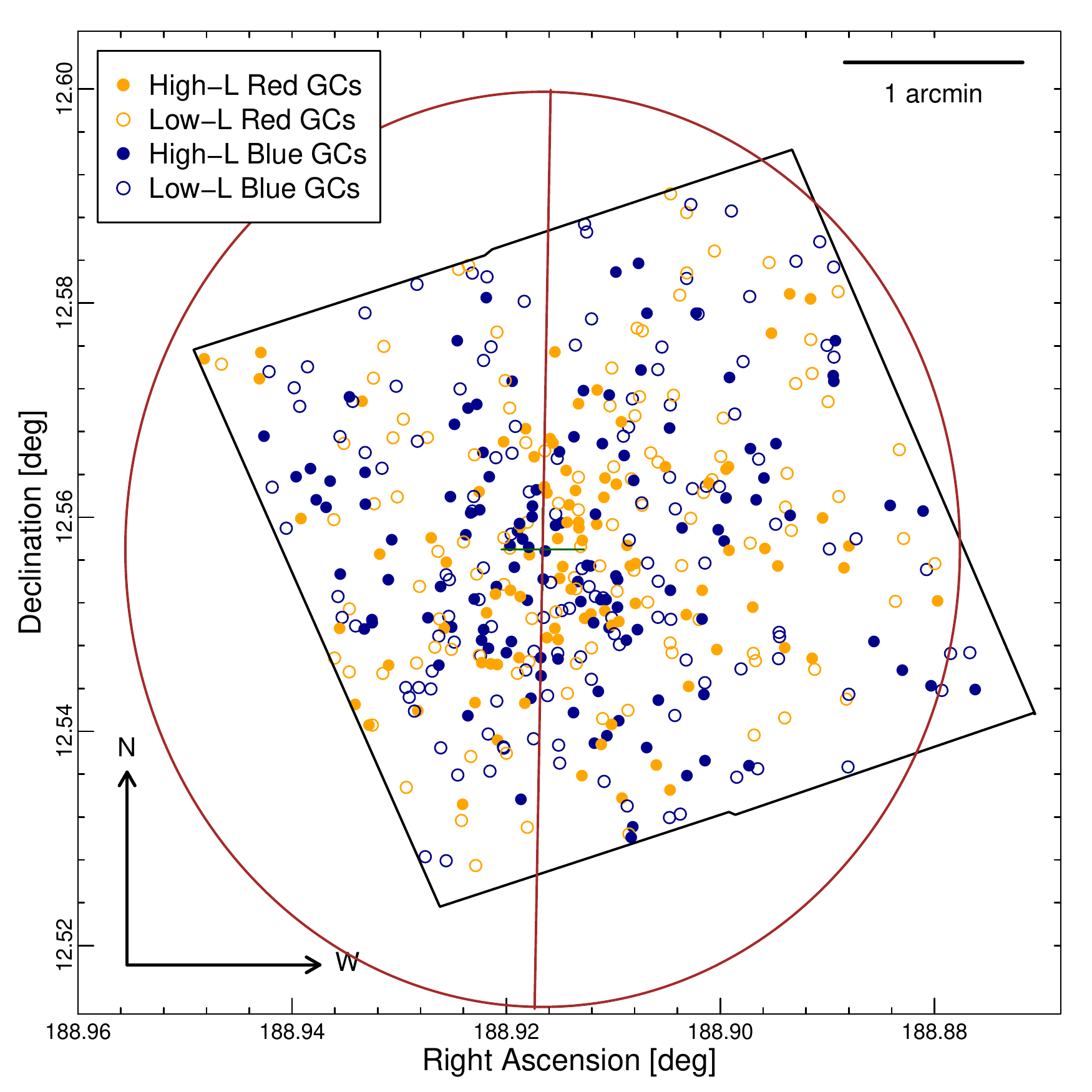}
	\includegraphics[height=5.5cm,width=5.5cm,angle=0]{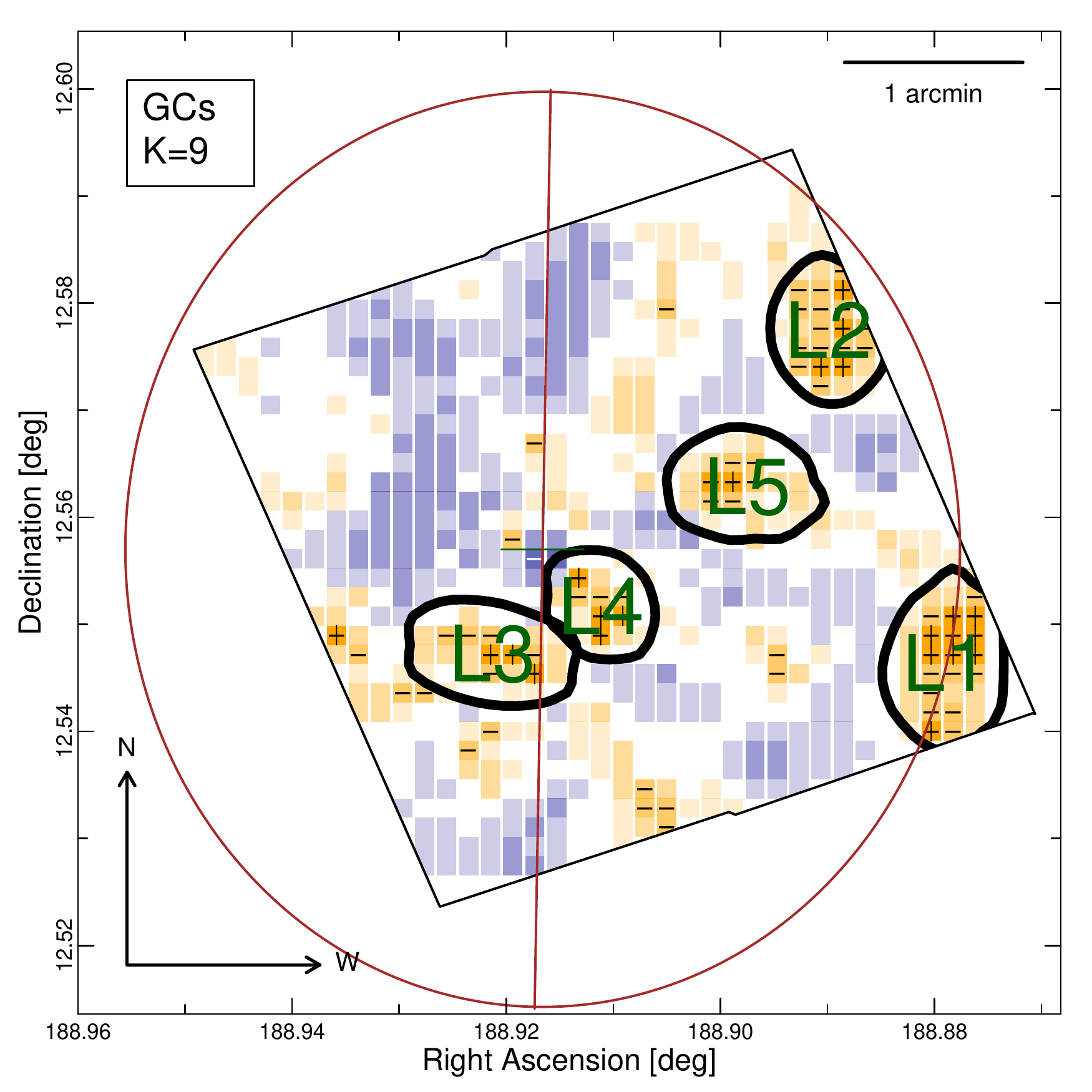}	
	\includegraphics[height=5.5cm,width=5.5cm,angle=0]{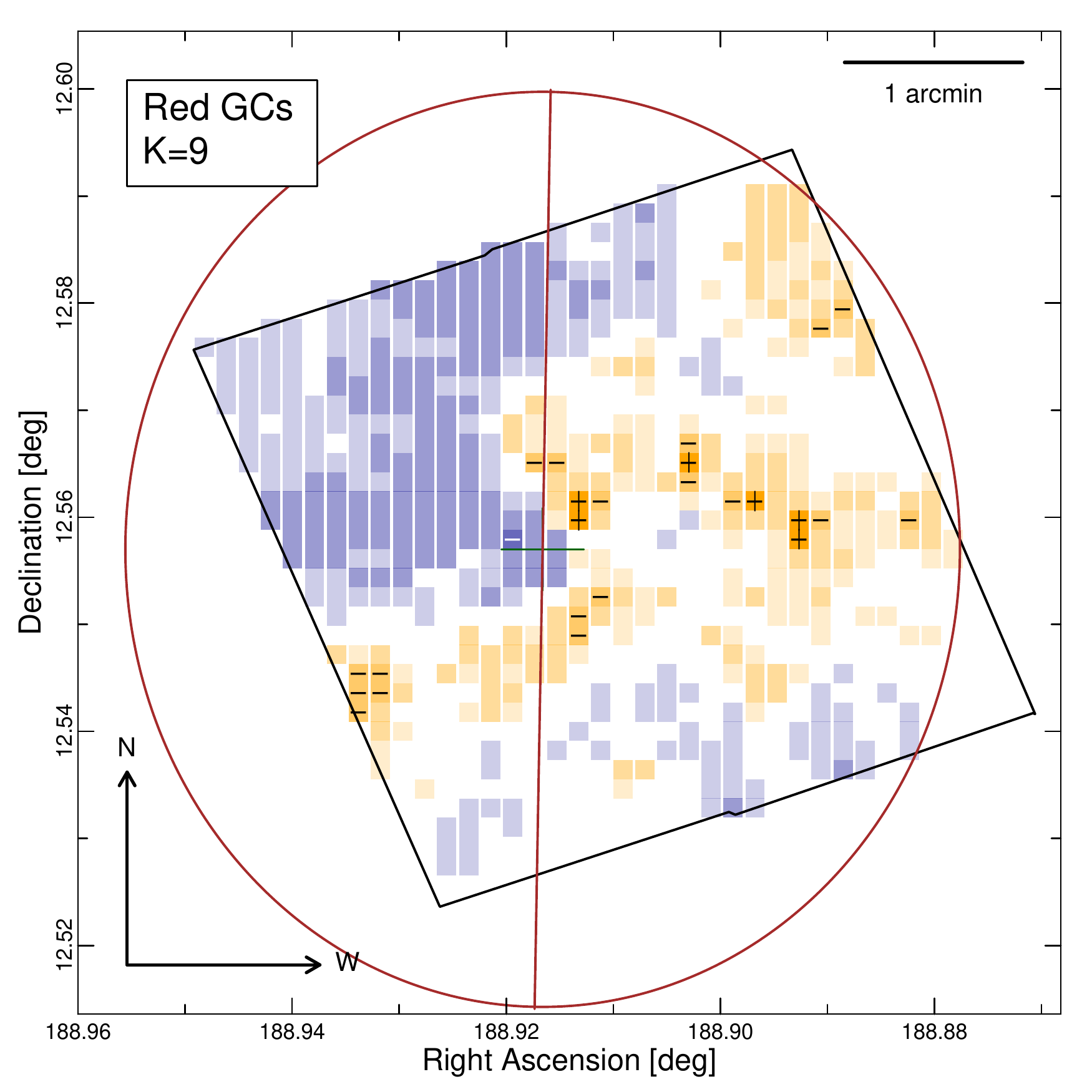}\\
	\includegraphics[height=5.5cm,width=5.5cm,angle=0]{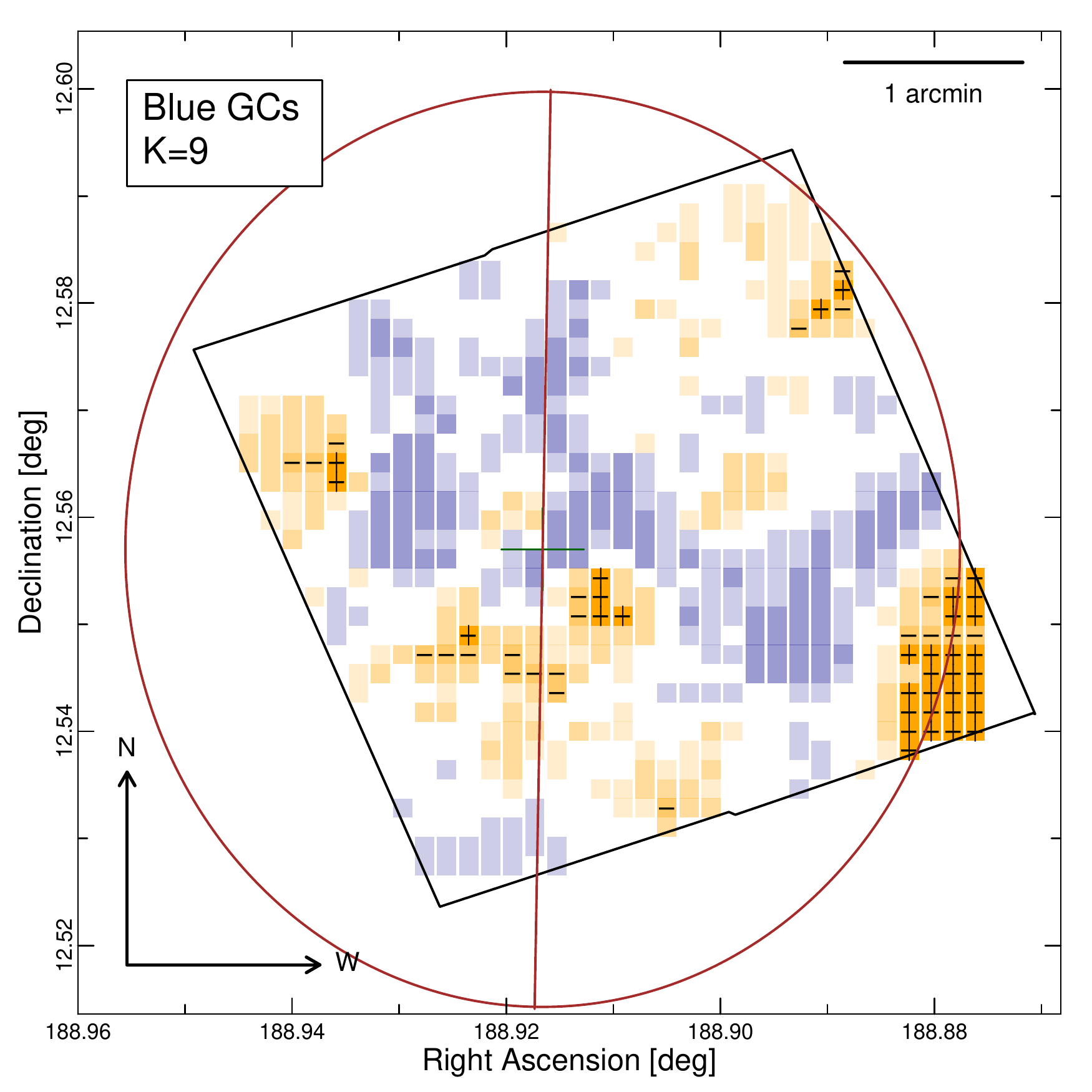}
	\includegraphics[height=5.5cm,width=5.5cm,angle=0]{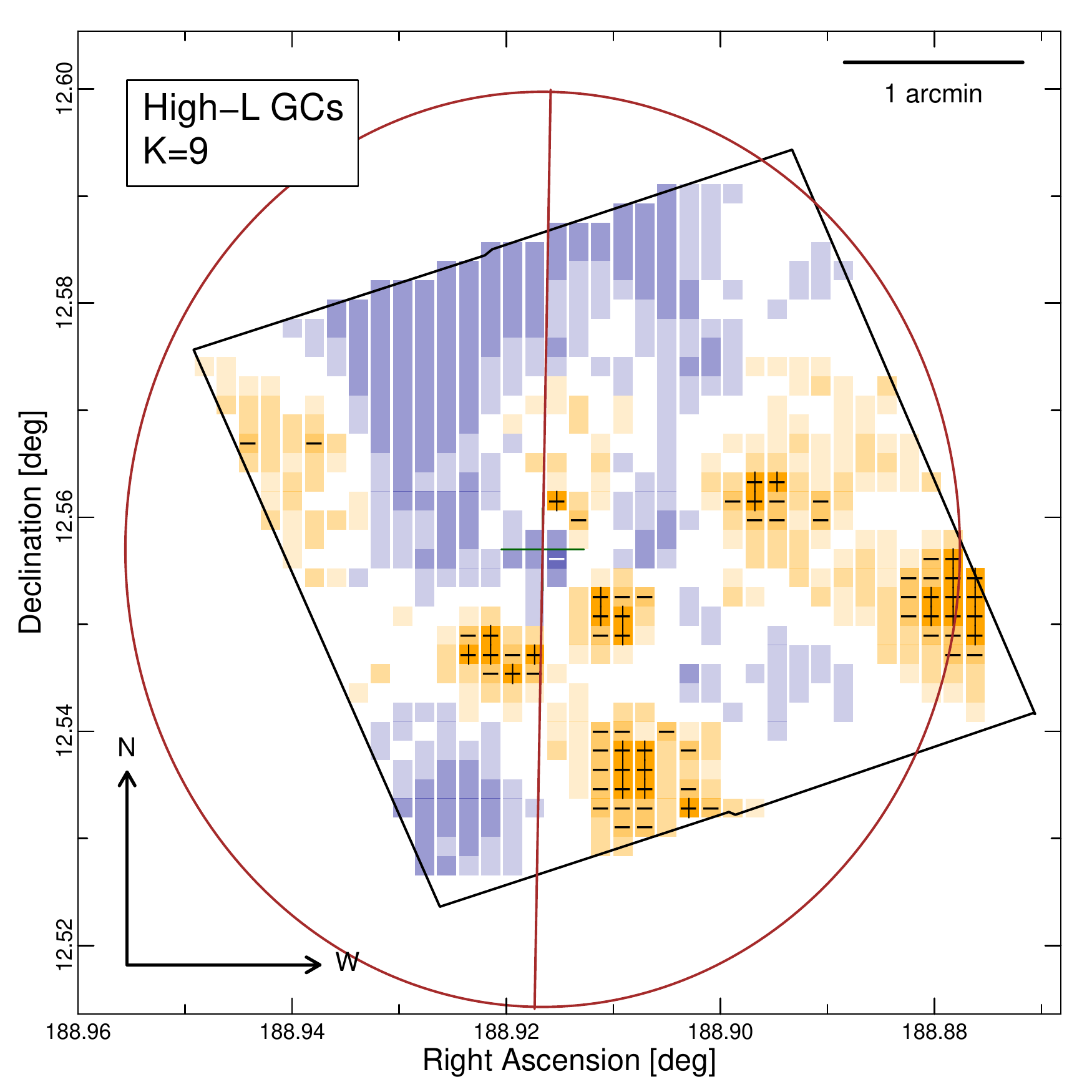}
	\includegraphics[height=5.5cm,width=5.5cm,angle=0]{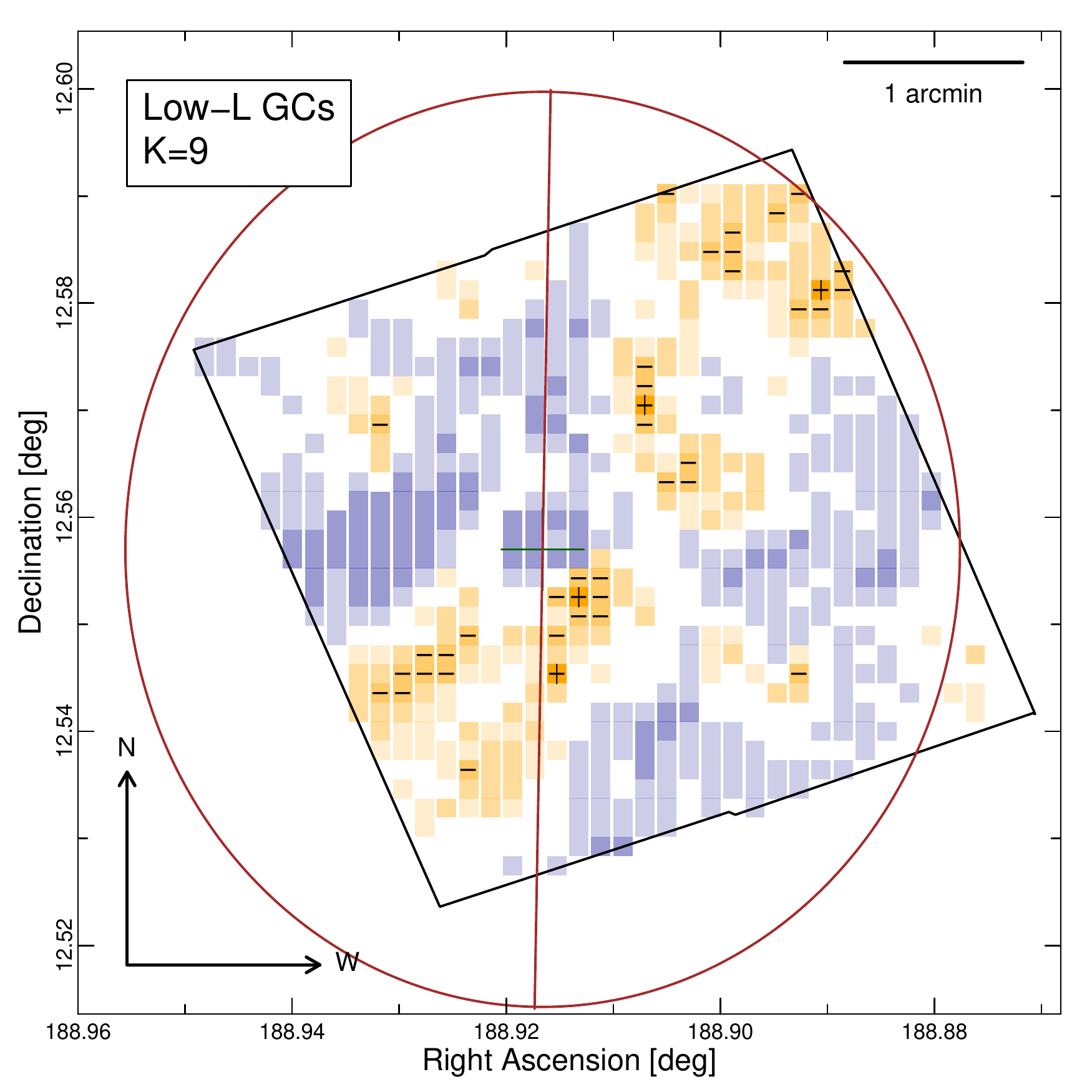}		
	\caption{Scatterplot of the position of the GCs in NGC4552 and residual maps obtained for 
	$K\!=\!9$ for the whole sample of GCs and the two color and luminosity GC classes. Refer 
	to Figure~\ref{fig:ngc4472} for a description of each panel.}
	\label{fig:ngc4552}
\end{figure*}

NGC4552 (upper-mid panel of
Figure~\ref{fig:ngc4552}) displays two large structures in the W side of 
the field. L1 ($\sim\!8.7\sigma$), located 
in the S-W corner, extends parallel to the $D_{25}$ ($\sim\!3.6\!r_{e}$) ellipse for $\sim0.9\arcmin$ and  
is prominent in blue and 
high-L GCs. L2 ($\sim\!7.3\sigma$), slightly 
S of the N-W corner of the field, within $D_{25}$, is clearly visible in all classes, except for 
the high-L GCs 
(Figure~\ref{fig:ngc4552}, lower mid panel). Two less significant structures L3 ($\sim\!6.4\sigma$) 
and L4 ($\sim\!5.7\sigma$) (visible in all classes of GCs), 
indicate the presence of a significant over-density close to the center of the 
galaxy. The fifth most significant residual structure L5 ($\sim\!4.8\sigma$) in the W half of the galaxy at 
galactocentric distance $\sim\!1\arcmin$, suggests the existence of a long, spatially coherent elongated 
sequence of structures extending from the N-W corner of the field to the center, possibly 
encompassing also L2, L3 and L4. 

{~\cite{ferrarese2006} report the presence of two thin dust filaments extending $\sim\!6\arcsec$ from the center 
towards N-NW direction and of patches of dust within a radial distance of $11\arcsec$. While these dusty regions 
could influence the reconstruction of the spatial distribution of GCs around the positions of these structures, their 
size is too small to affect them significantly.   
}

\section{Discussion}
\label{sec:discussion}	

The results presented in this paper confirm that localized spatial structures in the 2D distributions of 
GC systems are common. In the ten brightest Virgo cluster elliptical galaxies we have found GC 
over-density structures with a range of amplitudes, sizes, and shapes. We measure linear dimensions 
ranging from $\sim\!1$ to $\sim\!20$ kpc and azimuthal sizes ranging from 
$\sim\!10^{\circ}$ to $\sim\!90^{\circ}$. We observe GC structures with both simple, roughly circular 
shapes, and more complex forms. Below we discuss the characteristics of these structures, 
within the constraints provided by the existing limited observational coverage, and in the context of 
accretion of satellite galaxies~\citep[e.g.][]{bonfini2012,dabrusco2013,dabrusco2014a,dabrusco2014b}.
We use these results to set simple observation-based 
constraints to the size of the accreted satellite galaxies, that may have originated these structures.

\subsection{Location and shapes of GC over-density structures}
\label{subsec:positions}

\begin{figure*}
	\includegraphics[height=10cm,width=8cm,angle=0]{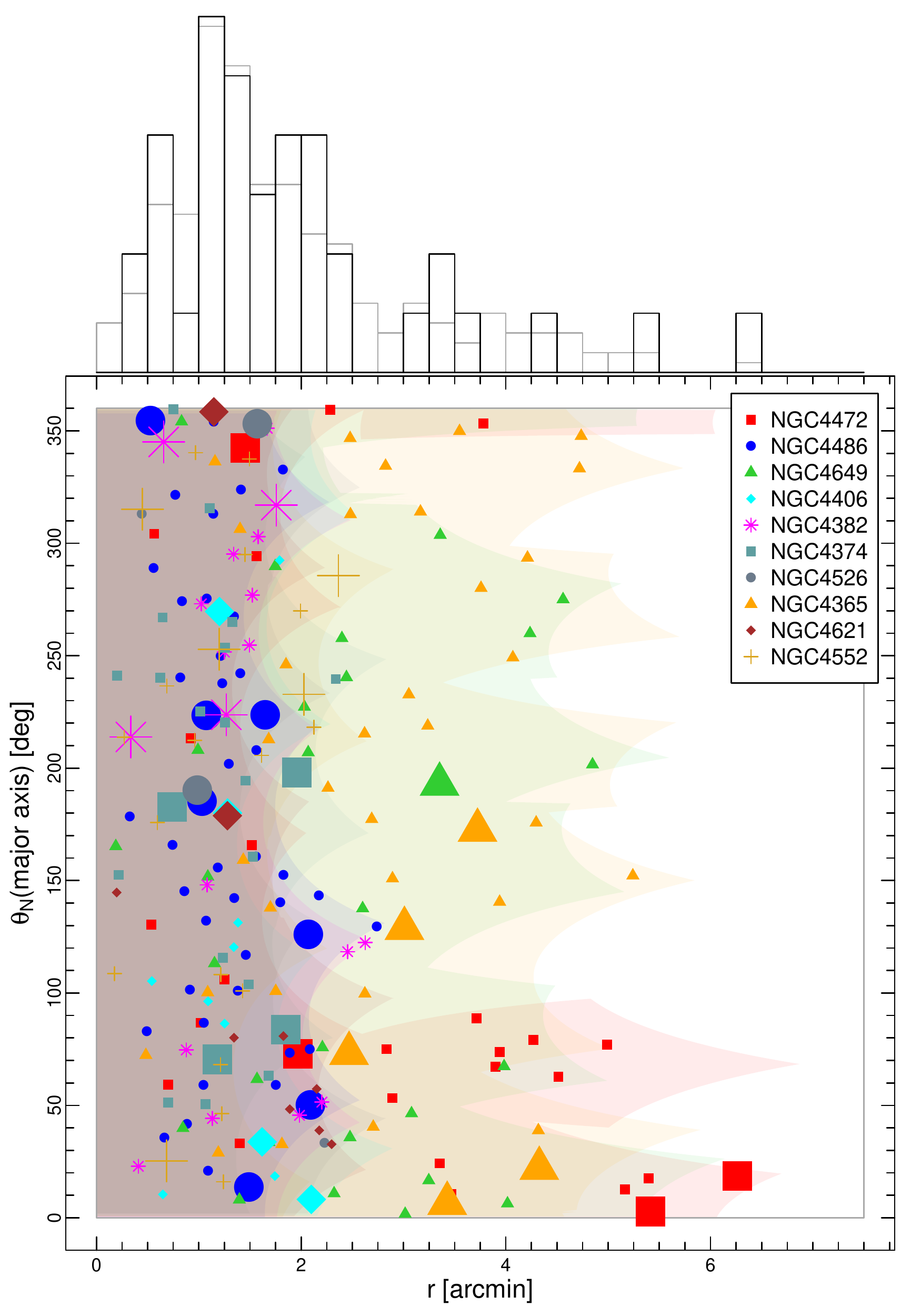}
	\includegraphics[height=10cm,width=8cm,angle=0]{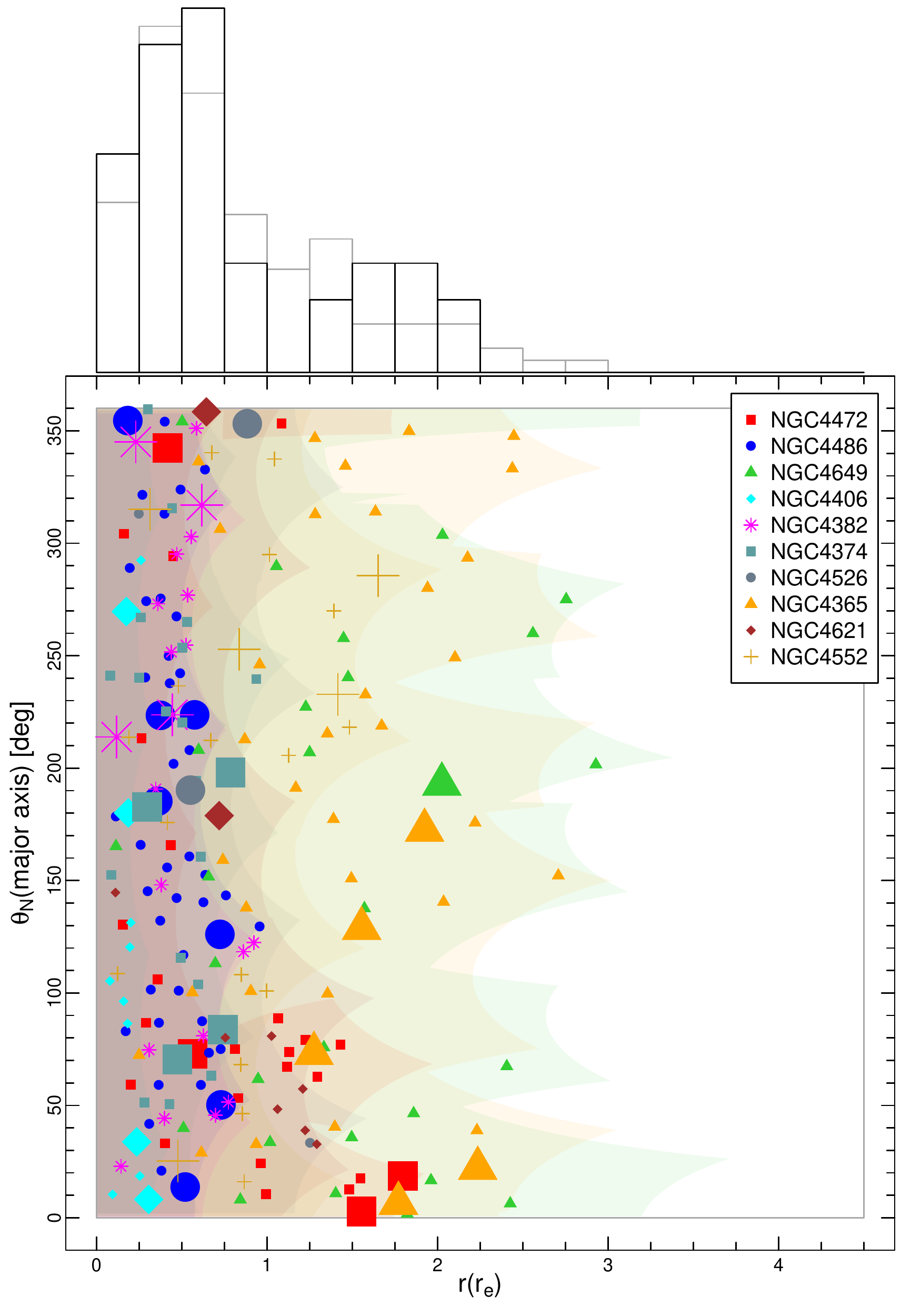}	
	\caption{Distribution of the over-density structures shown in 
	Table~3 in the galactocentric radius vs azimuthal angle plane 
	(measured counter-clockwise 
	from the N direction of the major axis of the galaxy). In the left panel, the galactocentric 
	radius is expressed as projected angular distance, while in the right panel we use units of 
	the effective radii $r_{e}$ of each galaxy as estimated by~\cite{ferrarese2006} (Table~1). 
	In both panels, the shaded background regions represent the areas of the plots that can be accessed 
	with the GC sample used in this paper for each galaxy. These regions follow the same 
	color-coding used for the points. In both plots, the large points represent the significance-weighted 
	central positions of the medium and large GCs structures discussed in Section~\ref{subsec:positions},
	and the gray and black marginal histograms show respectively the normalized distributions of 
	the galactocentric radii of all GCs structures (gray) and large only GCs structures only (black).}
	\label{fig:structures}
\end{figure*}

Most of the over-density structures are located between $\sim\!1\arcmin$ and 2\arcmin\ (0.2-0.8 $r_{e}$) 
from the center of the 
host galaxy where all the galaxies in our sample are reasonably well covered by the ACSVCS 
(Figures~\ref{fig:structures},~\ref{fig:percentage_areas}). 
While the decline in the number of structures detected at smaller galactocentric radii may be real, we 
cannot exclude the effects of either incompleteness of the GC catalog in the central area of the 
galaxies, or the decreased ability of our method to select structures in compact regions with very 
high GC density. The decrease in the frequency of structures for radii larger than 2\arcmin\ is definitely 
caused by the decrease of the galactic area covered by the ACSVCS observations. 
For $r\!>\!2.5\arcmin$ only NGC4472, NGC4649 and NGC4365 have adequate 
HST coverage. Figure~\ref{fig:percentage_areas} illustrates the large scatter in the fraction of areas 
covered for the different galaxies in our sample (at 1$r_{e}$S the fractional
coverage varies between 0\% and 100\% per galaxy, with $\sim35\%$ of the total area covered) 
and the lack of homogeneous coverage for galactocentric 
radii larger than 0.5$r_{e}$. 

In Figure~\ref{fig:structures} we show the spatial distribution of the over-density structures summarized in 
Table~3. The significance-weighted central coordinates of the structures 
are plotted in the plane generated by their galactocentric radius, expressed as projected angular distances 
and in units of the effective radii of each galaxy, and azimuthal angle, measured counter-clockwise from the 
N direction of the major axis of the galaxy. We also show the areas of the galactocentric radius vs azimuthal angle
plane that can be accessed with the GCs
samples used in this paper as shaded colored background. From Figure~\ref{fig:structures}, it is
evident that within the annulus $\sim\!1\arcmin$ - 2\arcmin\, where coverage is good for most galaxies, the smaller 
structures appear randomly azimuthally distributed in the $[0^{\circ}, 360^{\circ}]$ interval, suggesting that some of 
them might be the result of stochastic fluctuations, as discussed in Section~\ref{sec:results}. The medium 
and large structures (the largest size points in Figure~\ref{fig:structures}) cluster along the 
major and minor axes ($\theta_{N}\!\leq\!30^{\circ}$ and $\theta_{N}\!\geq\!330^{\circ}$ or 
$150^{\circ}\!\leq\!\theta_{N}\!\leq\!210^{\circ}$). We explore this effect further in Figure~\ref{fig:large_structures_plane}, 
where we plot the medium and large structures (see Section~\ref{sec:results} and Table~4) 
in the galactocentric radius vs azimuthal angle plane. 

The most extended large structure 
along the radial direction is C1, in NGC4649, which spans the interval of galactocentric distances 
$[0.5, 2.8]\ r_{e}$. Other large structures that extend mostly along the radial 
direction are I1 and I2, in NGC4621, from 0.2 $r_{e}$ to 1.1 $r_{e}$
and from 0.1 $r_{e}$ to 1.2 $r_{e}$ respectively, and G1 and G2 in NGC4526. All these structures 
are located along the 
major axis of their host galaxy in the whole range of radial distances covered by the ACSVCS observations. 
The distribution in angular position $\theta_{N}$ of the 
``large'' structures (Figure~\ref{fig:large_structures_plane}) shows that 9 out of 14 of them 
are located near the major axis of 
the host galaxies. When the areas of the structures are considered, the larger structures 
are preferentially located along the major axes of the host galaxies, accounting for $\sim\!86\%$ of the total 
``structure area''. This results appears consistent with the conclusion of~\cite{wang2013}, reporting preferential 
alignment of GC system with the major axis in Virgo galaxies. However, 5 large structures 
(B1, D1, F1, L1, L2) follow the minor axis. This minor axis effect was not previously reported. It
will need to be considered when trying to explain these results with theoretical simulations.The morphology of 
the host galaxy may play a role in these preferential orientations: the GC structures G1, G2, I1 and I2 observed 
in the galaxies NGC4526 and NGC4261 (classified as S0\_3\_(6) and E4, respectively, 
by~\citealt{ferrarese2006}) are almost perfectly aligned along the major axis of their galaxies. 

The large coherent structures may be suggestive of the disruption and merging of a sizable companion galaxy. 
For at least NGC4649, stellar kinematics suggest a major dry (dissipationless) merger, 
with a lenticular galaxy as progenitor~\citep{arnold2014}. However, we also note that the large structures 
may be slightly more prominent for the red GC subsystems. These large structures are on average 1.6 times 
more significant in the red GCs residual maps than the medium structures ($\sim\!3.5\sigma$/pixel 
for large structures to $\sim\!2.2\sigma$/pixel for medium structures), suggesting some dissipative 
merging-related GC formation~\citep[e.g.,][]{bekki2002}. 

\subsection{Observational evidence of the assembly history}
\label{subsec:observational_evidence}

For  all the galaxies in our sample with the exception of NGC4374, NGC4526 and NGC4621, evidence of
past merging or accretion of companions has been reported in the literature.

NGC4472 shows a fairly regular diffuse stellar distribution with elliptical isophotes~\citep{ferrarese2006} 
within few $r_{e}$, but it has been long known to be
interacting with its companion dwarf galaxy UGC7636~\citep{mcnamara1994}. 
Its active accretion history has been confirmed by~\cite{janowiecki2010}, 
which have observed a complex, extended shell system in the diffuse stellar light, suggesting 
radial accretion of a companion galaxy. 

NGC4486 has regular isophotes but the kinematics
of its GC system indicates that it is still accreting massive companions~\citep{romanowsky2012}; 
further evidence of stripping of low-luminosity dwarfs along radial orbits come from the structures
in its diffuse stellar light at large galactocentric radii~\citep{janowiecki2010}.

NGC4649 does not show significant
structures in its stellar light distribution which might be associated to the GC over-density structure 
C1~\citep{dabrusco2014a}. However, kinematic evidence based on the Planetary Nebulae (PN) 
and GC systems, suggests that a merger has contributed to 
the assembly of this galaxy~\citep{teodorescu2011,das2011,coccato2013}. In particular,~\cite{das2011} 
showed that PNs and GCs might belong to separate dynamical systems at radii larger than 12 
kpc.~\cite{arnold2014} used kinematics of the 
diffuse stellar light to discover disk-like outer rotation that might have originated from a massive lenticular
galaxy progenitor. 

NGC4406 also has small-scale streamers that can be explained by stripping of 
dwarf companions~\citep{janowiecki2010}; furthermore, observations of the cold dust emission 
in the surrounding area has confirmed that NGC4406 has stripped materials from the nearby
spiral galaxy NGC4438 by tidal interaction~\citep{gomez2010}. 

NGC4382 has boxy 
isophotes~\citep{ferrarese2006} and presents abundant fine structure in its diffuse stellar 
light~\citep{schweizer1990}, suggesting that it has undergone a major merger in the past 4-7 Gyr. 

NGC4374, whose optical isophotes are very regular in the ACSVCS images~\citep{ferrarese2006}, 
shows a rapidly rotating central disk~\citep{bower1998} but does not present evidence
of kinematically decoupled components and it lacks significant tidal structures~\citep{janowiecki2010}. 
~\cite{arnold2014} found that the outer region of the galaxy show little to none rotation, in agreement
with the expectations for a low rotator.

NGC4526 is a lenticular galaxy hosting a large dusty disk that extends along the major axis. Its isophotes 
become clearly boxy at $\sim10\arcsec$ before becoming disky at $\sim70\arcsec$ 
radius~\citep{ferrarese2006}. 

NGC4365 is known for its kinematically
decoupled core rotating along the major axis, which could be result of a major
merger involving a spiral galaxy~\citep{davies2001}. This galaxy is in the process of stripping
GCs from the companion NGC4342~\citep{blom2014}. 

NGC4621 shows a thin faint stellar disk in the ACSVCS images, but further evidence of a 
recent major merger or accretion event is missing. 

NGC4552 has regular morphology in the diffuse stellar light within the 
$D_{25}$~\citep{ferrarese2006}, but~\cite{janowiecki2010} detected a complex structure of 
shells and plumes that is suggestive of a complex accretion history involving either multiple
small accretions or a major merger.

\subsection{GC structures in color and luminosity classes}
\label{subsec:colorluminosity}

Color and luminosity differences in the GC over-density structures are noticeable
in several galaxies (Section~\ref{sec:results}, Table~4). While some of 
these differences may be due
to statistical noise, some are significant and may pose constraints on the origin and
evolution of the structures. Among the medium and large GC over-density structures, spatial
segregation has been detected for red/blue GC classes in H1 (NGC4365), and for high-L/low-L 
GCs in A1 (NGC4472), C1 (NGC4649), E2 (NGC4382) and L5 (NGC4552). We selected the GC 
structures with spatial segregation by comparing the significances of GC structures in each of the color 
(red/blue) and luminosity (high-L/low-L) classes. The structures for which the difference of significance 
was larger than 3$\sigma$ were visually inspected to ascertain the presence of spatial segregation.
In NGC4472, the luminosity segregation in the 
large structures A1 suggests an incoming tangential GC stream that then turns into the 
radial direction. In this stream, the low-luminosity GCs are prevalent in the inner radial direction, 
while the more luminous GCs are mostly found in the section of the structure tangential to the $D_{25}$. 
This evidence suggests that dynamical friction~\citep{fall2001,binney2008} can be ruled out as the 
only mechanism responsible for the observed segregation, as it would act more vigorously
on the luminous, more massive clusters that should be observed at smaller galactocentric distances than
the less massive, low-L GCs. Kinematic observations are needed to shed light on 
the origin of the spatially segregated high- and low-L components of A1. We stress that, lacking 
observational confirmation or a detailed explanation of the physical mechanisms responsible for the 
observed properties of this GC structure, the two components in A1 could have originated in 
distinct events.

\begin{figure}
%	\centering
	\includegraphics[height=8cm,width=8cm,angle=0]{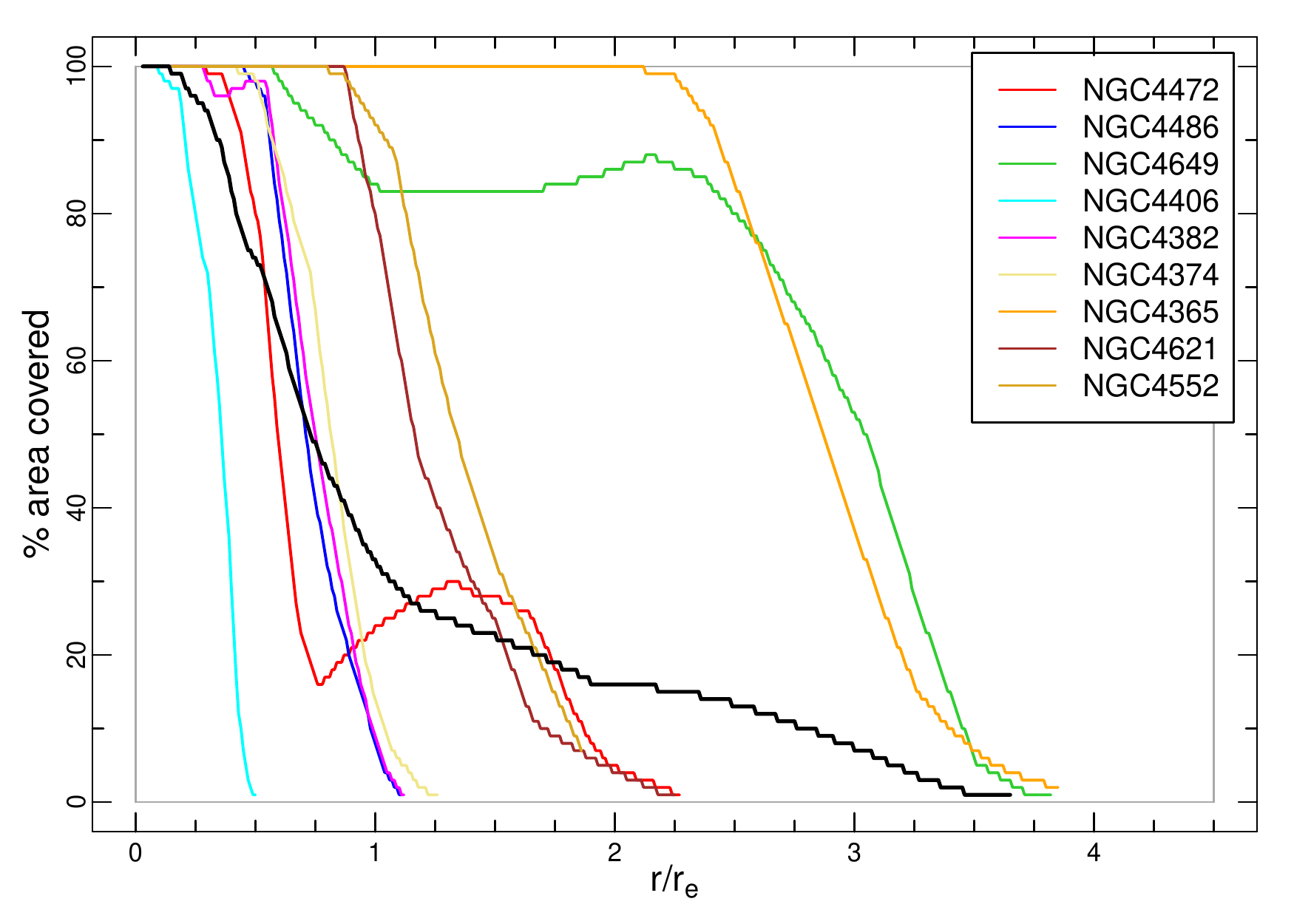}
	\caption{Fraction of the area of each galaxy as a function of the galactocentric distance in 
	units of the effective radii $r_{e}$ covered by the footprints of the HST ACS observations
	used to extract the catalogs of GCs employed in this paper. The black line represents the
	fraction of total area covered by the observations for the entire sample of galaxies.}
	\label{fig:percentage_areas}
\end{figure}

\begin{figure*}
	\includegraphics[height=12cm,width=16cm,angle=0]{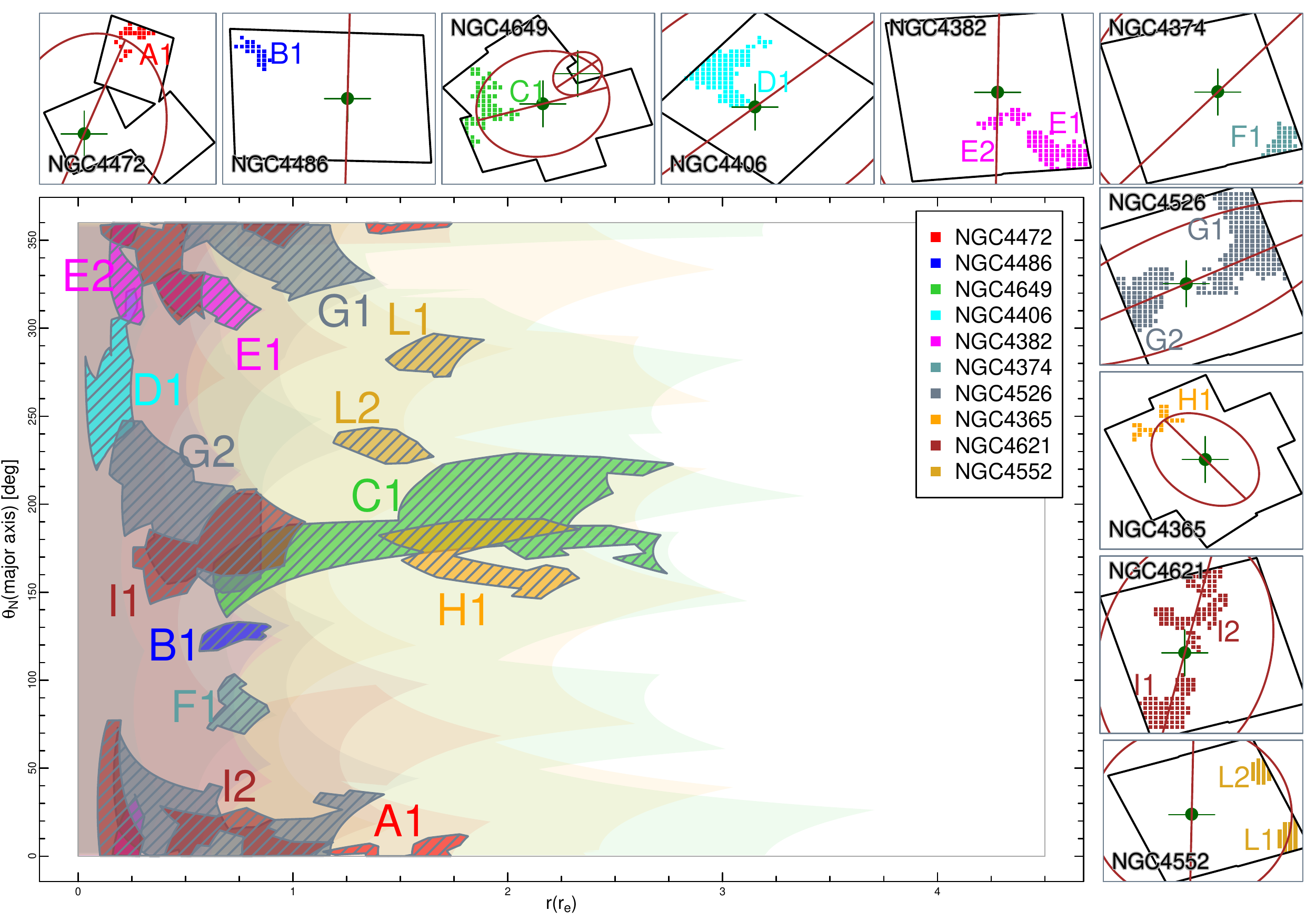}
	\caption{Boundaries of the ``large'' (more than 19 pixels) GC over-density structures in the 
	galactocentric distance (expressed in units of the effective radii $r_{e}$) vs azimuthal angle plane.
	The plots in the margins show the ``large'' structures in the R.A. vs Dec. plane. 
	The background regions are color-coded based on their membership to the galaxies of our sample.}
	\label{fig:large_structures_plane}
\end{figure*}

The spatial segregation of high-L/low-L GCs observed in C1 (NGC4649) 
could be tentatively explained as a result of differential dynamical evolution of the GC color classes 
formed during the major merger experienced by the host galaxy, as discussed 
by~\cite{dabrusco2014a}. Color differences in the distribution of the GC over-density structures 
may be related to the prevalent GC population of merging satellite galaxies, if these are the results
of dry mergers. Gas-rich mergers may increase the localized red GC population.  
We note that the highly incomplete spatial coverage of our sample (see Figure~\ref{fig:percentage_areas}) 
does not allow us to assemble a full picture of either the shapes or the characteristics of these GC features.

\subsection{GC structures and local galaxy density}
\label{subsec:galaxydensity}

\begin{figure*}
	\includegraphics[height=12cm,width=16cm,angle=0]{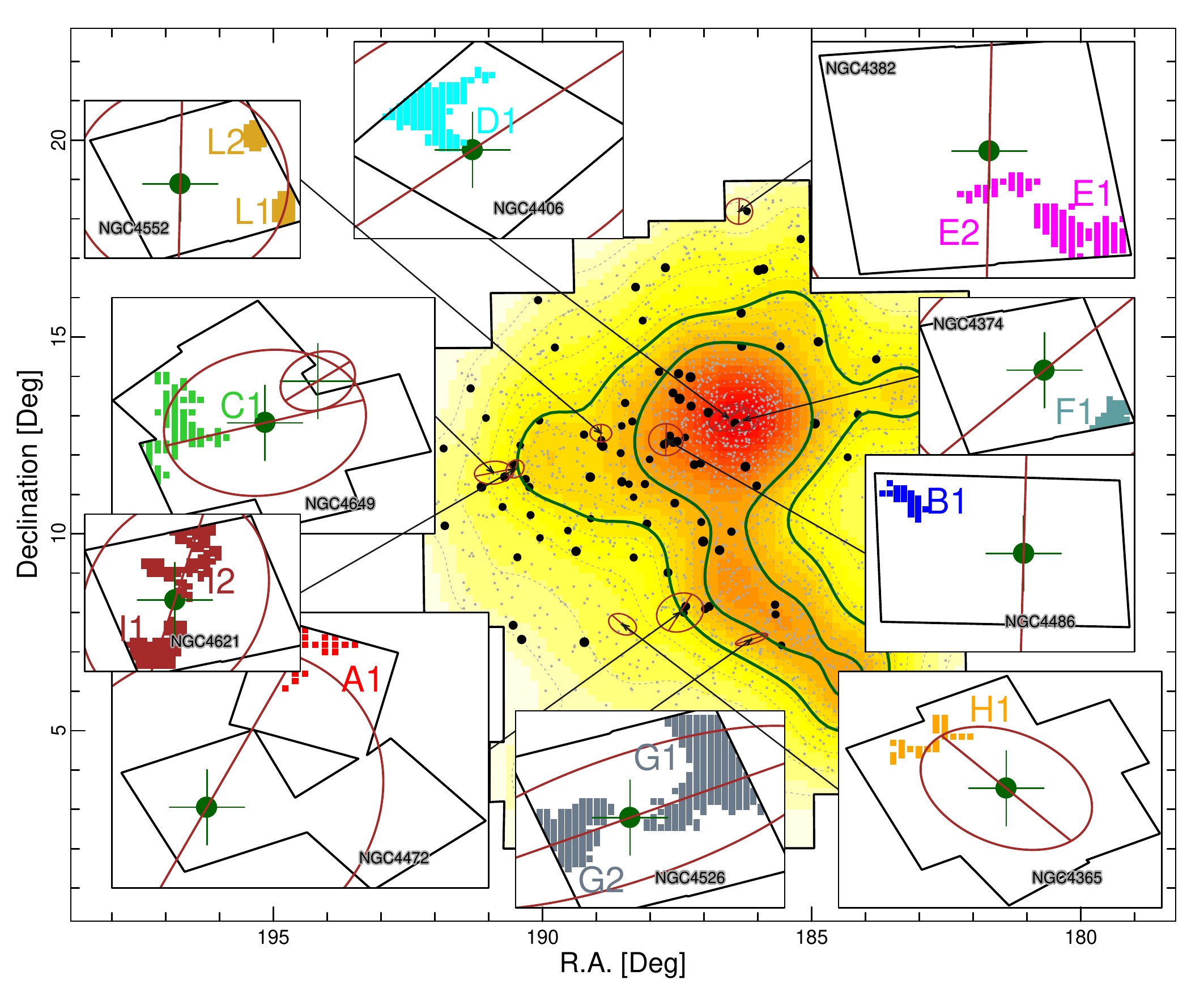}
	\caption{Positions of the ten Virgo cluster galaxies whose GC spatial distribution has been
	investigated in this paper (the $D_{25}$ isophotes of the galaxies, scaled for visibility, are also 
	plotted). On the background, the surface density of confirmed and candidate 
	Virgo cluster galaxies calculated from the~\cite{binggeli1985} catalog is shown. The two green 
	lines are associated, from the center to the outskirts of the cluster, to the 50\% ($\sim40$ galaxies/deg$^{2}$)
	and 33\% ($\sim27$ galaxies/deg$^{2}$) fractions 
	of the peak density. The insets display the location of the large GC structures within each host galaxy.}
	\label{fig:spatial_map_large_structures}
\end{figure*}

Figure~\ref{fig:spatial_map_large_structures} shows the location of the ten galaxies
studied in this paper in the region of the sky 
covered by the Virgo cluster. We compare their position with the local galaxy density, which we
have evaluated from the catalog of confirmed and candidate Virgo galaxies of~\cite{binggeli1985}. 
This catalog is complete to $B_{T}\sim\!18$ mag and covers $\sim\!140\ \mathrm{deg}^2$.
The large GC structures detected in each galaxy, color-coded based on their membership, 
are shown in the insets of Figure~\ref{fig:spatial_map_large_structures} for reference. 
We calculated
the average fraction of the observed galaxy areas belonging to medium and large GC structures for 
both higher- and lower-density regions of the Virgo cluster using the data shown 
in Table~\ref{tab:galaxies_structures}.
Only 2.1\% of the total observed area of the three galaxies (NGC4406, NGC4486 and 
NGC4374) located within the inner green line in Figure~\ref{fig:spatial_map_large_structures}, 
which is the isodensity contour corresponding to 50\% of the peak density of the Virgo cluster 
galaxies, is occupied by large GC structures (16.9\% if both medium and large structures are
considered). The fraction of the area of the remaining seven host galaxies located outside of the 
50\% isodensity contour which is occupied by large structures grows to 8.5\%, while it remains
almost constant (16.6\%) if both medium and large GC structures are considered. While these
numbers seem to indicate that galaxies in the highest-density regions of Virgo tend to have
smaller GC structures than galaxies located elsewhere, we note that the uneven spatial coverage
of ACSVCS is a source of uncertainty in these conclusions. 
For instance, if we consider only the five galaxies outside of the outer green line 
(the isodensity contour corresponding to the 33\% of the peak density), the fractions of 
area occupied by large and large plus medium GC structures become 2.8\% and 
15.2\%, respectively. This inversion of the trend, though, is likely caused by 
the incomplete spatial coverage of the sample: NGC4649 and NGC4365, the only two galaxies
almost completely observed within the $D_{25}$, are both located outside of the 33\%
contour, and their presence ``dilutes'' the average fractional area occupied by large structures.
Another possible source of bias is the uneven distribution of galaxy morphologies in 
higher- and lower-density regions: lenticulars and ellipticals with significant ellipticity
are mostly located in the lower-density regions, while two of the three galaxies within the 50\%
isodensity contour (NGC4486 and NGC4374) show negligible to small ellipticity (E0 and E1, 
respectively). 

As discussed in Section~\ref{subsec:positions}, the large GC structures 
are on average more significant ($\sim1.6$ times) in the red GCs than in the general GC population. When 
split according to the environment density and normalized by the area of the structures, the large GC
structures are even more significant in the residual maps generated by red GCs 
in galaxies in low-density regions than the large structures in medium and high-density areas of the cluster. 
In particular, the large GCs structures associated to host galaxies outside of the 
33\% of the peak density region are, on average, $5.3$ times more significant than GC structures in all GCs, while the
same excess probability decreases to $3.5$ and $1.2$ for large structures in the medium and high-density
regions.

Our findings are in agreement with two different recent accretion histories for the galaxies
in the higher- and lower-density regions of the cluster. As suggested by hydrodynamical
simulations~\citep[Millennium simulation, see][]{springel2005} and observations of the local
clusters of galaxies~\citep[for instance, for Coma cluster][]{smith2012}, major merging events 
occurred earlier in the high-density regions at the center of the clusters. For instance, according to the 
Millennium simulation, an average galaxy 
currently observed at small clustercentric distance joined a halo whose size is typical of 
the Virgo cluster 8.9 Gyr ago, corresponding to redshift $z\!=\!1.3$. At a distance of 2 Mpc, 
the same galaxy reached a cluster-like environment only 4.8 Gyr ago ($z\!=\!0.5$)~\citep{smith2012}.
Observationally, mergers in the distant past have depleted of similar-sized companion 
the environment of the largest 
galaxies, which are currently observed in the deepest potentials. 
Therefore, only recent accretions of smaller-than-average galaxies (satellites) can be expected
in the densest areas of clusters. Instead, major mergers (or wet mergers with larger mass ratios) 
may have occurred in the more recent past in lower density areas, producing the large scale, spatially 
extended GC structures mostly located along the major axis of the host galaxies, that we observe. 

\subsection{Constraints on potential progenitors of the GC structures}
\label{subsec:satellite}

The $\Lambda$CDM model of hierarchical galaxy formation~\citep{white1978,dimatteo2005} 
predicts continuous galaxy evolution via merging and accretion of satellites. Supporting
evidence are provided by recent observations of the local Universe, including the Sagittarius 
Stream~\citep{ibata1994}; Milky Way companions undergoing tidal disruption~\citep{belokurov2006}, and 
streams and dwarf galaxies in the halo of M31~\citep{mcconnachie2009}. The kinematics of the GC 
system of NGC4486 also 
suggest active accretion of satellite galaxies~\citep{romanowsky2012}, as do the presence of GC 
streams in a few ellipticals (e.g.,~\citealt{strader2011,blom2012a}).
Other results~\citep{bonfini2012} and our previous work on the 2D GC distributions of E 
galaxies~\citep{dabrusco2013,dabrusco2014a,dabrusco2014b} 
as well as this work suggest
that GC systems may bear the imprint of these effects, as fossil remnants of accretion events of 
small dwarf galaxies~\citep{penarrubia2009}. 

Based on our results, we assume that the structures we have detected in the 2D 
GC distributions of the ten most luminous Virgo ellipticals are the remnants of 
accreted galaxies, detected over the smooth, relaxed GC distribution of the main galaxy.
A discussion of the caveats to this assumption can be found 
in Section~\ref{subsec:caveats}. Using the excess number of GCs in the detected over-density 
structures, we can set 
lower limits on the masses of their parent galaxies, by using the relations of~\cite{harris2013} between the 
number of GCs and the dynamical mass of elliptical galaxies. We stress that these estimates are lower 
limits, because GCs may be destroyed or reduced in these encounters~\citep[see][for an extensive 
review on the dynamical evolution of GCs]{fall2001}. Moreover, detailed numerical simulations of the 
evolution of GCs in these scenarios are lacking. 

We considered all the GC medium and large over-density structures of the ten Virgo cluster
galaxies in our sample (Table~5). Among these structures, we 
will focus, in particular, on the over-density structures that are clearly identifiable also in the residual maps 
obtained from the blue GCs, because metal-poor GCs could have been either produced early in the assembly history 
of the host galaxy or subsequently added through accretion of low-mass, metal-poor satellite 
galaxies~\citep{brodie2006,peng2008}. These structures are A1, A3, B1, B2, B3, B4, B5, D2, E1,E2, E4, 
F1, F2, F3, G1, G2, H1, H2, H3, H4, H5, L1, L2, L3, L4 (the other structures for which the satellite properties 
have been calculated for comparison only are marked with an asterisk in 
Table~5). 

To estimate the number of GCs in each over-density structure, we used both the observed radial 
density profile $N_{\mathrm{exc,GC}}$ (which is a measure of the azimuthally averaged radial GC 
distribution in a galaxy) and the minimal radial density profile 
$N_{\mathrm{exc,GC}}^{\mathrm{(MRD)}}$\footnote{The minimal radial density (MRD) profile 
is defined as the density profile obtained by setting the density in each radial bin equal to the lowest density observed in that radial bin. 
In physical terms, this assumption translates to assuming that all positive residuals in the 
residual map of GCs can be interpreted as accreted GC systems from distinct satellite galaxies.}, 
obtained by using the minimum value for each given azimuth 
(Table~5). These two different definitions of the density 
radial profiles provide reasonable estimates of the minimum and maximum size of the GC systems 
in the structures, and were both used to evaluate the range of possible values of the basic properties 
of the accreted galaxies. We then corrected these estimates for incompleteness using the GC luminosity 
functions determined with the ACSVCS survey data for Virgo cluster galaxies by~\cite{jordan2007}. 
The average additive corrections are $\sim\!0.6$ and $\sim\!0.9$ GCs per density structure for the 
two radial density profiles used, respectively. We then applied the relations 
$N_{\mathrm{GCs}}\!=\!M_{\mathrm{dyn}}^{1.04\pm0.03}$ and 
$N_{\mathrm{GCs}}\!=\!(r_{e}\sigma_{e})^{1.29\pm0.03}$~\citep{harris2013}, where $M_{\mathrm{dyn}}$ 
is the dynamical mass~\citep{wolf2010} defined as:

\begin{equation}
	M_{\mathrm{dyn}}\!=\!\frac{4r_{e}\sigma_{e}^{2}}{G}
\end{equation}

\noindent which approximates the total baryonic mass of the galactic bulge within the effective 
radius $r_{e}$, where the contribution of the dark-matter halo is negligible~\citep{graves2010};  
$\sigma_{e}$ is the central velocity dispersion of the galaxy derived from spectroscopic observations
of the bulge available in the literature~\citep[see][and references therein]{harris2013}.~\cite{harris2013} 
found that $r_{e}\sigma_{e}$ strongly correlates with the size of the GC population. 

The results (Table~5) show that under our assumptions, the average fraction of mass 
contributed by each accretion event (among the GC structures with no asterisk) 
is $\sim\!1.3\%$. The largest fraction of dynamical mass of the host galaxy accreted through 
one satellite is $6.9\%$ ($10.1\%$ for MRD profile) for G2 in NGC4526 
(assuming that each GC structure corresponds to one accreted satellite). The largest fraction 
of total mass gained by the host galaxy through all the accretion events associated to the 
observed GC structures is $\sim\!12.2\%$ ($\sim\!17.8\%$ for MRD profile) for NGC4526 (G1 and G2)).
Among the other galaxies, the total fraction of dynamical mass contributed by all accretion events 
exceeds $4\%$ only for NGC4365 for the observed density profile; two more galaxies exceed 
this limit when using the MRD profiles (NGC4365 and NGC4621). The estimated dynamical masses of the 
accreted satellite galaxies span the $[\sim\!8.5,\sim\!10.2]$ $\log{M/M_{\sun}}$ for the observed
radial density profiles and $[\sim\!8.3,\sim\!10.3]$ $\log{M/M_{\sun}}$ for MRD profiles. Considering 
that these are lower limits, they are consistently larger that the typical masses of dwarf elliptical 
galaxies (dEs) $M\!\sim\!10^{7}\!M_{\sun}$~\citep{mateo1993}. Using the relation 
$N_{\mathrm{GCs}}\!=\!M_{\mathrm{dyn}}^{0.37\pm0.09}$ obtained from~\cite{harris2013} for 
dEs, the estimated masses of the accreted galaxies, on average, are only $\sim\!20\%$ smaller 
than the masses obtained with the relation for all elliptical galaxies, and still significantly more 
massive than typical dEs. This fact may represent another indication that these galaxies 
have undergone multiple mergers, which is a reasonable hypothesis since they are members of the Virgo 
clusters, and some of them even in sub-groups in Virgo (see Section~\ref{subsec:galaxydensity}). 

\begin{figure*}
	\includegraphics[height=12cm,width=16cm,angle=0]{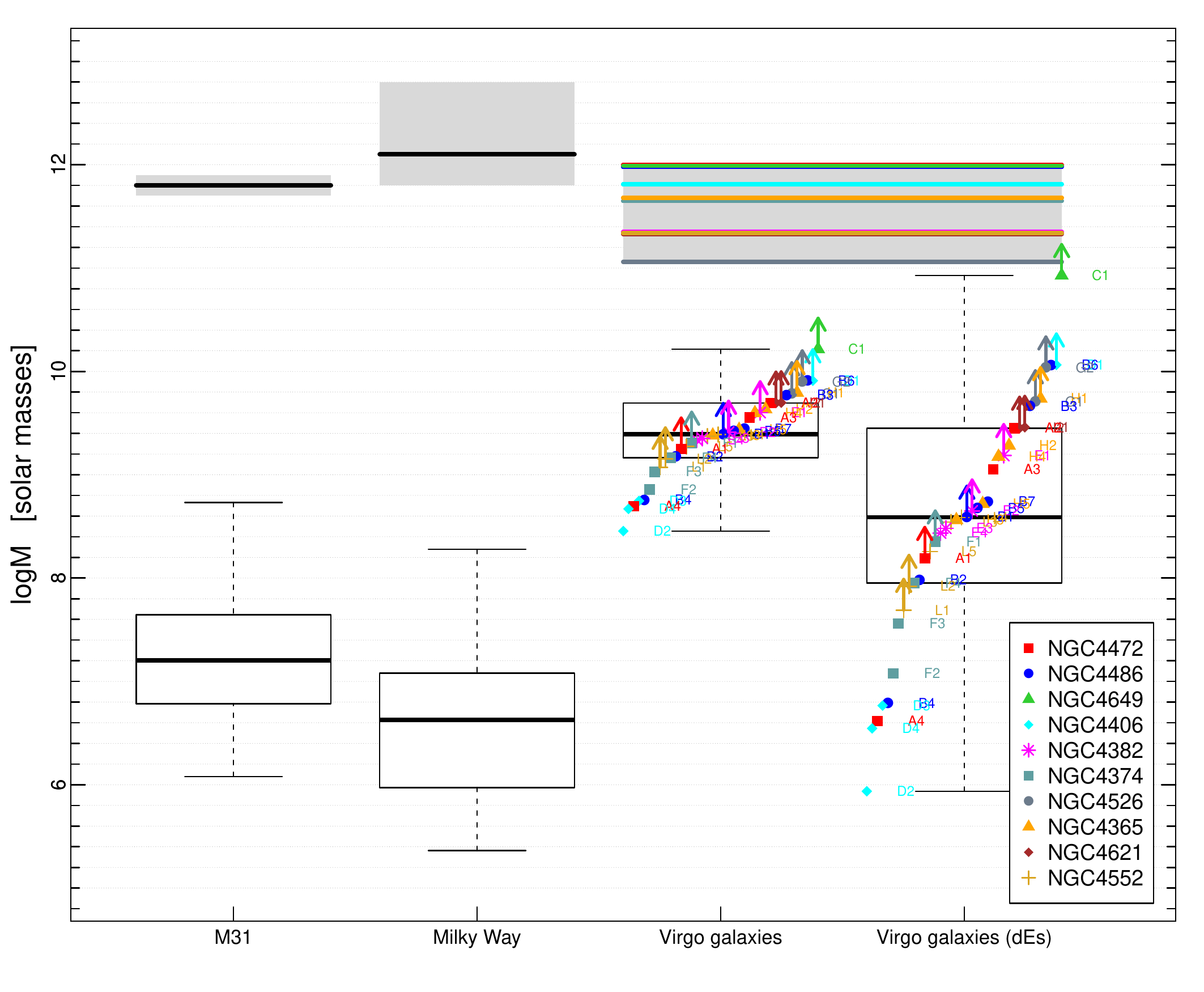}
	\caption{Comparison of the distribution of dynamical 
	masses of the satellites of the Virgo galaxies estimated
	with the method discussed in this paper, with the distributions of masses of the satellite galaxies of 
	M31 and the Milky Way collected by~\cite{mcconnachie2012}. The two boxplots on the 
	right represent the distribution of inferred masses of the Virgo satellite galaxies assuming that
	the GC over-density structures are due to accretion events. These are determined with the 
	two relations $N_{\mathrm{GCs}}\!=\!M_{\mathrm{dyn}}^{1.04\pm0.03}$ and 
	$N_{\mathrm{GCs}}\!=\!M_{\mathrm{dyn}}^{0.37\pm0.09}$ determined by~\cite{harris2013} using 
	all elliptical galaxies of the sample and only dEs, respectively. The gray areas above the boxplots
	represent the range of mass values for the host galaxies (the masses estimates for M31 are 
	from~\citealt{evans2000}, the interval of masses for the Milky Way is from~\cite{mcmillan2011}, 
	and the single Virgo cluster galaxies are dynamical masses from~\citealt{harris2013}). The estimated
	masses of the individual satellites for the Virgo cluster galaxies are 
	represented as points with the same symbols and color-coding used in the other plots of the paper. 
	Arrows are used for GC over-density structures that are likely to have been produced by major
	mergers (as discussed in Section~\ref{subsec:positions}), and represent lower limits 
	of the masses of the galaxies involved. The horizontal displacement of the points is only shown 
	for the sake of clarity.}
	\label{fig:boxplot_masses}
\end{figure*}

Figure~\ref{fig:boxplot_masses} compares the estimates of the 
distribution of dynamical masses of the satellites of the Virgo cluster galaxies 
(from Table~5), with the distributions of masses of the Milky Way~\citep{mcmillan2011} and 
M31 satellites~\citep{mcconnachie2012}. While we have used for all the M31 satellites 
the estimates of the dynamical masses within the observed
half-light radius available in~\citep{mcconnachie2012}, different definitions have been used to 
determine the masses of the MW satellites, that usually yield mass estimates than can be 
systematically larger than the dynamical masses for the same systems. Even within the uncertainties 
and the possible systematic differences in the masses of the host galaxies, our estimates point to 
larger satellites in the Virgo galaxies.
Considering the similar mass of the MW, M31 and the Virgo ellipticals, environment may play a 
role. As shown, the mass of each accreted objects is a small fraction of the total mass of the main 
galaxy. Therefore, one would expect only mild morphological and kinematic disturbances (assuming a 
dry merger) that would not be easily discernible from the light of the galaxy. Furthermore, one would 
expect a correlation between the mass of the accreted satellite and the number of excess GCs. However, 
such a correlation is easily diluted by multiple cannibalism events, the time since coalescence, and the 
geometry of the interaction. Simulations of the evolution of the GC populations for a period of several Gyr 
after a dry merger are needed to address these issues.
 
In Section~\ref{subsec:galaxydensity}, we reported that the GC structures observed in host
galaxies located in lower-density regions of the Virgo cluster are more significant
in the distribution of red GCs than in all GCs, than structures observed in galaxies located 
in higher-density regions. We argued that this evidence, in agreement with the predictions of cosmological
simulation of the assembly history of galaxies in cluster, indicates that recent major mergers
are mostly responsible for the properties of the GC spatial distribution in the outskirts of Virgo, 
while the innermost galaxies, recently, have only accreted smaller companion galaxies. The 
fraction of dynamical mass contributed to the host galaxy by a single progenitor
(shown in Table~5) does not exceed 5\% (10\% for MRD profiles) for the five galaxies 
(NGC4472, NGC4649, NGC4382, NGC4365 and NGC4621) located outside of the 33\% isodensity 
contour shown in Figure~\ref{fig:spatial_map_large_structures}. No single GC structure accounts for
25\% of the mass of the final galaxy, which is the threshold used to label major 
mergers (i.e., mergers where the progenitors have 3:1 or lower mass ratio). Even assuming that 
all observed GC structures originate from a single accretion event, the 
total fraction of host galaxy dynamical mass contributed by the progenitors (including also GC structures marked
with asterisks in Table~5) for these five galaxies are 1.7\% for NGC4472, 2.2\% for NGC4649, 3.7\% for NGC4382,
5.9\% for NGC4365 and 4.6\% for NGC4621 (with MRD profiles the fractions are 1.2\%, 19.2\%, 
1.6\%, 6.5\% and 7.9\% respectively). The disagreement between our discussion in Section~\ref{subsec:galaxydensity}
and these estimates depends mostly on two factor. Firstly, in our simple model
that employs either the observed or MRD profiles to calculate the excess number of GCs for each GC
structure, the dynamical evolution of the GCs is not taken into account. The changes in the mass 
function of GCs due to internal mechanism (stellar evolution and two-body relaxation) and to interactions 
with the host galaxy potential~\citep[gravitational shocks and dynamical friction;][]{fall2001} all contribute to 
lower the estimated number of excess GCs and, based 
on the~\cite{harris2013} relations, to underestimate the dynamical mass of the progenitors. The second factor is
represented by the fact that major and/or wet mergers are usually conducive of new star formation and, 
consequently, the formation of new GCs. The use of the relations derived by~\cite{harris2013} implies a
``passive'' scenario where GCs are donated from the progenitors to the main galaxy; it
does not take into account the effect of the merger on the GC system of the resulting galaxy. Moreover, 
the uncertainties on the GC specific frequencies for galaxies of large luminosity and different morphology 
that are typically involved in wet/major mergers and that can
affect the scatter and accuracy of the $N_{GC}$ - $M_{\mathrm{dyn}}$ relation (see discussion 
in~\citealt{harris2013}), further undermines the validity of the estimated dynamical masses in the massive
tail of the mass distribution of the progenitors.
It is also worth mentioning that major mergers are not the only possible way of forming new GCs, as 
wet mergers with mass ratios larger than 3:1 may also produce significant amounts of star formation and, 
in turn, new young, metal-rich GCs.

Detailed simulations of the formation and evolution of the GC populations in mergers of different
types would be crucial to reconcile the tension between the assembly histories discussed in 
Section~\ref{subsec:galaxydensity} and the inferences about the progenitor galaxies made in this Section, 
and would contribute to paint a more realistic and complex views of the evolution of GCs and their
host galaxies in environment like the Virgo cluster.
 
 %\begin{table*}
\begin{sidewaystable*}[p]
	%\small
	\centering	
	\vspace{6cm}
	\resizebox{\textwidth}{!}{	
	\begin{threeparttable}			
	\caption{Properties of the accreted galaxies associated to the ``medium'' and ``large'' 
	GC over-density structures.}
	\begin{tabular}{lccccccccccc}
	\tableline
	Galaxy			&	Structure	& $N_{\mathrm{exc,GCs}}$	 &	$N_{\mathrm{exc,GCs}}^{(\mathrm{MRD})}$			& 	
						$\log{(M_{\mathrm{dyn}}/M_{\sun})}$		&	$\log{(M_{\mathrm{dyn}}/M_{\sun})}^{(\mathrm{MRD})}$	& 
						$\log{(M_{\mathrm{dyn}}/M_{\sun})}^{(\mathrm{dEs})}$	&	$\log{(M_{\mathrm{dyn}}/M_{\sun})}^{(\mathrm{MRD,dEs})}$	& 
						$r_{e}\sigma_{e}$								& 	$r_{e}\sigma_{e}^{(\mathrm{MRD})}$ 					&	
						$\%(M_{\mathrm{Dyn}})$							&	$\%(M_{\mathrm{Dyn}})^{(\mathrm{MRD})}$				\\
					&	(a)	&	(b)	&	(c)	&	(d)	&	(e)	&	(f)	&	(g)	&	(h)	&	(i)	&	(l)		&	(m)	\\	
	\tableline
	NGC4472	(M49)&  A1 & 8.1 & 18.7 & 9.3$\pm$0.2 & 9.6$\pm$0.4 & 8.2$\pm$1.6 & 9.2$\pm$3.1 & -1.2$\pm$0.1 & -1.0$\pm$0.2 & 0.4 & 0.2\\
			&  A2$^{*}$ & 23.2 & 31.0 & 9.7$\pm$0.4 & 9.8$\pm$0.5 & 9.5$\pm$3.5 & 9.8$\pm$4.2 & -0.9$\pm$0.3 & -0.8$\pm$0.3 & 0.7 & 0.6\\
			&  A3 & 16.6 & 24.0 & 9.6$\pm$0.3 & 9.7$\pm$0.4 & 9.1$\pm$2.8 & 9.5$\pm$3.6 & -1.0$\pm$0.2 & -0.9$\pm$0.3 & 0.5 & 0.3\\
			&  A4$^{*}$ & 2.1 & 6.0 & 8.7$\pm$0.1 & 9.1$\pm$0.1 & 6.6$\pm$0.3 & 7.8$\pm$1.2 & -1.7$\pm$0.1 & -1.3$\pm$0.1 & 0.1 & 0.1\\
	NGC4486	(M87)&  B1 & 11.3 & 14.5 & 9.4$\pm$0.3 & 9.5$\pm$0.3 & 8.6$\pm$2.1 & 8.9$\pm$2.6 & -1.1$\pm$0.2 & -1.1$\pm$0.2 & 0.3 & 0.1\\
			&  B2 & 6.8 & 8.6 & 9.2$\pm$0.2 & 9.3$\pm$0.2 & 8.0$\pm$1.3 & 8.3$\pm$1.7 & -1.3$\pm$0.1 & -1.2$\pm$0.1 & 0.2 & 0.1\\
  			&  B3 & 27.7 & 36.5 & 9.8$\pm$0.5 & 9.9$\pm$0.6 & 9.7$\pm$3.9 & 10.0$\pm$4.6 & -0.8$\pm$0.3 & -0.7$\pm$0.4 & 0.8 & 1.0\\
  			&  B4 & 2.5 & 5.1 & 8.8$\pm$0.1 & 9.1$\pm$0.1 & 6.8$\pm$0.4 & 7.7$\pm$1.0 & -1.6$\pm$0.1 & -1.4$\pm$0.1 & 0.1 & 0.1\\
  			&  B5 & 12.1 & 14.6 & 9.4$\pm$0.3 & 9.5$\pm$0.3 & 8.7$\pm$2.2 & 8.9$\pm$2.6 & -1.1$\pm$0.2 & -1.0$\pm$0.2 & 0.3 & 0.1\\
  			&  B6$^{*}$ & 38.7 & 49.7 & 9.9$\pm$0.6 & 10.0$\pm$0.7 & 10.1$\pm$4.7 & 10.4$\pm$5.4 & -0.7$\pm$0.4 & -0.6$\pm$0.4 & 1.1 & 2.4\\
  			&  B7$^{*}$ & 12.8 & 16.3 & 9.4$\pm$0.3 & 9.5$\pm$0.3 & 8.7$\pm$2.3 & 9.0$\pm$2.8 & -1.1$\pm$0.2 & -1.0$\pm$0.2 & 0.4 & 0.1\\
	NGC4649	(M60)&  C1$^{*}$ & 80.3 & 107.4 & 10.2$\pm$0.8 & 10.3$\pm$1.0 & 10.9$\pm$6.8 &11.3$\pm$7.7&-0.5$\pm$0.5 & -0.4$\pm$0.6 & 2.2 & 19.2\\
	NGC4406	(M86)&  D1$^{*}$ & 38.8 & 53.1 & 9.9$\pm$0.6 & 10.0$\pm$0.7 & 10.1$\pm$4.7 & 10.4$\pm$5.6 & -0.7$\pm$0.4 & -0.6$\pm$0.4 & 1.7 & 4.2\\
 	 		&  D2 & 1.2 & 2.1 & 8.5$\pm$0.1 & 8.7$\pm$0.1 & 5.9$\pm$0.1 & 6.6$\pm$0.3 & -1.9$\pm$0.1 & -1.7$\pm$0.1 & 0.1 & 0.1\\
 		 	&  D3$^{*}$ & 2.4 & 3.3 & 8.7$\pm$0.2 & 8.9$\pm$0.1 & 6.8$\pm$0.4 & 7.1$\pm$0.6 & -1.7$\pm$0.2 & -1.5$\pm$0.2 & 0.1 & 0.1\\
 	 	    	&  D4$^{*}$ & 2.0 & 3.5 & 8.7$\pm$0.2 & 8.9$\pm$0.1 & 6.5$\pm$0.3 & 7.2$\pm$0.6 & -1.7$\pm$0.1 & -1.5$\pm$0.1 & 0.1 & 0.2\\
	NGC4382	(M85)&  E1 & 18.6 & 22.7 & 9.6$\pm$0.4 & 9.7$\pm$0.4 & 9.2$\pm$3.0 & 9.4$\pm$3.5 & -1.0$\pm$0.2 & -0.9$\pm$0.3 & 2.2 & 1.2\\
 	 		&E2 & 11.8 & 15.0 & 9.4$\pm$0.3 & 9.5$\pm$0.3 & 8.7$\pm$2.2 & 8.9$\pm$2.6 & -1.1$\pm$0.2 & -1.0$\pm$0.2 & 1.5 & 0.4\\
 		 	&E3$^{*}$ & 10.3 & 12.6 & 9.4$\pm$0.2 & 9.4$\pm$0.3 & 8.5$\pm$2.0 & 8.7$\pm$2.3 & -1.2$\pm$0.2 & -1.1$\pm$0.2 & 1.2 & 0.2\\
  			&E4 & 9.9 & 12.6 & 9.3$\pm$0.2 & 9.4$\pm$0.3 & 8.4$\pm$1.9 & 8.7$\pm$2.3 & -1.2$\pm$0.2 & -1.1$\pm$0.2 & 1.2 & 0.2\\
	NGC4374	(M85)&  F1 & 9.2 & 9.5 & 9.3$\pm$0.2 & 9.3$\pm$0.2 & 8.4$\pm$1.8 & 8.4$\pm$1.8 & -1.2$\pm$0.1 & -1.2$\pm$0.1 & 0.5 & 0.1\\
  			&	F2 & 3.2 & 3.3 & 8.9$\pm$0.1 & 8.9$\pm$0.1 & 7.1$\pm$0.5 & 7.1$\pm$0.6 & -1.6$\pm$0.1 & -1.5$\pm$0.1 & 0.2 & 0.2\\
  			&F3 	& 4.7 & 5.6 & 9.0$\pm$0.1 & 9.1$\pm$0.1 & 7.6$\pm$0.9 & 7.8$\pm$1.1 & -1.4$\pm$0.1 & -1.4$\pm$0.1 & 0.3 & 0.2\\
  			&F4$^{*}$ & 6.6 & 7.5 & 9.8$\pm$0.5 & 9.2$\pm$0.2 & 8.0$\pm$1.3 & 8.1$\pm$1.5 & -1.3$\pm$0.1 & -1.3$\pm$0.1 & 0.4 & 0.1\\	
	NGC4526	&  G1 &  28.8 & 47.7 & 9.3$\pm$0.2 & 10$\pm$0.7 & 9.7$\pm$2.4 & 10.3$\pm$4.6 & -0.8$\pm$0.3 & -0.7$\pm$0.2 & 5.3 & 7.7\\
  			&  G2 & 37.9  & 58.3 & 9.9$\pm$0.5 & 10.1$\pm$0.7 & 10.3$\pm$4.1 & 10.5$\pm$4.7 & -0.7$\pm$0.2 & -0.6$\pm$0.1 & 6.9 & 10.1\\
	NGC4365	&  H1 & 29.5 & 47.2 & 9.8$\pm$0.5 & 10.0$\pm$0.7 & 9.7$\pm$4.1 & 10.3$\pm$5.2 & -0.8$\pm$0.3 & -0.7$\pm$0.4 & 2.1 & 4.1\\
 	 		&H2 & 20.1 & 28.2 & 9.6$\pm$0.4 & 9.8$\pm$0.5 & 9.3$\pm$3.2 & 9.7$\pm$4.0 & -0.9$\pm$0.3 & -0.8$\pm$0.3 & 1.2 & 1.0\\
  			&H3 & 11.0 & 15.0 & 9.4$\pm$0.3 & 9.5$\pm$0.3 & 8.6$\pm$2.1 & 8.9$\pm$2.6 & -1.1$\pm$0.2 & -1.0$\pm$0.2 & 0.7 & 0.2\\
  			&H4 & 18.4 & 27.9 & 9.6$\pm$0.4 & 9.8$\pm$0.5 & 9.2$\pm$3.0 & 9.7$\pm$3.9 & -1.0$\pm$0.2 & -0.8$\pm$0.3 & 1.2 & 1.0\\
  			&H5 & 12.5 & 15.7 & 9.4$\pm$0.3 & 9.5$\pm$0.3 & 8.7$\pm$2.3 & 9.0$\pm$2.7 & -1.1$\pm$0.2 & -1.0$\pm$0.2 & 0.7 & 0.2\\
	NGC4621	&I1	& 23.4 & 36.5 & 9.7$\pm$0.4 & 9.9$\pm$0.4 & 9.4$\pm$2.2 & 10$\pm$3.6 & -0.9$\pm$0.3 & -0.7$\pm$0.1 & 2.3 & 3.5\\
  			&I2	& 23.2 & 44.6 & 9.7$\pm$0.4 & 10$\pm$0.3 & 9.5$\pm$2.5 & 10.2$\pm$3.6 & -0.9$\pm$0.3 & -0.7$\pm$0.1 & 2.3 & 4.4\\
	NGC4552	(M89)&  L1 & 5.3 & 0.8 & 9.1$\pm$0.1 & 8.3$\pm$0.1 & 7.7$\pm$1.0 & 5.4$\pm$0.1 & -1.4$\pm$0.1 & -2.0$\pm$0.2 & 0.1 & 0.1\\
  			&L2 & 6.4 & 6.8 & 9.2$\pm$0.2 & 9.2$\pm$0.2 & 7.9$\pm$1.3 & 8.0$\pm$1.3 & -1.3$\pm$0.1 & -1.3$\pm$0.1 & 0.7 & 0.2\\
  			&L3 & 10.8 & 12.4 & 9.4$\pm$0.3 & 9.4$\pm$0.3 & 8.5$\pm$2.0 & 8.7$\pm$2.3 & -1.2$\pm$0.2 & -1.1$\pm$0.2 & 1.2 & 0.2\\
  			&L4 & 11.2 & 13.0 & 9.4$\pm$0.3 & 9.5$\pm$0.3 & 8.6$\pm$2.1 & 8.8$\pm$2.4 & -1.1$\pm$0.2 & -1.1$\pm$0.2 & 1.3 & 0.3\\
  			&L5$^{*}$ & 8.5 & 9.8 & 9.3$\pm$0.2 & 9.3$\pm$0.2 & 8.3$\pm$1.7 & 8.4$\pm$1.9 & -1.2$\pm$0.1 & -1.2$\pm$0.1 & 1.0 & 0.1\\
	\tableline
	\end{tabular}
	\begin{tablenotes}[para]
 Ê Ê Ê Ê Ê	\item {Column description:}\\
 Ê Ê Ê Ê Ê	\item[a] GC structure label\\
 Ê Ê Ê Ê Ê	\item[b] number of excess GCs evaluated with the observed radial density profiles $N_{\mathrm{exc,GCs}}$\\
		\item[c] number of excess GCs evaluated with the minimal radial density (MRD) profiles $N_{\mathrm{exc,GCs}}^{(\mathrm{MRD})}$\\
		\item[d] dynamical masses of the accreted galaxies based on the observed radial density profiles $\log{(M_{\mathrm{dyn}}/M_{\sun})}$\\
		\item[e] dynamical masses of the accreted galaxies based on the MRD profiles $\log{(M_{\mathrm{dyn}}/M_{\sun})}^{(\mathrm{MRD})}$\\		
		\item[f] dynamical masses of the accreted galaxies based on the observed radial density profiles $\log{(M_{\mathrm{dyn}}/M_{\sun})}^{(\mathrm{dEs})}$
		assuming the~\cite{harris2013} relation for dEs\\		
		\item[g] dynamical masses of the accreted galaxies based on the MRD profiles $\log{(M_{\mathrm{dyn}}/M_{\sun})}^{(\mathrm{MRD,dEs})}$
		assuming the~\cite{harris2013} relation for dEs\\	
		\item[h] $r_{e}\sigma_{e}$ parameter based on the observed radial density profiles\\	
		\item[i] $r_{e}\sigma_{e}^{(\mathrm{MRD})}$ parameter based on the MRD profiles\\
		\item[l] fraction $\%(M_{\mathrm{Dyn}})$ of the total dynamical mass of the host galaxy contributed by the distinct accreted satellite galaxies
		(observed radial density profiles)\\		
		\item[m] fraction $\%(M_{\mathrm{Dyn}})^{(\mathrm{MRD})}$ of the total dynamical mass of the host galaxy contributed by the distinct accreted satellite galaxies
		(MRD profiles)\\			
	\end{tablenotes}	
	\end{threeparttable}	
	}		
%	}
	\label{tab:fraction_mass_accreted_galaxies}
%\end{table*}
\end{sidewaystable*}

\subsection{Caveats}
\label{subsec:caveats}

Our inferences on the properties of the progenitors of the GC structures 
(Section~\ref{subsec:satellite}) are based on the hypothesis 
that all the structures originated by either accretion of satellite galaxies or major mergers.
While this may be is an oversimplification of the multiple, diverse mechanisms of GC formation at
work in massive early-type galaxies~\citep{brodie2006}, it may apply at least for some of the GC structures 
we have detected (see Section~\ref{subsec:observational_evidence}). Below we discuss the possible 
issues that may affect our conclusions.

\subsubsection{Projection effects}
\label{subsubsec:projection}

A concern is the possibility that the observed GC 
structures are not physical structures, but may results from 
superposition of GCs, due to projection effects. Such bias cannot be excluded {\it a priori} 
without knowledge of the real three dimensional distribution of GCs around the host galaxy. It is 
nonetheless reasonable to assume that the importance of projection effects increases towards 
the center of the galaxy. For instance,~\cite{woodley2011} investigated
the presence of physical groups of GCs in NGC5128, a giant local elliptical galaxy, and argued that 
the chance of projection yielding chance two-dimensional groupings of all GCs is high within $10\arcmin$
($\sim\!2 r_{e}$). In our sample, because of the limited ACSVCS spatial coverage (see Figure~\ref{fig:percentage_areas}), 
we cannot exclude that some of the GC structures may be caused or enhanced by projection effects.
For instance, streamer-like structure can be generated by three-dimensional GC spatial distributions 
following the triaxial stellar morphology shown by some elliptical galaxies.
However, the risk of detecting unphysical GC structures is mitigated by the 
fact that most of the large and medium GC structures at distance $d\!\leq\!2 r_{e}$ have complex 
morphologies that cannot be explained with simple projection effects, assuming a smooth three-dimensional
spatial distribution of GCs. Moreover, pure projection effects cannot explain
the striking differences observed in the properties of the same GC structures detected in distinct color 
and luminosity classes (see Section~\ref{sec:results}), even when we take into account the differences in the radial 
profile of red and blue GCs. 

\subsubsection{Dynamical times}
\label{subsubsec:dynamicaltimes}

Another possible issue with our interpretation of the origin of GC structures observed 
is related to their dynamical evolution. The GC structures located in the outer regions of their
host galaxies (A1 and A4 in NGC4472, C1 in NGC4649 and H1 in NGC4365) might be relatively long-lived,
as relaxation times are longer at large galactocentric radii. However, several observations have 
uncovered evidence of currently ongoing
or recent accretion of satellites in the halos of these galaxies (see Section~\ref{subsec:observational_evidence}).  
The GC structures located closer to the center of the host galaxies, assuming that they 
are real physical over-densities, ought to be the results of recent events in the assembly history of the 
host galaxy in order to be still visible today, given that relaxation times are much shorter at small 
galactocentric radii. This simple reasoning rules out the possibility that these
structures have formed in the outskirts of the galaxies and drifted towards their observed positions 
under the sole effect of the dynamical friction of the host galaxy stellar field. 

We can estimate the time 
required for a single GC orbiting the host galaxy to reach the observed (projected) galactocentric distance
from an arbitrary distance under the only effect of dynamical friction. 
We assume a standard mass for a single GC $M_{GC}\!=\!10^{7} M_{\Sun}$, consistent with 
the observed average absolute magnitudes of GCs~\citep{strader2012}, relative stellar circular velocity 
$v_{\mathrm{circ}}\!\sim\!100\ \mathrm{km} \mathrm{s}^{-1}$, which is the typical measured values for 
Virgo cluster galaxies at
the largest radius where reliable estimates are available~\citep{cortes2015} and, as initial distance $r_{i}\!=\!1 r_{e}$. 
The angular distances of the closest dwarfs were converted to physical distances 
using the galaxy distances by~\cite{blakeslee2009}. 
The dynamical friction time $\Delta t_{\mathrm{fric}}$ of the GC infall has been calculated using 
equation (7-26) of Binney \& Tremaine (2008), which express the time required for a GC to reach 
the center of the host galaxy as:

\begin{equation}
	t_{\mathrm{fric}}\!=\! \frac{1.17}{\ln{\Lambda}}\frac{r_{i}^{2}v_{\mathrm{circ}}}{GM}
\end{equation}

The final distance $r_{f}$ of the GC is set to the observed galactocentric distance of the GC 
over-density, which is a lower
limit of the real three-dimensional distance from the center of the galaxy. 
The $\Delta t_{\mathrm{fric}}$ times range in the 6-30 Gyr interval for
all GC structures in our sample (the structures A1 and A4 in NGC4472, C1 in NGC44649, H1, H2, H3, 
H4 and H5 in NGC4365 and L1 and L2 in NGC4552 were excluded because located
at galactocentric distances larger than $1\ r_{e}$). In the case of spatially extended structures, 
the final position has been chosen to coincide with the center of mass of the structures. The uncertainties 
$\sigma(\Delta t_{\mathrm{fric}})$ can be as large as $\sim50\%$ of the estimated values, with average error 
$\bar{\sigma}(\Delta t_{\mathrm{fric}})\!\sim\!22\%$. 
The large times required for these structures to have reached the observed positions under the effect of 
dynamical friction alone are incompatible with the typically short relaxation times in the center of the galaxy
that would have disrupted the GC structures and prevented the observation of the GC overdensity 
structures. Assuming a fixed deprojected galactocentric distance $r\!=\!0.8\ r_{e}$, the dynamical friction times 
are, on average, $\sim$ 13\% shorter (up to $\sim40\%$ in one case), making at least some of the observed GC 
structures compatible with the effect of internal dynamical evolution only.
However, while drifting towards the center of the galaxy under the effect of dynamical friction, the orbits of the 
GC belonging to a structure would lose coherence due to the differential effect of tidal forces. For 
instance,~\cite{prieto2008} found that GCs formed in a satellite halo moving on an initial circular orbit
within the satellite halo, deviate significantly from the original orbit in a $\sim\!10$ Gyr timeframe (shorter 
decoherence times are expected if the satellite galaxy has been tidally disrupted before 5 Gyr). GCs
formed in the main halo and scattered by accreted satellite halos would deviate significantly 
from their original circular orbits within $\sim 6$ Gyr~\citep{prieto2008} from their formation.
For these reasons, internal dynamical evolution {\it alone} can be mostly ruled out as explanation 
of the presence of the GC structures near the center of the galaxies. 

A possible formation mechanism for these GC structures is that they are the coherent remains of the
GC systems of satellite galaxies falling towards the center of the main galaxy at a  
large relative velocity. Another possibility is that 
recent, local formation of new GCs has occurred as a result of a merger. 
While~\cite{cote1998} and~\cite{pipino2007} found that the radial distribution, specific frequency 
and metallicity of the GC systems of massive elliptical 
galaxies can be understood by requiring a large fraction of GCs to have formed in 
satellite galaxies and stripped by the accreting galaxy, the formation of young, 
metal-rich GCs during mergers is believed to be another significant factor
shaping of the observed GC systems of elliptical galaxies~\citep{bekki2003,brodie2006,wang2013}. 
A detailed modeling of the dynamics of the GC structures progenitors, which is out
of the scope of this paper, should also take into account the physical 
mechanisms acting on the GCs during their crossing of the accreting galaxy potential. 
These mechanisms would introduce differential effects in the spatial distribution 
of the observed GCs depending on the relative geometry of the orbit followed by the GC 
system and the potential of the host galaxy.

\subsubsection{Phase-space counterparts of spatial GC structures}
\label{subsubsec:phasespace}

The physical existence of over-density structures in the projected spatial distribution of GCs 
can be confirmed if they are also characterized by a
specific motion, inherited by the accreted host galaxy or the merger progenitors, which is distinct
from the systemic bulk motion of the main host 
galaxy~\citep[e.g.][]{romanowsky2012,coccato2013,blom2012a}.
The lack of spectroscopic data and 
kinematic measurements that would confirm the
nature of all the GC structures in our sample will only be overcome as new observations become 
available. However, we stress that the kinematic identification of GC structures
is a sufficient but not necessary
condition to ascertain the existence of physical groups of GCs originally associated to an accreted 
satellite galaxy. The kinematic method can only  
select GC structures with motions significantly different from the systemic
velocity of the host galaxy along the line-of-sight direction. GC structures mostly moving
transversally to the line-of-sight are missed. Our approach could detect structures no
matter what the direction of motion.

\section{Summary and Conclusions}
\label{sec:conclusions}

We have studied the 2D distributions of the GC systems of the ten most massive Virgo 
galaxies, using the GC catalogs extracted from the ACSVCS data~\citep{jordan2009}, and GC catalogs
extracted from analogous ACS data by~\citep{strader2012} and~\citep{blom2012b}. We have 
estimated the statistical significance of localized spatial structures in these GC 2D distributions, using 
the KNN method augmented by MC simulations~\citep{dabrusco2013}. We have applied
this method to the 2D distribution all the GCs in each galaxy and also to the classes of red, blue, 
low-L and high-L GCs (Section~\ref{sec:results}). Our results can be summarized as follows:

\begin{itemize}
	\item GCs over-density structures are common in the sample of galaxies studied. We observed
	a total of 229 GC structures in the residual maps of the spatial distribution of all GCs in the ten 
	galaxies, and 42 significant, spatially extended GC structures classified as ``medium'' or ``large''
	structures, i.e. structures formed by groups of 5 to 19 adjacent pixels or formed by 20 or more pixels 
	with total significance larger
	than $4\sigma$, respectively (see Section~\ref{sec:results}). Most of these GC structures
	are also visible in the residual maps for red/blue and/or high-L/low-L GC subsamples. The
	physical sizes of the large and intermediate GC structures range between $\sim1$ to $\sim25$
	kpc.
	\item The spatial distribution of the small over-density structures is 
	radially and azimuthally homogeneous (between 1\arcmin and $2\arcmin$). However, with 
	the present data we cannot exclude that some of these small structures may be due to statistical
	noise.
	\item The medium and large structures may be related to the accretion of smaller satellite galaxies. 
	This effect is stronger when the area of the GC structures is considered. However, the
	spatial coverage of our sample is far from complete at larger radii. 
	\item The large, coherent spatial structures, which tend to be aligned along the major
	axis of the host galaxies and are clearly visible in the spatial distribution of red GCs, could
	be the result of major mergers, as predicted by simulations~\citep[see][]{bekki2002}. 
	The medium-size structures may be related to the accretion of smaller satellite galaxies. 
	\item The smaller spatial structures (accounting for only $\sim\!14\%$ of the total area of the 
	GC over-density structures observed) are observed to be more likely to be located along the 
	direction of the minor axis of the host galaxy than the large structures than on the major axis. 
	No significant differences are observed in their color and luminosity distributions.
	\item Large, spatially complex GC structures, mostly aligned along the major axis of the host galaxies, 
	are more likely to be found in low-density regions within the Virgo cluster than in the
	high-density areas. While the red large GCs structures are, on average, more significant
	than structures in the spatial distribution of all GCs (Section~\ref{subsec:positions}), this effect is stronger
	in the lower-density environment, suggesting the relatively recent mergers have 
	determined the GC spatial distribution in these galaxies. Additional observations will be necessary to 
	rule out observational biases.
	\item Using the total number of 
	excess GCs associated to the large and medium GCs structures as a proxy of the size of the 
	whole GC system, 
	we have calculated the range of dynamical masses of the satellites using the relations 
	from~\cite{harris2013}. We found that masses of the satellites estimated with our method are
	significantly larger than the observed masses of the satellites in the Milky Way and M31, suggesting
	that the environment plays an important role in the assembly of galaxies by accretion of satellites. 
 	The estimated masses of the progenitors are too small to be compatible with 
	assembly histories involving recent major mergers, especially in the case of the host galaxies 
	discussed in Section~\ref{subsec:galaxydensity}. We argue that, in case of major mergers, 
	these disagreements could be reconciled with a more detailed description of the evolution of 
	GC systems and the relation between size of the GC system and the properties of the host galaxy. 
	We also note that non-equal mass wet mergers with 
	larger mass ratios could also be responsible for significant GC formation.
\end{itemize}

We have proposed a simple mechanism responsible
for the formation of these GC structures that invokes either small accretion of satellite galaxies or, 
in the case of the largest, spatially extended structures, major mergers, and used it to derive fundamental 
properties of the potential progenitor galaxies (Section~\ref{subsec:satellite}).

Our results show that 2D over-density structures in the spatial distribution 
of GCs in massive early-type galaxies are a common phenomenon, but we 
are currently limited by the lack of deep, high spatial resolution data reaching the outskirts of the host
galaxies. New HST observations complementing the ACSVCS data in the outer regions 
of the Virgo galaxies are necessary to fully determine the characteristics of the 2D spatial 
distribution of GCs. Moreover, GCs 
catalogs covering larger galactocentric baseline would be useful to compare our results with 
data from other tracers of the dynamics and evolution of the host 
galaxy, like GCs and Planetary Nebulae kinematics. The events in the assembly history of the 
host galaxy that, according to our model, may 
have generated the observed GC structures, should also leave evident 
signatures in the dynamics of these systems. Unfortunately, kinematic data for 
homogeneous, large samples of GCs located within the inner region of the host galaxies that 
could validate our assumptions are difficult to obtain. 

Constraining the evolution of massive galaxies also requires synergy between observations and
simulations. Fully cosmological simulations 
of galaxy mergers are now being extended to resolve GC-sized objects~(Mayer 2014, private 
communication) and to determine the effect of hierarchical evolution on the spatial distribution of the GC 
systems. This new
generation of simulations will provide a unique testbed for the formation and evolution of GCs with 
unprecedented resolution and physical detail, producing GC parameters directly comparable 
with the observations. The comparison of the features of the observed 2D GC structures with the
simulated data will assess how the observed 2D GC structures can be used to infer past merging/accretion 
history of the host galaxies. The study of the GC over-density structures and its comparison with 
cosmological simulations that include GC formation will represent a powerful,
game-changing combination of tools to advance the understanding of how galaxies grow and evolve.

\acknowledgements

The authors thank the anonymous referee for insightful comments that have helped to improve the 
manuscript. The authors also thank Marta Volonteri for useful comments to the manuscript. R. D'A. acknowledges 
the SI competitive research grant "A New Astro-Archeology 
Probe of the Merging Evolution of Galaxies" for support. R. D'A.'s work  
was also partially supported by the {\it Chandra} X-ray Center (CXC), which is 
operated by the Smithsonian Astrophysical Observatory (SAO) under NASA contract NAS8-03060.
AZ acknowledges funding from the European Research Council under the European 
Union's Seventh Framework Programme (FP/2007-2013)/ERC Grant Agreement n. 617001.

\begin{appendix}

\section{Comparison with other density estimation methods}

In order to confirm the effectiveness of our method and its ability to detect over-density structures in the 
projected spatial distribution of GCs, we have compared the residual maps 
obtained with our KNN-based technique with residual maps derived with a multidimensional 
Gaussian Mixture Model (GMM) technique~\citep{dempster1977} and the Kernel Density Estimation 
(KDE) method~\citep{rosenblatt1956,parzen1962}.

The GMM method approximates the density function of the observed points with 
a combination of multiple, independent, multi-dimensional Gaussian distributions, where each component
represents a different group or subpopulation. In this paper, we have used the {\it Mclust} 
implementation~\citep{fraley2012} of the GMM technique, written in {\it R}~\citep{R2014}. In this 
implementation, the number of mixture model Gaussian
components can be either specified by the user or determined automatically by the algorithm with an EM
approach. {\it R} is an open-source free statistical environment developed under the GNU 
GPL ({\tt \url{http://www.r-project.org}}). An application of the {\it Mclust} algorithm to the problem 
of the detection of substructures in the velocity distribution of a sample of galaxies is discussed 
by~\cite{einasto2012}. 

The KDE method is a non-parametric 
density estimation technique which reconstructs the unknown density function of a random variable 
using a kernel with a typical smoothing parameter. This parameter, the bandwidth, is estimated locally 
and depends on the local topology of the distribution of points whose density function is reconstructed. 

We have applied the KNN, GMM and KDE methods for the determination of the density maps to the 
observed distribution of ACSVCS GCs in NGC4472 (see Section~\ref{subsec:ngc4472}), and 
to a simulated distribution of GCs where distinct components are visible. Density 
maps of the two GCs distributions have been determined in both cases. The other steps for the calculation 
of the residual maps from the density of the observed GCs and the model simulated GC distribution were 
unchanged. 

\subsection{Spatial distribution of GCs in NGC4472}

\begin{figure*}[]
	\centering
	\includegraphics[height=5.5cm,width=5.5cm,angle=0]{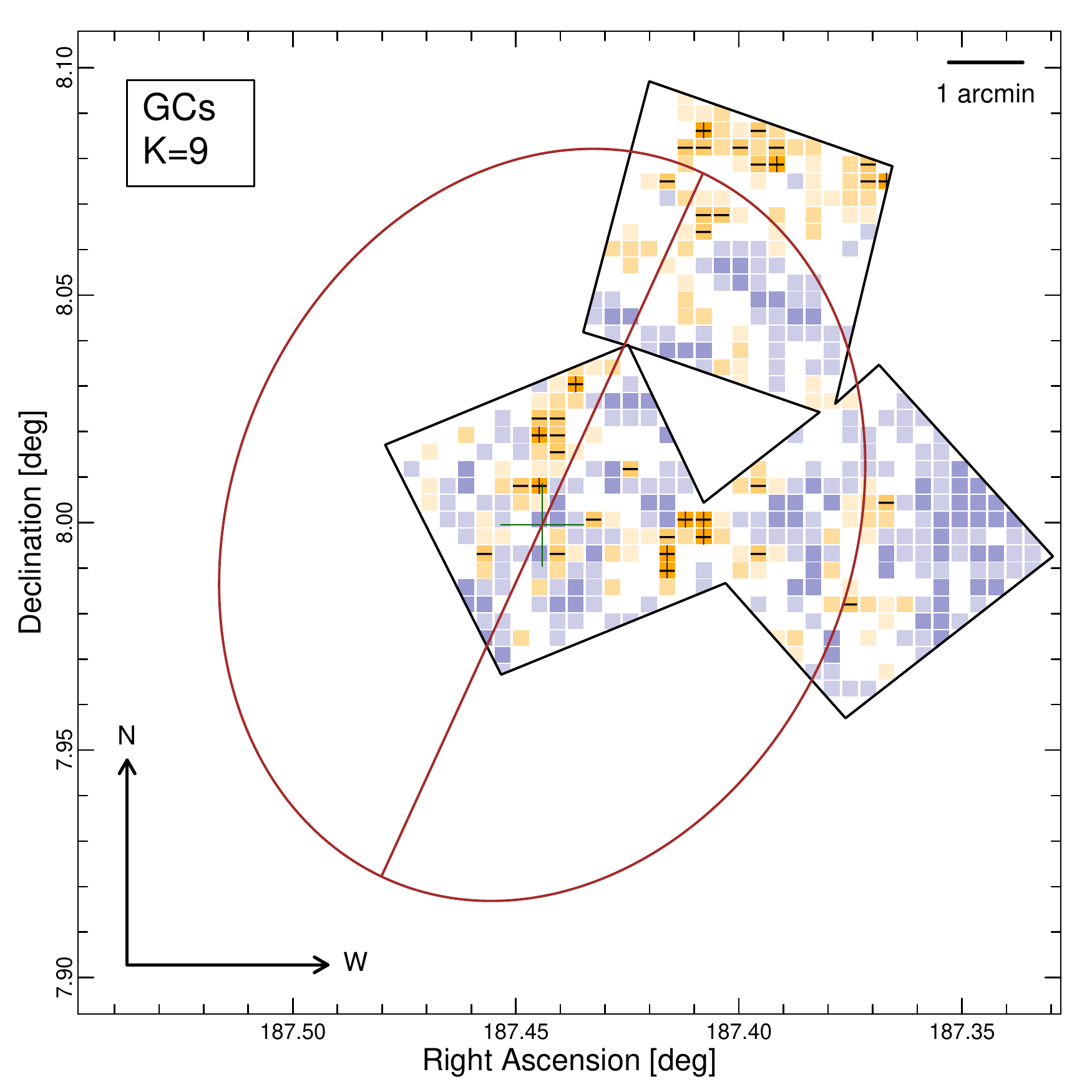}
	\includegraphics[height=5.5cm,width=5.5cm,angle=0]{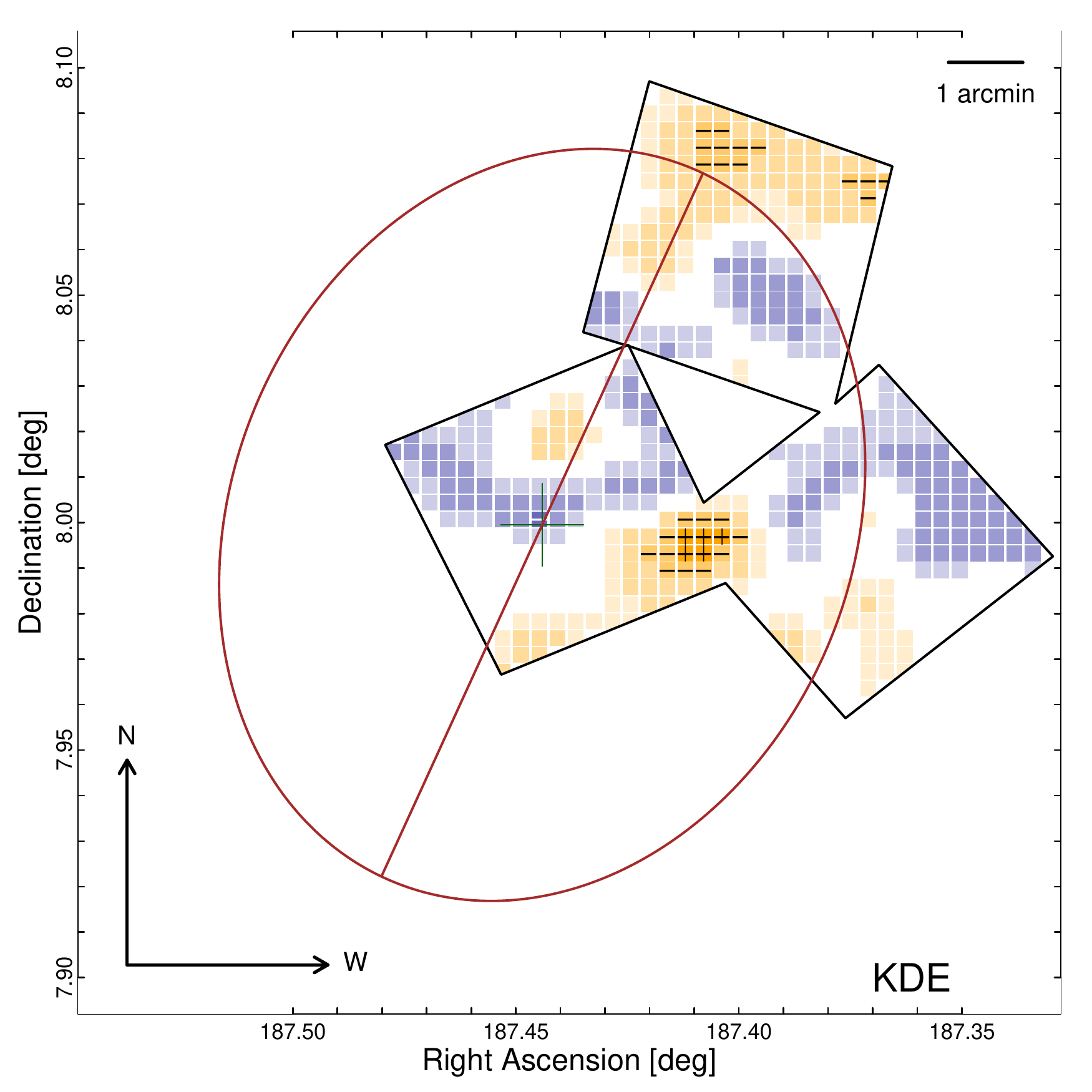}\\	
	\includegraphics[height=5.5cm,width=5.5cm,angle=0]{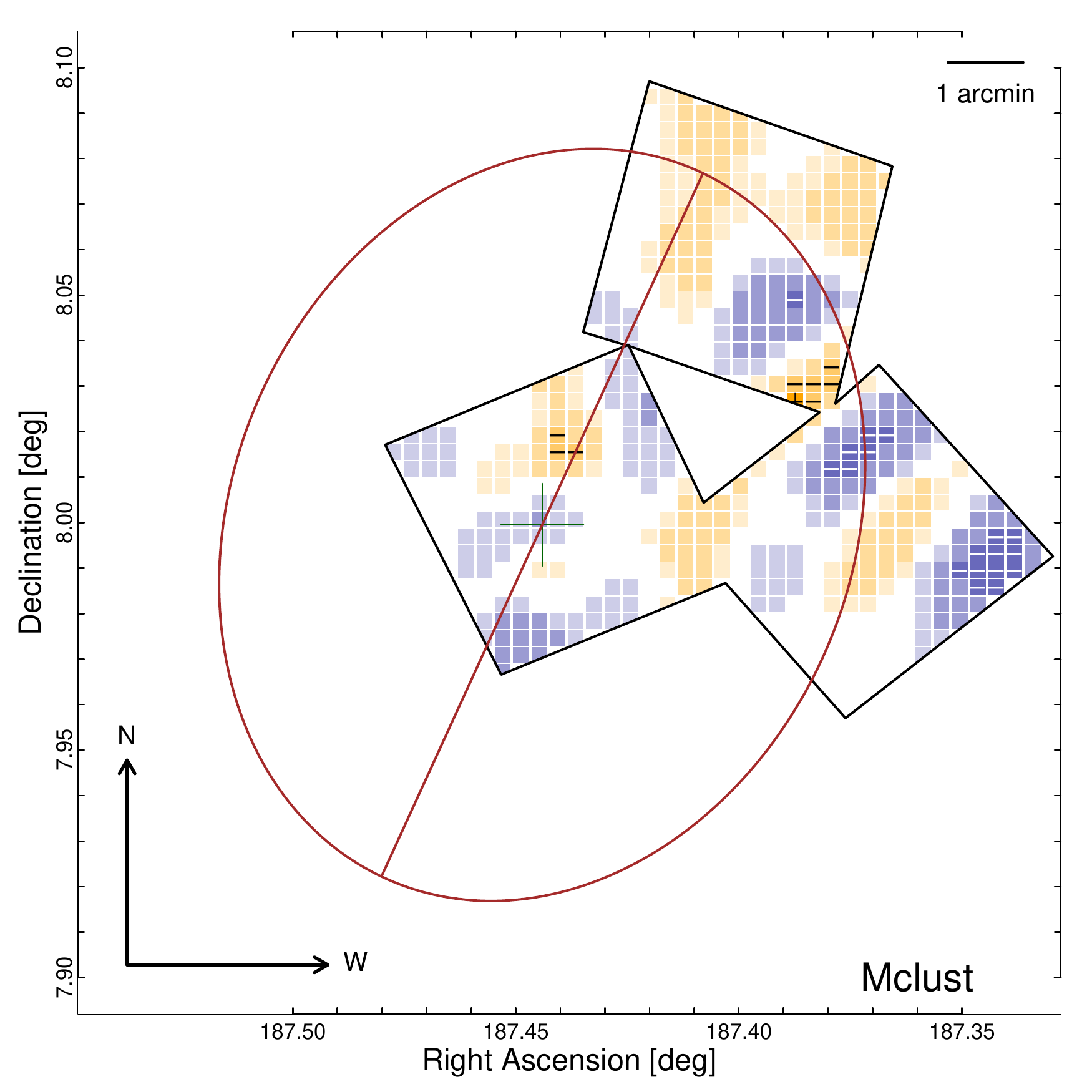}
	\includegraphics[height=5.5cm,width=5.5cm,angle=0]{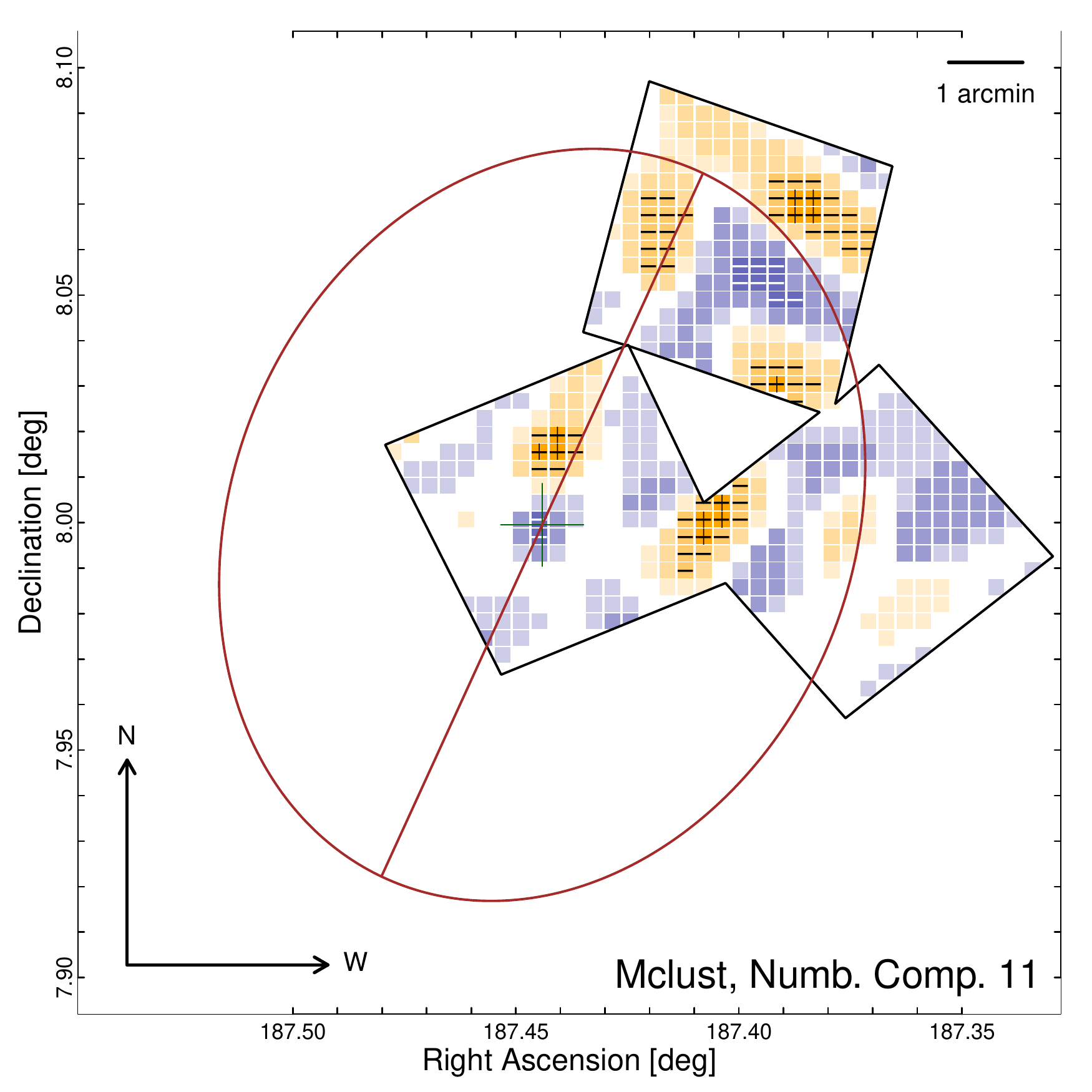}
	\caption{{ Residual maps of the GC distribution in NGC4472 obtained with different density estimation
	methods: upper left) KNN with $K\!=\!9$; upper right) KDE; lower left) {\it Mclust} with number of 
	mixture model components automatically determined; lower right) {\it Mclust} with optimal number
	of components fixed to 11.}}
	\label{fig:methods_ngc4472}
\end{figure*}

The results of the application of the three density estimation techniques to the spatial 
distribution of the observed  
GCs distribution in NGC4472 are shown in Figure~\ref{fig:methods_ngc4472}. The upper left plot 
shows the residual maps obtained with the KNN method ($K\!=\!9$) described in 
Section~\ref{subsec:ngc4472}, the upper right plot is the residual map obtained using the KDE, 
while the lower panels show the residual maps obtained using the {\it Mclust} method where the
number of components of the mixture model is determined automatically by the algorithm (left), 
or fixed to the optimal value\footnote{We have derived the residual maps for number of 
components varying between 3 and 30, and selected the map the most closely resembles
the KNN residual map}(right). The four residual maps are in substantial agreement on the overall 
number and positions of the GC structures detected by the KNN method, while the spatial extension 
varies significantly. The KDE and {\it Mclust} methods produce ``smoother'' and more spatially 
extended structures than the KNN method. 

\subsection{Simulated spatial distribution of GCs}

We have also applied the KNN, KDE and {\it Mclust} methods to a simulated distribution of 
GCs (Figure~\ref{fig:methods_simulation}) consisting in three different components: an 
azimuthally smooth distribution of GCs with radial positions randomly drawn from 
a power-law peaking in the center of the galaxy; three clusters, modeled as two-dimensional 
Gaussians with different variances along the R.A. and Dec. axes, radial and 
azimuthal positions relative to the center of the galaxy, and one ``streamer'', modeled as a 
uniform azimuthal and radial distribution of GCs within a circular sector with 40$\deg$ aperture. 
The total number of simulated GCs is 1750, split in 1500 GCs belonging to the smooth distribution, 
100 GCs in the ``streamer'' and 50 GCs for each cluster. The total number of GCs and the number 
of GCs in each spatial component have been set to yield average GC densities and GC density
contrasts in the over-density structures comparable to the observed values in the Virgo galaxies 
investigated in this paper (see Table~\ref{tab:galaxies} for comparison). As a result, the number
of excess GCs per pixels in each spatial structure of the simulated distribution is similar to 
the average number of excess GCs in the structures observed in the galaxies investigated. 
The spatial distribution of the simulated components are color-coded in the upper left panel in 
Figure~\ref{fig:methods_simulation}. As in the case of the NGC4472 GC system, the residuals maps 
produced with the three methods are similar: the two clusters E and N-W of the center of the 
galaxy are recovered in each map as well as the streamer (which, in the case of the KDE, has a 
rounded shape reflecting the use of a Gaussian as kernel function). The third cluster located 
just S of the center of the galaxy is associated to lesser over-density structures in the KNN and 
{\it Mclust} residual maps but it is missed by the KDE. Both the KDE and {\it Mclust} maps show
artifacts structures located in the corners of the field and due to border effects, which are missing 
in the map produced by the KNN. 

\subsection{Conclusions}

\begin{figure*}[]
	\centering
	\includegraphics[height=5.5cm,width=5.5cm,angle=0]{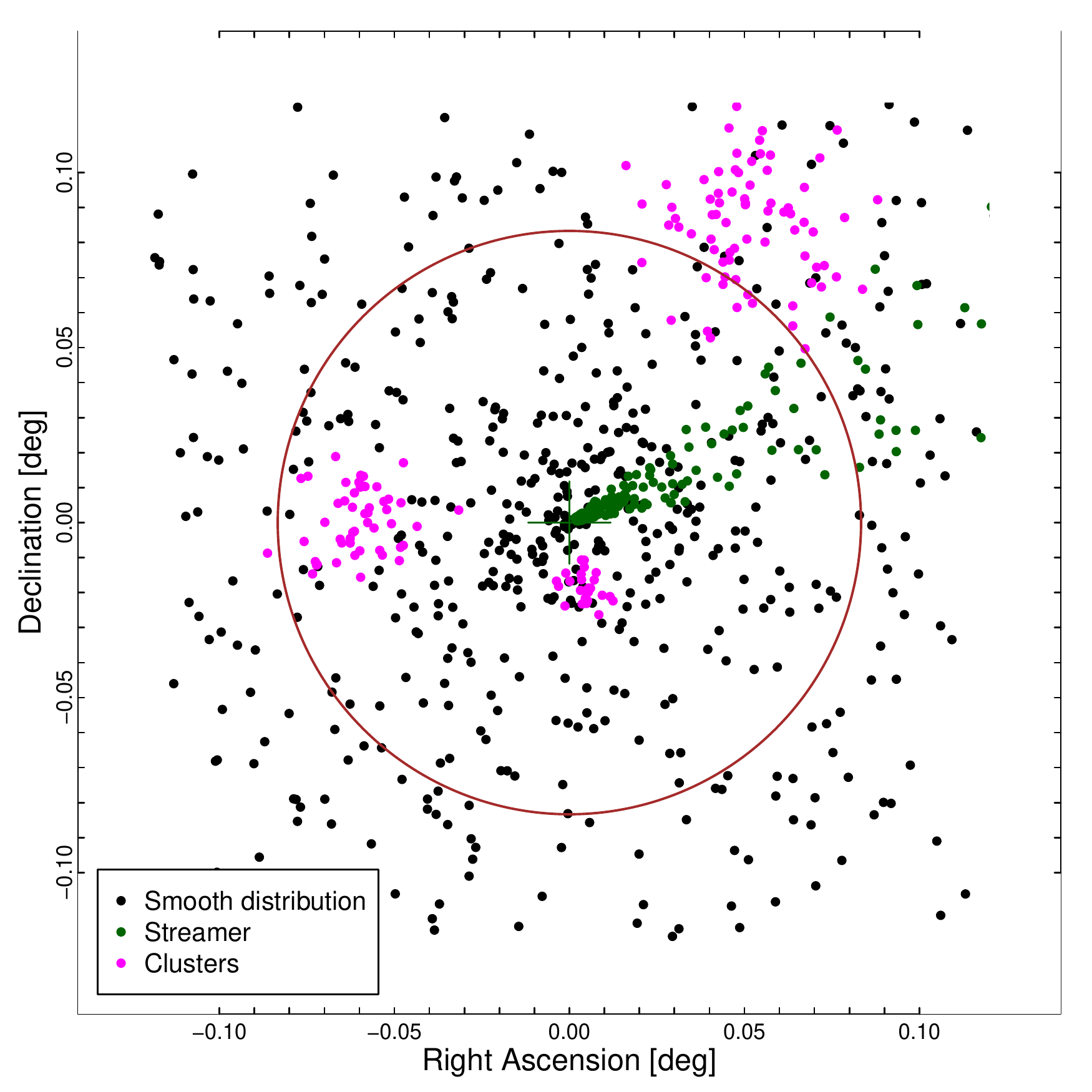}
	\includegraphics[height=5.5cm,width=5.5cm,angle=0]{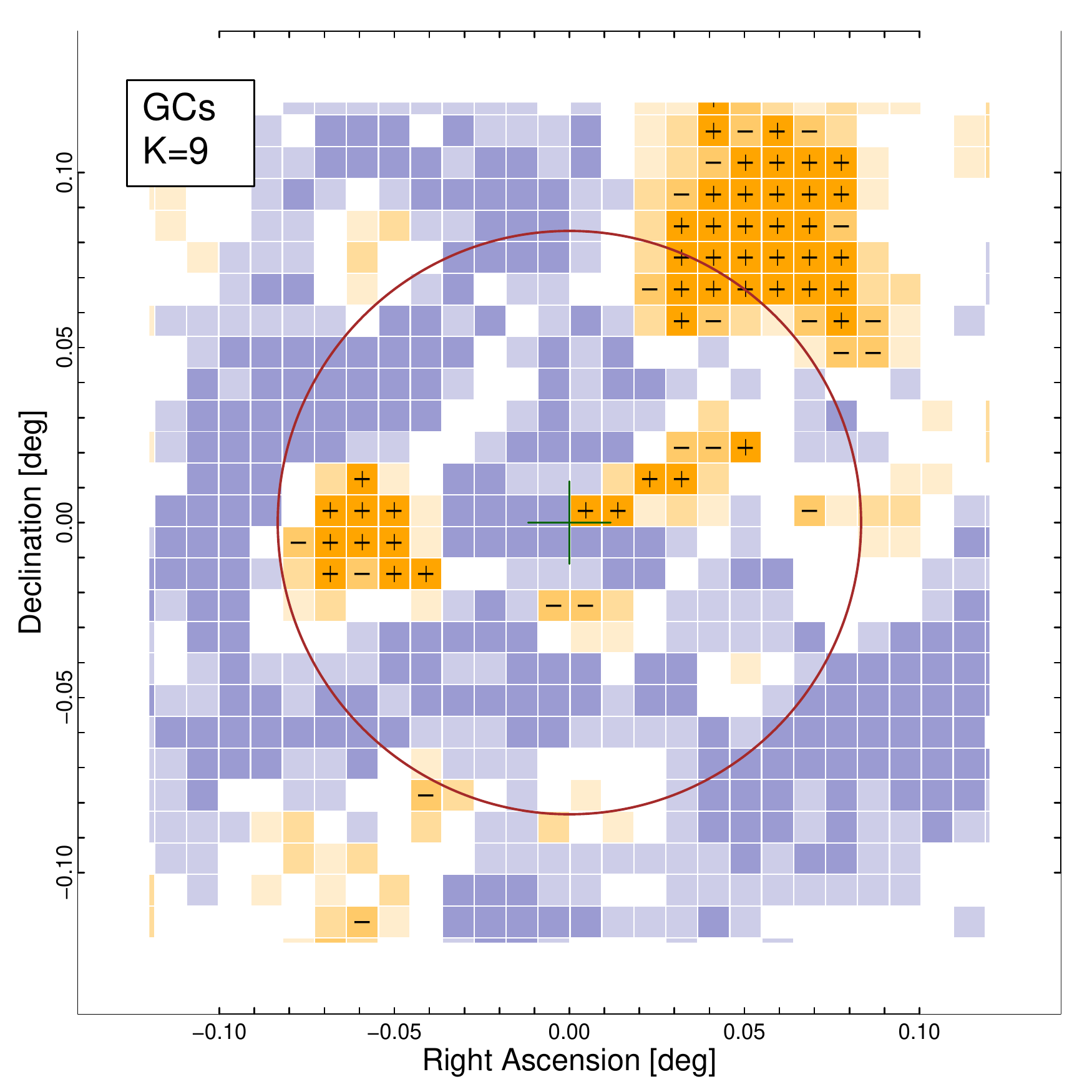}\\	
	\includegraphics[height=5.5cm,width=5.5cm,angle=0]{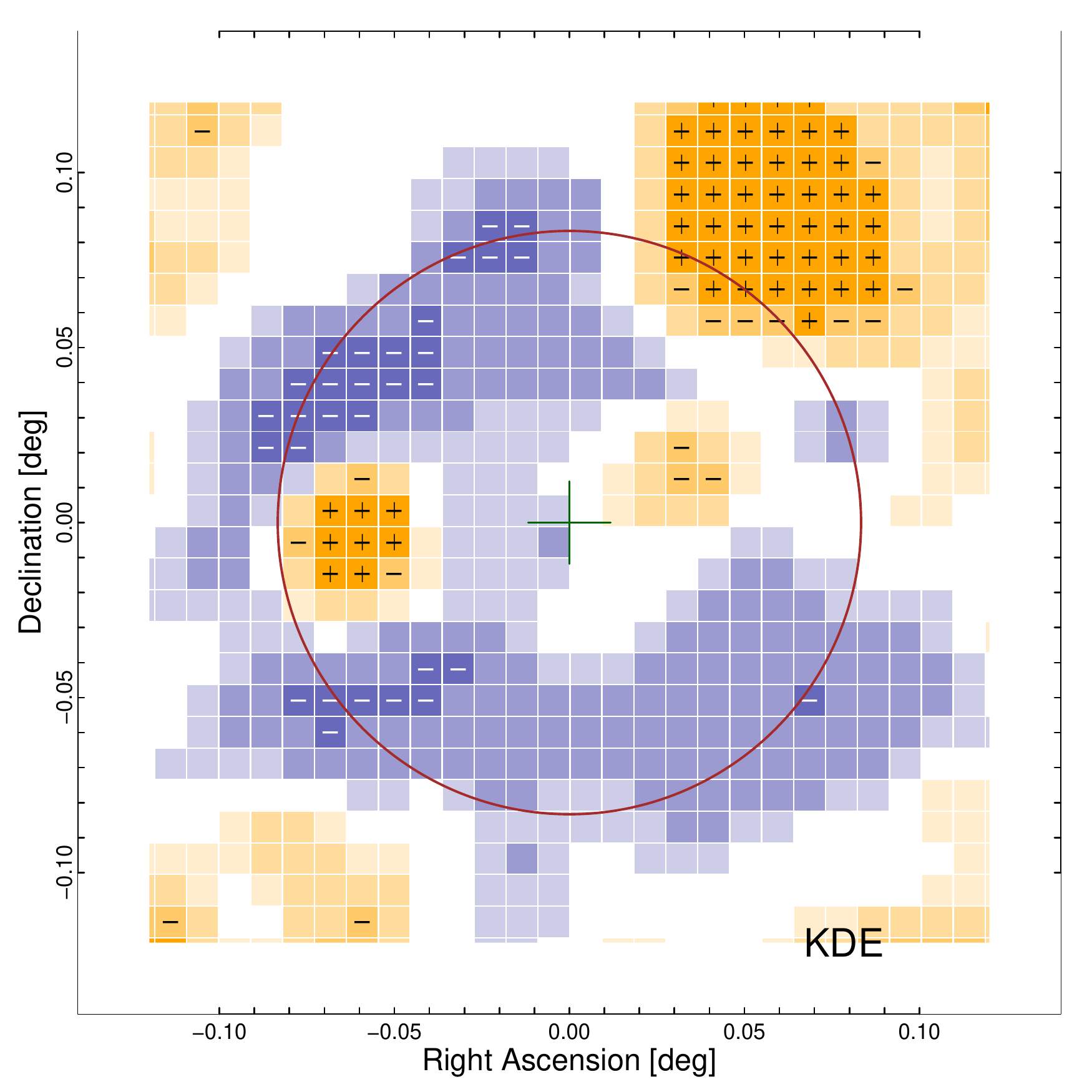}
	\includegraphics[height=5.5cm,width=5.5cm,angle=0]{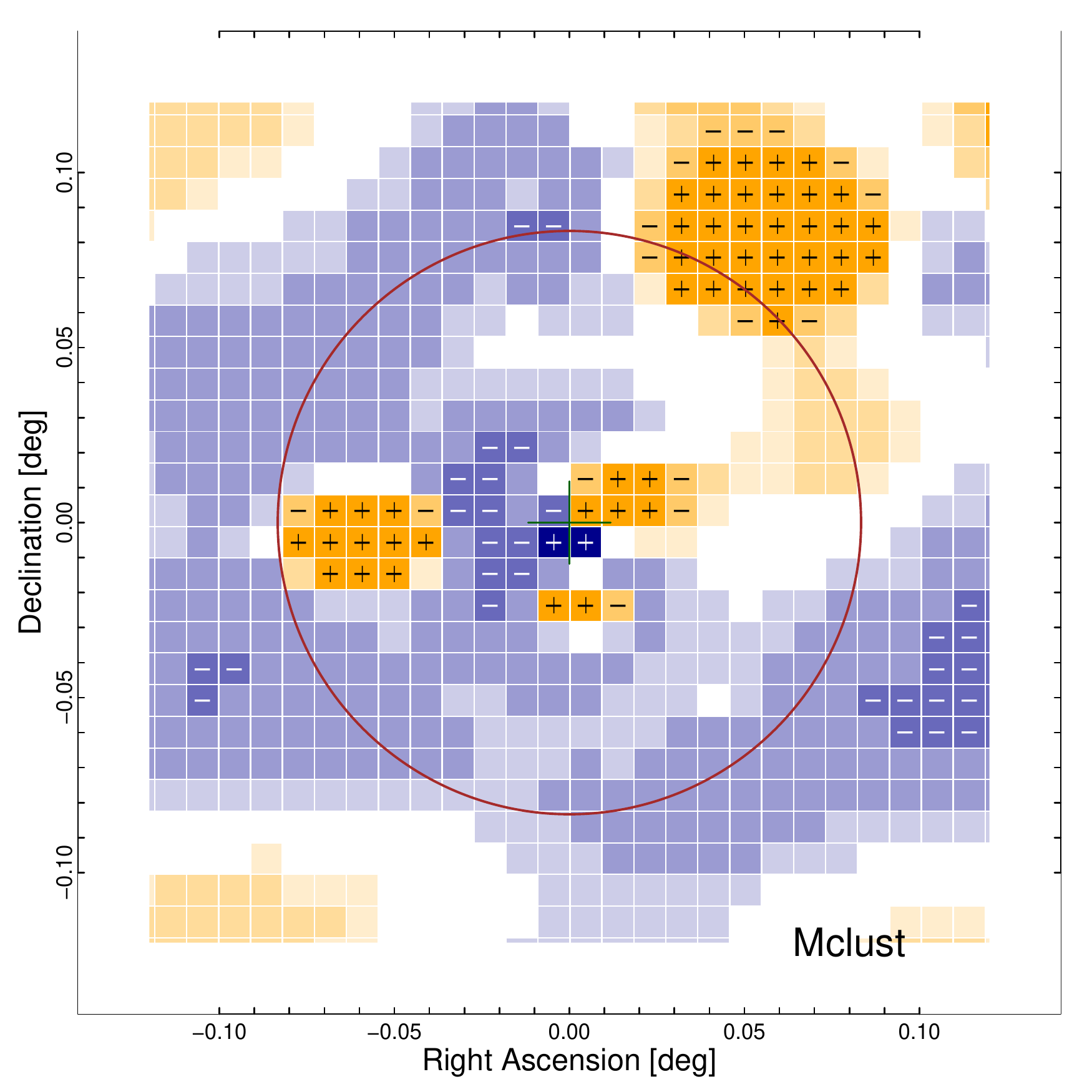}
	\caption{{ Scatterplot of a simulated distribution of GCs with three distinct components (upper left) and 
	residual maps obtained with the $K\!=\!9$ KNN method (upper right), the KDE (lower left) and the {\it Mclust}
	method with number of mixture model components automatically determined (lower right).}}
	\label{fig:methods_simulation}
\end{figure*}

The three methods recover the same GC structures in both examples, with similar spatial and
statistical properties. The shapes of the over-densities reconstructed with the GMM and KDE 
techniques, in general, have smoother boundaries than those of the KNN-determined structures
because the two former methods tend to spread the small-scale over-densities (often driven by 
GC excesses in one or few pixels) over larger areas. Moreover, GMM and KNN ``bridge over'' 
sharp pixel-by-pixel changes in the GC density, and the reconstructed final shapes are not 
representative, on spatial scales of few pixels, of the real geometry of the structures. The significances 
of the structures have been measured independently 
in the residual maps shown in Figures~\ref{fig:methods_ngc4472} and~\ref{fig:methods_simulation}
as described in Section~\ref{sec:method}. Overall, the significances of all GC structures in both GC
distributions differ by less than 1.8$\sigma$ and 0.8$\sigma$ respectively, with negligible overall bias in the 
their distribution. The overall similarity of the significances indicates that the KNN method  
provides robust estimates of the real statistical significances of the GC structures. The small fluctuations
measured are likely due to the different spatial extensions of the GC structures detected in the different 
residual maps. For instance, the lack of $\geq3\sigma$ pixels in the map of the NGC 4472 GCs spatial 
distribution in Figure~\ref{fig:methods_ngc4472} produced with the unconstrained {\it Mclust} 
algorithm is balanced by the larger areas occupied by the structures. The combined effect of singe pixels 
significance and total extension and shape of the structures produced 
similar statistical significances for all methods tested. 

Our conclusions is that, overall, the properties
(location, significance and shape) of the GC residual structures obtained with the KNN method, 
are robust and in substantial agreement with the features of the GC structures detected using 
different density estimation methods.

\end{appendix}

{}
\end{document}